\documentclass[fleqn,usenatbib]{mnras}

\usepackage[T1]{fontenc}

\DeclareRobustCommand{\VAN}[3]{#2}
\let\VANthebibliography\thebibliography
\def\thebibliography{\DeclareRobustCommand{\VAN}[3]{##3}\VANthebibliography}

\usepackage{graphicx}	
\usepackage{amsmath}	
\usepackage{amssymb}	

\usepackage{tikz}
\usetikzlibrary{shapes.geometric, arrows, decorations.markings}

\usetikzlibrary{fit,backgrounds,calc}

\tikzstyle{Step} = [rectangle, rounded corners, minimum width=3cm, minimum height=1cm,text centered, draw=black, text width=3cm, fill=red!25]
\tikzstyle{Hypothesis} = [ellipse, minimum width=1.5cm, minimum height=1cm,text centered, draw=black, fill=blue!25]
\tikzstyle{Model} = [diamond, minimum width=1cm, minimum height=1cm,text centered, draw=black, fill=green!25]
\tikzstyle{Data} = [rectangle, rounded corners, minimum width=1cm, minimum height=1cm,text centered, draw=black, text width=3cm, fill=red!25]
\tikzstyle{Prediction} = [ellipse, minimum width=1cm, minimum height=1cm, text centered, draw=black, fill=blue!25]
\tikzstyle{Article} = [rectangle, minimum width=1cm, minimum height=1cm, text centered,  fill=yellow!25]

\tikzstyle{arrow} = [thick,->,>=stealth]

\newcommand{\Msol}{\ensuremath{M_{\odot}}}
\newcommand{\sole}{\ensuremath{_{\odot}}}

\usepackage[greek, english]{babel}

\usepackage{mathtools, cuted}

\usepackage{scalerel} 

\def\scaleint#1{\vcenter{\hbox{\scaleto[3ex]{\displaystyle\int}{#1}}}}
\def\bs{\mkern-18mu} 

\newcommand{\orcid}[1]{\href{https://orcid.org/#1}{\includegraphics[width=10pt]{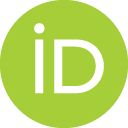}}}

\usepackage{newtxtext,newtxmath}

\newcommand{\describecolours}{\textit{Blue:} Best ICM-optimised profile, with an \texttt{idPIE} model. \textit{Red:} Best lens model inferred profile. }

\title[Strong lensing reconstruction of the ICM]{A full reconstruction of two galaxy clusters intra-cluster medium with strong gravitational lensing} 

\author[Joseph F. V. \textsc{Allingham} et al.]{
Joseph F. V. \textsc{Allingham}\orcid{0000-0003-2718-8640}$^{1,2}$\thanks{E-mail: \href{mailto:ajoseph@campus.technion.ac.il}{ajoseph@campus.technion.ac.il}},
C\'eline \textsc{B\oe{}hm}\orcid{0000-0002-5074-9998}$^{1}$,
Dominique \textsc{Eckert}\orcid{0000-0001-7917-3892}$^{3}$,
Mathilde \textsc{Jauzac}\orcid{0000-0003-1974-8732}$^{4,5,6,7}$, 
\newauthor
David \textsc{Lagattuta}\orcid{0000-0002-7633-2883}$^{4,5}$, 
Guillaume \textsc{Mahler}\orcid{0000-0003-3266-2001}$^{4,5,8}$, 
Matt \textsc{Hilton}\orcid{0000-0002-8490-8117}$^{6,7}$, 
Geraint F. \textsc{Lewis}\orcid{0000-0003-3081-9319}$^{1}$,
\newauthor 
and Stefano \textsc{Ettori}\orcid{0000-0003-4117-8617}$^{8,9}$
\\
$^{1}$School of Physics, A28, The University of Sydney, NSW 2006, Australia;\\ 
$^{2}$Physics Department, Technion, 3200003 Haifa, Israel;\\
$^{3}$Department of Astronomy, University of Geneva, ch. d’Écogia 16, CH-1290 Versoix Switzerland;\\
$^{4}$Centre for Extragalactic Astronomy, Durham University, South Road, Durham DH1 3LE, UK;\\
$^{5}$Institute for Computational Cosmology, Durham University, South Road, Durham DH1 3LE, UK;\\
$^{6}$Wits Centre for Astrophysics, School of Physics, University of the Witwatersrand, Private Bag 3, 2050, Johannesburg, South Africa;\\
$^{7}$Astrophysics Research Centre, School of Mathematics, Statistics \& Computer Science, University of KwaZulu-Natal, Durban 4041, South Africa;\\
$^{8}$STAR Institute, Quartier Agora - All\'ee du six Ao\^ut, 19c B-4000 Li\`ege, Belgium;\\
$^{9}$INAF - Osservatorio di Astrofisica e Scienza dello Spazio di Bologna, via Piero Gobetti 93/3, 40129 Bologna, Italy;\\
$^{10}$INFN, Sezione di Bologna, viale Berti Pichat 6/2, 40127 Bologna, Italy.
}

\date{Accepted 2024 January 08. Received 2023 December 22; in original form 2023 September 13}

\pubyear{2024}

\begin{document}
\label{firstpage}
\pagerange{\pageref{firstpage}--\pageref{lastpage}}
\maketitle

\begin{abstract}
Whilst X-rays and Sunyaev-Zel'dovich observations allow to study the properties of the intra-cluster medium (ICM) of galaxy clusters, their gravitational potential may be constrained using strong gravitational lensing.
Although being physically related, these two components are often described with different physical models.
Here, we present a unified technique to derive the ICM properties from strong lensing for clusters in hydrostatic equilibrium.
In order to derive this model, we present a new universal and self-similar polytropic temperature profile, which we fit using the X-COP sample of clusters.
We subsequently derive an analytical model for the electron density, which we apply to strong lensing clusters MACS\,J0242.5-2132 and MACS\,J0949.8+1708.
We confront the inferred ICM reconstructions to \textit{XMM-Newton} and ACT observations.
We contrast our \textit{analytical} electron density reconstructions with the best canonical $\beta$-model.
The ICM reconstructions obtained prove to be compatible with observations.
However they appear to be very sensitive to various dark matter halo parameters constrained through strong lensing (such as the core radius), and to the halo scale radius (fixed in the lensing optimisations).
With respect to the important baryonic effects, we make the sensitivity on the scale radius of the reconstruction an asset,
and use the inferred potential to constrain the dark matter density profile using ICM observations.
The technique here developed should allow to take a new, and more holistic path to constrain the content of galaxy clusters.
\end{abstract}

\begin{keywords}
galaxies: clusters: intracluster medium, gravitational lensing: strong, X-rays: galaxies: clusters, galaxies: clusters: individual: MACS\,J0242.5-2132, galaxies: clusters: individual: MACS\,J0949.8+1708, hydrodynamics
\end{keywords}



\section{Introduction}

In the last decades, tremendous progress has been achieved in gravitational lensing observations \citep[see][for a review]{Kneib_2011}; from the first mass reconstruction of Abell\,370 \citep{1987hrpg.work..467H, 1988A&A...191L..19S} all the way to the \textit{Hubble} Frontier Fields \citep[HFF,][]{Lotz_2017}, the Beyond the Ultra-deep Frontier Fields and Legacy Observations \citep[BUFFALO,][]{Steinhardt_2020}, and the numerous \textit{JWST} lensing surveys \citep[see e.g. UNCOVER,][]{2022arXiv221204026B}.
As a result, our understanding of this indirect observation of dark matter has improved, yet leaving open problems to discussions, such as the total matter (baryons \& dark matter) potential distribution in galaxy clusters \citep[][]{2018MNRAS.476.2086L, 2021arXiv210913284R}, in particular in their outskirts \citep{2017MNRAS.471L..47T}, or the size of clusters themselves \citep[see e.g.][]{2018ApJ...864...83C, 2020MNRAS.499.1291T, 2021MNRAS.508.1777B, 2023MNRAS.521.3981A}.

In parallel, the total matter distribution at large radii is assumed to be traced by the ionised intergalactic medium, the signature of which is detectable in the X-rays, and thanks to the Sunyaev-Zel'dovich (SZ) effect.
While the projected lensing gravitational potential of galaxy clusters can be reconstructed from X-rays and SZ data through the Richardson-Lucy deprojection algorithm \citep[see][]{2013A&A...553A.118K, 2015A&A...584A..63S, Tchernin_2018}, the reverse path has not been explored yet.
Efforts to relate the two observables range from \citet{2010ApJ...720.1038B}, describing the intra-cluster medium (ICM, composed of ionised gas) density through a dark matter (DM) profile and inferring the potential using the X-ray observation, to CLUMP-3D \citep[see][]{2013MNRAS.428.2241S, 2017MNRAS.467.3801S}, exploiting the triaxial hypothesis to perform a joint ICM-lensing optimisation of the cluster physics, using disjoint models for the potential and the ICM thermodynamics.

We propose to predict the ICM thermodynamics using only the gravitational lensing-inferred potential.
This should light our way towards a more holistic understanding of the dark matter profile, galaxy clusters thermodynamics, and interplay between baryons and dark matter.
For instance, an offset between the lensing prediction and the X-ray observations could for instance hint towards interacting dark matter scenarios.
In order to establish such a comparison, we convert the strong lensing detected gravitational potential into a predictive ICM model.

Deriving such a model requires to understand the thermodynamics of galaxy clusters.
Making precise measurements from X-ray observations is limited by a series of assumptions, about e.g. the halo geometry \citep{2012MNRAS.421.1399B, 2017MNRAS.467.3801S}, or the dynamical state of the cluster \citep{2014ApJ...782..107N, 2016ApJ...827..112B}.
Multiple studies have shown the hydrostatic regime to be an acceptable description of the ICM for cool-core and relaxing clusters \citep{2013SSRv..177..119E, 2016ApJ...827..112B, 2018MNRAS.481L.120V, 2019A&A...621A..39E}. 
Conversely, recent mergers or dynamically disturbed systems present strong deviations to the hydrostatic equilibrium \citep{2013ApJ...767..116M}.
Moreover, galaxy clusters have followed a hierarchical model of formation, made of mergers and gravitational collapse. 
For this reason, their thermodynamics scales according to the cluster mass \citep{1986MNRAS.222..323K, 1998ApJ...495...80B}, which is confirmed by simulations \citep{1999ApJ...525..554F, 2005MNRAS.361..233B, 2005RvMP...77..207V} and observations \citep{2019A&A...621A..41G}.
These assumptions (hydrostatic equilibrium, self-similarity) are common in joint X-rays and SZ analyses \citep[cf.][]{Capelo_2012, 2019A&A...621A..41G, Ghirardini_2019_polyt}.

Using these assumptions, we adopted an effective polytropic temperature law in order to describe the thermodynamic model of the ICM of galaxy clusters \citep[following e.g.][]{2001MNRAS.327.1353K}.
\citet{Capelo_2012} predicted a constant $\Gamma \sim 1.2$ polytropic index for the ICM in hydrostatic equilibrium with a NFW density profile, and \citet{Ghirardini_2019_polyt} recovered this value for the outskirts of clusters, but found radiative cooling to bring this value to $\Gamma \sim 0.8$ in the centre.
In order to produce precise predictions \textit{a priori},
we conduct a study of the polytropic index on the X-COP sample of data \citep[described in][]{2017AN....338..293E}. 
X-COP is a sample of 12 massive clusters selected from the \textit{Planck} all-sky survey for which a deep X-ray follow-up with \textit{XMM-Newton} was conducted \citep[see][]{2019A&A...621A..41G, 2021A&A...650A.104C}.
The thermodynamic properties of the ICM (pressure, temperature, density) were recovered over a broad radial range, which makes this sample ideal to derive the relation between the various thermodynamic quantities.

In \citet{2023MNRAS.522.1118A}, we analysed two galaxy clusters and reconstructed their gravitational potential with \textsc{Lenstool} \citep[see][]{2007NJPh....9..447J} using strong gravitational lensing. 
Galaxy clusters MACS\,J0242.5-2132 and MACS\,J0949.8+1708, dynamically relaxed and relaxing respectively, provide the inputs to the ICM predictions for this work, and allow to justify the hydrostatic description of the ICM \citep{2016ApJ...827..112B}.
In this paper, we also test the $\beta$-model \citep[see][]{King_1966}, commonly used by the X-ray and SZ communities to describe the ICM density distribution.
We refer to it, and to the family of empirical models introduced in e.g. \citet{Vikhlinin_2006} as \textit{canonical}, in contrast to our models, which are derived analytically from the full matter density, using the Poisson and Euler equations, following the logic of \citet{2010ApJ...720.1038B}.
As our \textit{analytical} ICM models scale with the gravitational potential obtained with strong lensing, we directly work with the parameters of the lensing model.

After establishing the theoretical models, the quantitative ICM results are confronted to the \textit{XMM-Newton} and the ACT Data Release 5 millimetre-wave \citep[see][]{Naess_2020, 2021ApJS..255...11M}.
The quality of the reconstruction is tested with a MCMC on the density parameters, using these ICM data.

This article is structured as follows: the data are presented in Section\,\ref{sec:Data}; the strong lensing models are summarised in Section\,\ref{sec:lensing_models}; the theoretical possible models for the electron density, the temperature, the gas fraction, the X-ray surface brightness and the SZ effect are introduced in Section\,\ref{sec:Theory}; quantitative results for the density and temperature are presented in Section\,\ref{sec:quantitative_results}; the method to evaluate the quality of observational predictions follows up in Section\,\ref{sec:Method_ThePredictions}; ICM predictions and MCMC optimisation results using the ICM observations are detailed in Section\,\ref{sec:ICM-Optimised_Results}; a discussion on the limitations and possibilities of such a model is given in Section\,\ref{sec:Discussion}; and a summary and conclusion are provided in Section\,\ref{sec:Conclusion}.
We assume the $\Lambda$CDM cosmological model, with $\Omega_m=0.3$, $\Omega_{\Lambda}=0.7$, and $H_{0}=70$\,km.s$^{-1}$.Mpc$^{-1}$.

\section{Data}
\label{sec:Data}

\subsection{X-ray observations}

\subsubsection{MACS\,J0242 and MACS\,J0949}
\label{subsubsec:XMM-Newton_analysis}

The ICM is primarily probed with X-ray observations. We used the \textit{XMM-Newton} publicly available observations of the MACS\,J0242 and MACS\,J0949 in the 0.7-1.2\,keV band \citep[see][]{2021A&A...650A.104C}.
MACS\,J0242 was observed for a total of 70\,ks (OBSID:0673830101), and MACS\,J0949 for a total of 36\,ks (OBSID:0827340901). We analysed the two observations using \textsc{XMMSAS v17.0}, and the most up-to-date calibration files. We used the XMMSAS tools \texttt{mos-filter} and \texttt{pn-filter} to extract light curves of the observations and filter out periods of enhanced background, induced by soft proton flares. After flare filtering, the available clean exposure time is 61\,ks (MOS) and 53\,ks (PN) for MACS\,J0242, and 35\,ks (MOS) and 34\,ks (PN) for MACS\,J0949. 
The EPIC MOS filter maximises the signal-to-noise ratio, thus we used primarily these data.
We extract the X-ray data following the procedure detailed through \citet[][]{2019A&A...621A..41G}, and the hydrostatic mass through \citet{2022A&A...662A.123E}.

With the NASA tool \href{https://heasarc.gsfc.nasa.gov/cgi-bin/Tools/w3nh/w3nh.pl}{PIMMS}, we get access to the conversion constants from flux to counts per second $C^{\mathrm{count}}_{\mathrm{flux}}$ for both clusters: for MACS\,J0242 and MACS\,J0949, $C^{\mathrm{count}}_{\mathrm{flux}} = 2.087 \times 10^{11}$ and $2.084 \times 10^{11}$ counts.erg$^{-1}$.cm$^{2}$ respectively\footnote{The temperature of the ICM of both clusters being $> 3$\,keV, we can neglect the influence of temperature on the conversion constant.}.

\subsubsection{The X-COP clusters}

In order to tune our temperature and gas fraction models, we study a number of comparable clusters. The \textit{XMM} cluster outskirts project \citep[X-COP\footnote{\url{https://dominiqueeckert.wixsite.com/xcop}}, described in][]{2017AN....338..293E}
is ideal for this purpose: it gathers data from 12 massive clusters.
These clusters are comparable to MACS\,J0242 and MACS\,J0949, with $3 \times 10^{14} \Msol < M_{500} < 1.2 \times 10^{15} \Msol$, but are in the redshift range $0.04 < z < 0.1$, smaller than for MACS\,J0242 and MACS\,J0949, at redshifts $0.313$ and $0.383$ respectively.

\subsection{SZ observations}

It is possible to study the `imprint' of the ICM on the CMB through the SZ effect, which is seen as a deficit of CMB photons in the direction of clusters when observed at frequencies less than 217\,GHz.
With the Atacama Cosmology Telescope (ACT), we use the f090 and f150 `daynight' DR5 maps \footnote{\url{https://lambda.gsfc.nasa.gov/product/act/actpol_dr5_coadd_maps_get.html}}, centred on respective frequencies 97.8 and 149.6\,GHz \citep[see][]{2021ApJS..253....3H, 2021ApJS..255...11M}.

The BCG of MACS\,J0242 is a bright radio source, whose spectral energy distribution (SED) is detailed by \citet{10.1093/mnras/stv1518}.
At 1\,GHz, the source peaks at 1\,Jy, making the extraction of the ICM signal with the SZ effect impossible.
We thus only exploit the MACS\,J0949 data in this article.

\section{Strong Lensing Analyses of \texorpdfstring{MACS\,J0242 \& MACS\,J0949}{TEXT}}
\label{sec:lensing_models}

\begin{figure*}
\centering
\begin{minipage}{0.48\textwidth}
\centerline{\includegraphics[width=1\textwidth]{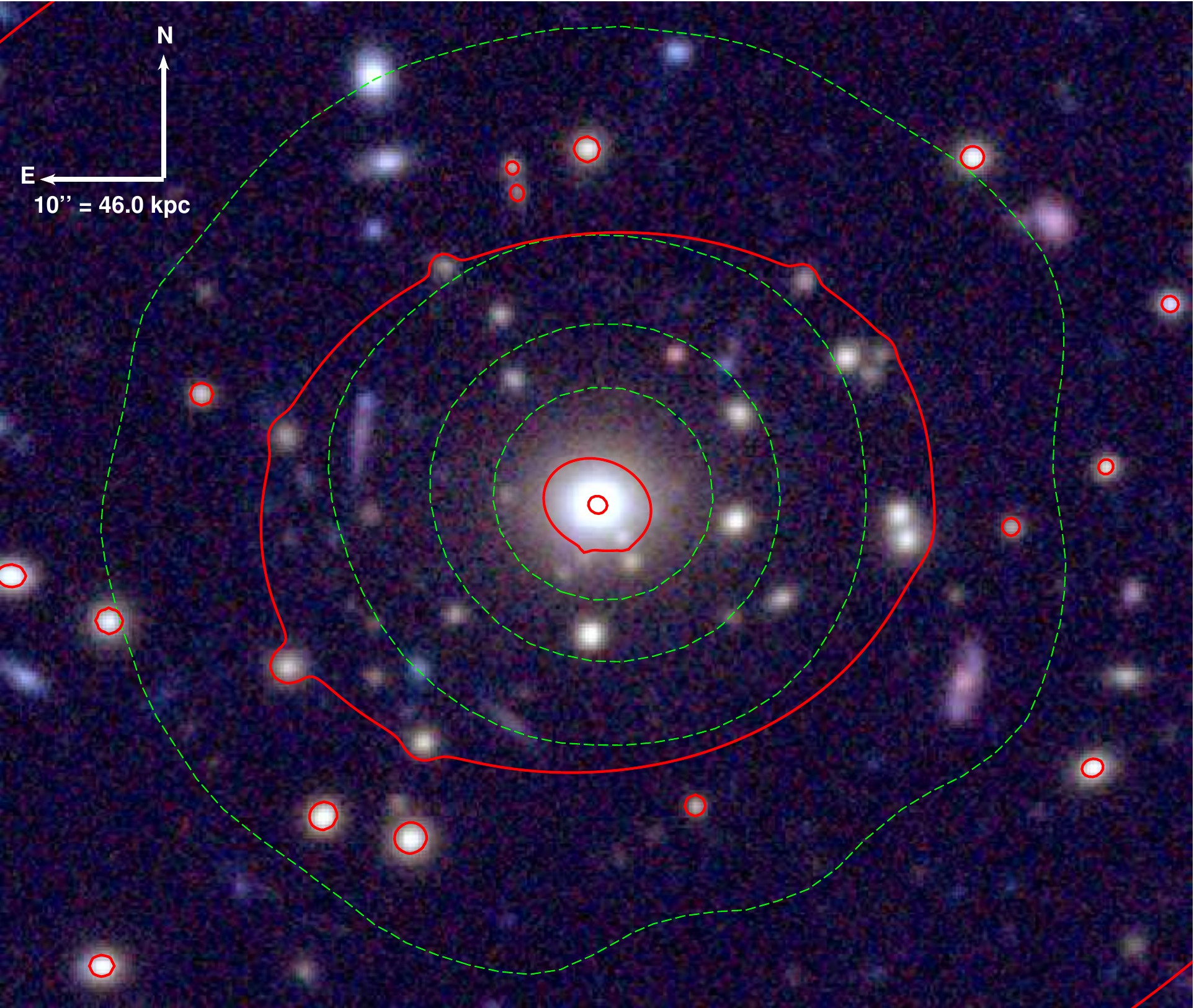}}
\end{minipage}
\hfill
\begin{minipage}{0.48\linewidth}
\centerline{\includegraphics[width=1\textwidth]{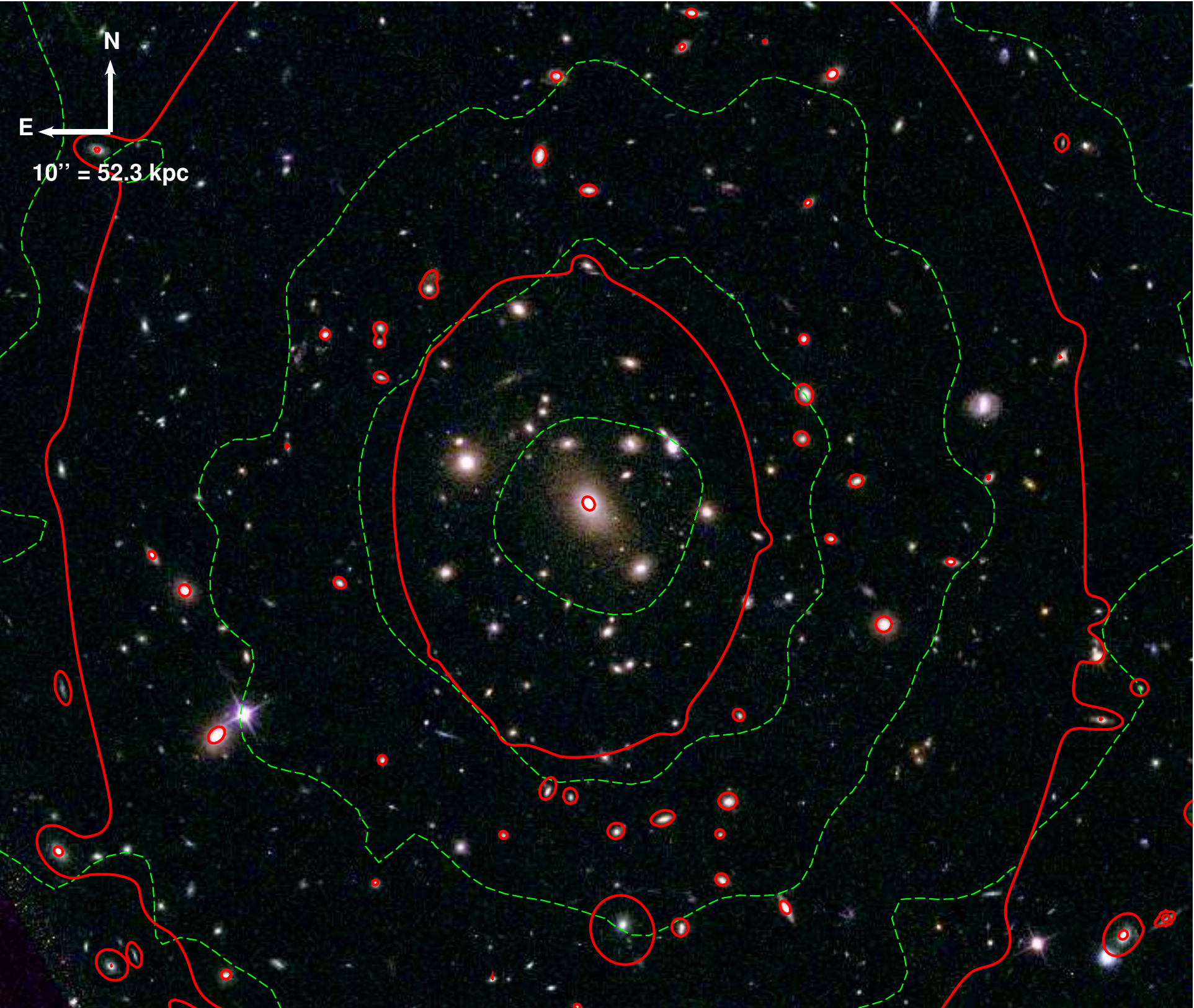}}
\end{minipage}
\caption{Composite RGB colour images of the two lensing clusters.
\textit{Left:} Composite DES colour image of MACS\,J0242. 
\textit{Right:} Composite colour \textit{HST} image of MACS\,J0949.
\textit{Green:} Hot gas distribution, obtained with \textit{XMM-Newton} observations.
\textit{Red:} Contours of equal density, inferred from lensing models.
}
\label{fig:astro_images}
\end{figure*}

\begingroup
\renewcommand{\arraystretch}{1.2}
\begin{table*}
	\centering
	\caption{Summary of the lensing reconstruction of galaxy clusters MACS\,J0242 and MACS\,J0949. We here list: (i) the galaxy clusters; (ii) number of galaxies in the cluster catalogue; (iii) number of multiply-lensed images detected; (iv) number of associated spectroscopic redshift measurements; (v) $rms$ deviation of predicted multiply-lensed images positions from their observed positions in the image plane; (vi) reduced $\chi^2$; (vii) projected mass enclosed within 200\,kpc (in $10^{14}\,\Msol$).}
	\label{tab:summary_lensing}
\makebox[\textwidth][c]{
	\begin{tabular}{lcccccc}
	    \hline
		\hline
		Galaxy cluster & $N_{\rm gal}$ & $N_{\rm im}$ & $N_{\mathrm{spec}-z}$ & $rms$ & $\chi^2_{\rm red}$ & $M (R < 200\,\mathrm{kpc})$ [$10^{14} \Msol$] \\
		\hline
        MACS\,J0242 & 57 & 18 & 18 & $0.39''$ & 0.86 & $1.67_{-0.05}^{+0.03}$\\
        MACS\,J0949 & 170 & 20 & 9 & $0.15''$ & 0.67 & $2.00_{-0.20}^{+0.05}$\\
		\hline
		\hline
	\end{tabular}
	}
\end{table*}
\endgroup

\begingroup
\renewcommand{\arraystretch}{1.2}
\begin{table*}
	\centering
	\caption{Best fit parameters of the strong lensing mass model for MACS\,J0242 and MACS\,J0949. We here list the central coordinates, $\Delta_{\alpha}$ and $\Delta_{\delta}$, in arcsec, relative to the centre, the ellipticity, $e$, the position angle in degrees, $\theta$, the core radius in kpc, $a$, the cut radius in kpc, $s$, and the velocity dispersion in km.s$^{-1}$, $\sigma$, for each component of the model.
	The centres are taken to be respectively $(\alpha_c, \delta_c) = (40.649555, -21.540485)$ deg and $(\alpha_c, \delta_c) = (147.4659012, 17.1195939) $ deg for MACS\,J0242 and MACS\,J0949.
	The asterisks highlight parameters which are fixed during the optimisation.
	$L^{\star}$ represents the cluster member galaxies catalogue, scaled with the Faber-Jackson scaling relation \citep{1976ApJ...204..668F}.
	MACS\,J0949 includes a southern dark matter clump O3.
	}
	\label{tab:best_model}
\makebox[\textwidth][c]{
	\begin{tabular}{lccccccc}
	    \hline
		\hline
		 & $\Delta_{\alpha}$ [arcsec] & $\Delta_{\delta}$ [arcsec] & $e$ & $\theta$ [deg] & $a$ [kpc] & $s$ [kpc] & $\sigma$ [km.s$^{-1}$]\\
		\hline
		\hline
		\multicolumn{8}{c}{MACS\,J0242} \\
		\hline
        DMH  & $-0.14_{-0.14}^{+0.09}$ & $0.14_{-0.18}^{+0.11}$ & $0.29_{-0.03}^{+0.04}$ & $17.9_{-1.8}^{+0.8}$ & $57.2_{-8.4}^{+6.0}$ & $1500^{\star}$ & $918.5_{-36.1}^{+29.0}$\\ 
        BCG & $0.04^{\star}$ & $-0.09^{\star}$ & $0.23^{\star}$ & $155.8_{-9.6}^{+10.8}$ & $0.30^{\star}$ & $177.6_{-58.0}^{+32.2}$ & $524.5_{-44.0}^{+58.8}$\\ 
        $L^{\star}$ &  &  &  &  & $0.03^{\star}$ & $5.6_{-1.8}^{+7.8}$ & $199.2_{-53.3}^{+30.7}$\\ 
		\hline
		\hline
		\multicolumn{8}{c}{MACS\,J0949} \\
		\hline
        DMH & $-1.94_{-2.84}^{+0.22}$ & $-0.67_{-0.67}^{+0.57}$ & $0.25_{-0.05}^{+0.40}$ & $92.4_{-1.3}^{+0.6}$ & $116.2_{-51.7}^{+24.1}$ & $1500^{\star}$ & $1236.1_{-310.6}^{+59.3}$\\ 
        BCG & $0^{\star}$ & $0^{\star}$ & $0.48^{\star}$ & $120.1^{\star}$ & $0.25^{\star}$ & $98.0_{-34.3}^{+153.7}$ & $253.7_{-18.5}^{+196.5}$\\ 
        Clump O3 & $4.80_{-0.46}^{+0.75}$ & $-60.13_{-1.42}^{+2.39}$ & $0.01_{-0.06}^{+0.29}$ & $128.6_{-27.5}^{+41.4}$ & $20.5_{-8.8}^{+31.6}$ & $232.5_{-119.9}^{+180.1}$ & $323.2_{-54.9}^{+120.2}$\\ 
        $L^{\star}$ & & & & & $0.15^{\star}$ & $23.1_{-2.1}^{+111.5}$ & $139.3_{-18.5}^{+25.8}$\\ 
		\hline
		\hline
	\end{tabular}
	}
\end{table*}
\endgroup

If distant background sources happen to be close to the line-of-sight between a massive galaxy cluster and an observer, the background image can be strongly lensed, to the point multiple images appear to the observer.
Using this gravitational lensing effect in the strong regime, we can precisely map the gravitational potential of the cluster, and its total -- baryonic and dark matter -- mass density.

In \citet{2023MNRAS.522.1118A}, we performed a reconstruction of the total matter density, $\rho_m$, of the two galaxy clusters MACS\,J0242 and MACS\,J0949.
These lensing models were obtained thanks to a combination of imaging with the \textit{Hubble Space Telescope} (\textit{HST}) and DES from the ground, together with spectroscopy obtained with the MUSE instrument at the Very Large Telescope (VLT).
Figure\,\ref{fig:astro_images} shows colour-composite images of the two clusters used in this work, together with the ICM distribution obtained using X-ray observations, and density contours from the strong-lensing analyses presented in \citet{2023MNRAS.522.1118A}.

With spectroscopy, we detected 6 and 2 systems of multiply-lensed images in MACS\,J0242 and MACS\,J0949 respectively, for a total of 18 and 9 images with spectroscopic redshifts. 4 additional systems were detected with imaging \textit{HST} observations in cluster MACS\,J0949, but do not have a redshift measurement.
Using a combination of photometry and spectroscopy, we identified 57 and 170 cluster member galaxies respectively.
We performed the strong lensing optimisation with \textsc{Lenstool} \citep[][]{2007NJPh....9..447J}, using the multiply-imaged systems to invert the lens equation.
We have assumed the potential of a galaxy cluster to be a superposition of dPIE potentials.
We modelled each cluster with a large-scale dark matter halo (DMH), a brightest cluster galaxy (BCG), and a $L^{\star}$ catalogue of elliptical galaxies, scaled using the Faber-Jackson relationship \citep{1976ApJ...204..668F}.
Additionally, we introduced in MACS\,J0949 a clump in the south of the halo, to explain multiply-lensed images in this region.

Tables\,\ref{tab:summary_lensing} and \ref{tab:best_model} present respectively the summary of the lensing information available for each cluster, and the best-fit parameters of the strong-lensing models obtained for the different potentials of each galaxy clusters. 
The average distance between the multiple images predicted with the lensing models and the observations is $0.39''$ and $0.15''$, and the reduced $\chi^2$, $\chi^2_{\rm red} = 0.86$ and $0.67$, for clusters MACS\,J0242 and MACS\,J0949 respectively, indicating a good quality reconstruction.
The enclosed mass within 200\,kpc of the cluster centre were respectively $M(R<$200\,kpc$) = 1.67_{-0.05}^{+0.03}\times 10^{14}\Msol$ for MACS\,J0242, and $M(R<$200\,kpc$) = 2.00_{-0.20}^{+0.05}\times 10^{14}\Msol$ for MACS\,J0949.
Cluster MACS\,J0242 is found to be dynamically relaxed, with a peaked central density, while MACS\,J0949 presents a flatter density distribution in the core ($R \in [10, 100]$\,kpc), and is still relaxing, but not strongly disturbed.

The inferred 3D density profiles were well fit by NFW profiles (see Section\,\ref{subsec:full_matter_density_profiles}). 
For MACS\,J0242, we found the best fitting NFW parameters to be 
$\rho_S = 3.42 \times 10^{-25}$\,g.cm$^{-3}$
and $r_S = 209.9$\,kpc, for a reduced $\chi^2 = 1.11$.
For MACS\,J0949, the best fitting parameters are 
$\rho_S = 1.23 \times 10^{-25}$\,g.cm$^{-3}$, 
$r_S = 405.5$\,kpc, for a reduced $\chi^2 = 1.90$.
In this article, we only use the DMH and BCG potentials to represent the clusters' gravitational potential.

\section{Galaxy clusters: a theoretical description}
\label{sec:Theory}

This Section introduces the observables and models necessary to describe the physics of the ICM using gravitational lensing.
Section\,\ref{subsec:full_matter_density_profiles} introduces the two general full matter density profiles we use in this work;
Section\,\ref{subsec:ICM_density} presents the \textit{canonical} description for the ICM density;
Section\,\ref{subsec:fully_analytical_ne} shows the derivation of the \textit{analytical} ICM density using a temperature model and total matter density;
Section\,\ref{subsec:temperature_model} extends the common polytropic temperature density to the higher electron densities found in the centre of clusters;
Sections\,\ref{subsec:X-ray_def} and \ref{subsec:SZ_def} define the X-ray surface brightness and the SZ effect temperature contrast respectively, in order to make observable predictions.

\subsection{Galaxy clusters matter density models}
\label{subsec:full_matter_density_profiles}

The total matter density is modelled parametrically.
We here present two cases, a Navarro-Frenk-White (NFW) density profile, and a dual Pseudo-Isothermal Elliptical mass distribution (dPIE) density profile. 
The generalised NFW and Einasto profiles are described in Appendix\,\ref{sec:Additional_densities}.
For the discussion and illustration purposes, we discuss purely radial profiles, but in our reconstruction, we use the various geometrical parameters of the individual potentials inferred from lensing (position of the centre, ellipticity, position angle).

\subsubsection{Navarro–Frenk–White (NFW) profile}
The NFW profile \citep[introduced in][]{Navarro_1996} describes the DM density. We here approximate it to the total density distribution, $\rho_m$:
\begin{equation}
    \rho_m (r) = \rho_S \left\{\frac{r}{r_S} \left( 1 + \frac{r}{r_S} \right)^2 \right\}^{-1},
    \label{eq:NFW_density_matter}
\end{equation}
where $\rho_S$ is the density normalisation, and $r_S$, the scale radius. These are parameters different for each cluster. We assume the NFW profile to describe the total density with one profile for a single cluster.

\subsubsection{Dual Pseudo-Isothermal Elliptical mass distribution (dPIE) profile}
\label{sec:dPIE_description}

In \citet{1993ApJ...417..450K} and \citet{eliasdottir2007matter}, the dPIE 
profile scales as:
\begin{equation}
    \rho_m (r) = \rho_0 \left\{ \left[ 1 + \left( \frac{r}{s} \right)^2 \right] \left[ 1 + \left( \frac{r}{a} \right)^2 \right] \right\}^{-1},
\end{equation}
with the core radius, $a$, of the order of 100\,kpc for the dark matter halo, and a truncation radius, $s > a$.
Whilst this distribution is spherically symmetric, we also consider two other parameters:
a rotation angle, $\theta$, and an ellipticity, $e$ (see Appendix \ref{sec:ellipticity}).
This model is sometimes referred to as \textit{pseudo-Jaffe}, as in the review of \citet{2001astro.ph..2341K}.

Contrarily to the NFW density profile, the dPIE profile does not present any divergence in $r \to 0$, i.e. presents a finite density in the core.
For our lensing reconstruction, we used a large scale DMH modelled with a dPIE, and superposed it to individual profiles fitting individual cluster member galaxies.
In the case of the analysed clusters, their (relative) relaxation allows us to discard all individual (galaxy) potentials but that of the BCG.
The DMH and the BCG respectively govern the large and small radii total matter densities. 
This DMH and BCG superposition was well fitted by a NFW profile for both clusters (see Section \ref{sec:lensing_models}).

\subsection{Electron density}
\label{subsec:ICM_density}

We focus on the electron density $n_e$ because it can be derived from X-ray observations.
However, one could derive the ion or gas density too.
The number density of electrons, $n_e$, is related to the gas volume density, $\rho_g$ through:
\begin{equation}
    n_e (r) = \mathcal{F}_I (r) \frac{\rho_g (r)}{\mu_e m_a},
    \label{eq:electron_density_func_rhog}
\end{equation}
where $\mathcal{F}_I$ is the local ionisation fraction which is taken to be $1$, $m_a = 1$\,Da is the atomic mass constant, and $\mu_e \approx 1.15$ is the mean molecular weight of electron.

A number of models have been proposed to fit the electron density profile, which we refer to as \textit{canonical}.
One of the most complete model can be found in \citet{Vikhlinin_2006}.
As this model relies on a large set of parameters, unnecessary here, we shall focus on a simplified model, namely the (simple) $\beta$-model \citep[see][]{King_1966, 1976A&A....49..137C}:
\begin{equation}
    n_e (r) = n_{e,0} \left[ 1 + \left( \frac{r}{r_c} \right)^2 \right]^{- \frac{3}{2} \beta}, 
    \label{eq:beta_model}
\end{equation}
with $r_c$, the core radius, $n_{e,0}$ a density normalisation, and $\beta \in [0.5; 0.9]$, an empirical index.
We find this model to fit well both clusters' ICM distribution (see Sections\,\ref{subsubsec:m0242_beta} and \ref{subsubsec:m0949_beta}), thus showing a more complex model to be unnecessary here.

\subsection{A fully analytical electron density}
\label{subsec:fully_analytical_ne}

\subsubsection{General case}

We consider the total (DM and baryons) mass density as a sum of densities:
\begin{equation}
    \rho_m (r) = \sum_i \rho_{0, m, i} f_i (r),
    \label{eq:separate_rho_in_fi}
\end{equation}
which is constrained by the strong lensing analyses.
Each potential can be normalised differently, and the distributions, $f_i$, are assumed to be of the same type (here NFW or dPIE). 
Here, we write them as a sum of radial functions for simplicity.
In practice, every profile, $f_i$, has its own geometric parameters (central position, ellipticity, rotation angle), which we consider to be fixed from the strong lensing analysis.

Assuming integrability, we introduce:
\begin{equation}
\begin{split}
    g_i (r) &= \int_0^r \mathrm{d}s s^2 f_i (s),\\
    h_i (r) &= \int_0^r \mathrm{d}s s^{-2} g_i (s),
\end{split}
\label{eq:def_g_h_func}
\end{equation}
where integration constants are included.
Denoting $\Phi$ the Newtonian potential, integrating the gravitational Poisson equation, it reads:
\begin{equation}
 \Phi (r) = -4\pi G \sum_i \rho_{0, m, i} h_i(r).
\label{eq:general_grav_potential}
\end{equation}
The conservation of momentum in the Lagrangian formalism, i.e. the momentum Navier-Stokes equation for a perfect fluid \citep[viscosity neglected, see e.g.][]{1959flme.book.....L} reads:
\begin{equation}
    \rho_g \frac{\mathrm{D} \boldsymbol{v}}{\mathrm{D} t} = \rho_g \left[ \partial_t \boldsymbol{v} + \left( \boldsymbol{v} \cdot \boldsymbol{\nabla} \right) \boldsymbol{v} \right] = - \boldsymbol{\nabla} P_g + \rho_g \boldsymbol{\nabla} \Phi,
    \label{eq:Navier-Stokes_perfect_fluid}
\end{equation}
with $\boldsymbol{v}$ the velocity field, and $P_g$ the gas pressure.

As the pressure in galaxy clusters is of the order of $10^{10}$\,Pa, the plasma is thermalised, and therefore the temperature of the ions is equal to that of the electrons.
With equation (\ref{eq:electron_density_func_rhog}), we can write the number density of the electron $n_e$ as being directly proportional to the gas density $\rho_g$, as $\mu_e$ and $\mathcal{F}_I$ are assumed to be constant in a given cluster.
Using the ideal gas law, we rewrite equation (\ref{eq:Navier-Stokes_perfect_fluid}) in the purely radial case:
\begin{equation}
\begin{split}
    \frac{k_B}{\mu_g m_a} \frac{\partial_r \left( n_e T_e \right)}{n_e} + \partial_t v_r + v_r \partial_r v_r  = - 4 \pi G \sum_i \rho_{0, m, i} r^{-2} g_i (r),
    \label{eq:full_eq_int_once}
\end{split}
\end{equation}
where $T_e$ is the electron temperature, $\mu_g \approx 0.60$ mean molecular weight of the gas, and $k_B$ the Boltzmann constant.
As the galaxy clusters used in our study are not strongly perturbed, we work under the hypothesis of hydrostatic equilibrium.
Assuming we could decompose the velocity in its radial and temporal dependencies, one could then integrate numerically. 
As we do not have access to the ICM velocity resolution, we here
assume a polytropic temperature distribution and the stream to be hydrostatic, i.e. of constant velocity both in time and in all spatial directions \citep[see e.g.][]{Zaroubi_2001}. Thus, with $\partial_t v_r = \partial_r v_r =0$, we get:
\begin{equation}
    \frac{\partial_r \left( n_e T_e \right)}{n_e} = \epsilon \sum_i \rho_{0, m, i} r^{-2} g_i (r),
    \label{eq:full_eq_int_once_hydrostatic}
\end{equation}
where $\epsilon = - 4 \pi G \mu_g m_a / k_B$.

In order to reduce this expression, we define a general $\mathcal{J}$ function which contains information on the temperature profile as:
\begin{equation}
\mathcal{J} (n_e) = \int_0^{n_e} \frac{\mathrm{d} \left[ n T_e (n) \right]}{T_0 n},
\label{eq:J_gen_definition}
\end{equation}
where $T_0$ is a temperature normalisation (see Sect.\,\ref{subsec:temperature_model}).
More details on the definition of $\mathcal{J}$ are given in Sect.\,\ref{subsec:polytropic_index_scaling}, given a precise temperature model.

Separating variables and integrating equation (\ref{eq:full_eq_int_once_hydrostatic}), we obtain:
\begin{equation}
\mathcal{J} (n_e) = \frac{\epsilon}{T_0} \sum_i \rho_{0, m, i} h_i (r) = \frac{\mu_g m_a}{k_B T_0} \Phi(r),
\label{eq:fully_integrated_hydrostatic_equation}
\end{equation}
where the right-handed term stems from equation (\ref{eq:general_grav_potential}).
If we assume $\mathcal{J}$ to be bijective\footnote{A bijective function $f: \mathcal{X} \rightarrow \mathcal{Y}$ associates to each element of a set $\mathcal{X}$ an image of set $\mathcal{Y}$, and reciprocally. Therefore, a function is invertible if and only if it is bijective.} (which is justified given a temperature model in Sect.\,\ref{subsec:polytropic_index_scaling}), then inverting this equation simply provides $n_e (r)$:
\begin{equation}
    n_e (r) = \mathcal{J}^{-1} \left( \frac{\mu_g m_a}{k_B T_0} \Phi(r) \right).
    \label{eq:reverted_J_ne}
\end{equation}

Under reasonable physical hypotheses, we have provided a general, completely \textit{analytical} description of the electron density using only the gravitational potentials characterised with strong lensing observations. 
To allow this reconstruction to be independent from other observations, 
we need to pair this analytical model with an electron temperature model. 

\subsubsection{Case of a dPIE density}

In the case of a dPIE profile, we give:
\begin{equation}
    \begin{split}
    f (r) &= \left\{ \left[ 1 + \left( \frac{r}{s} \right)^2 \right] \left[ 1 + \left( \frac{r}{a} \right)^2 \right] \right\}^{-1},\\
    g (r) &= \frac{a^2 s^2}{a^2 - s^2} \left[ a \arctan \frac{r}{a} - s \arctan \frac{r}{s} \right],\\
    h (r) &=  \frac{a^2 s^2}{a^2 - s^2} \left[ \frac{s}{r} \arctan \frac{r}{s} - \frac{a}{r} \arctan \frac{r}{a} + \frac{1}{2} \ln \left( \frac{r^2 + s^2}{r^2 + a^2} \right) \right],
    \end{split}
    \label{eq:Integrated_func_dPIE}
\end{equation}
where $a$ and $s$ represent the core $r_{\mathrm{core}}$ and scale $r_{\mathrm{cut}}$ radii of the $i$-th dPIE potential respectively. Indices were avoided for clarity.
To avoid confusion, we write \texttt{idPIE} the $n_e$ distribution with $h_i$ given by a dPIE.

\subsubsection{Case of a NFW density}
In the case of a NFW potential (see equation \ref{eq:NFW_density_matter}), we can rewrite the different integrals given equation (\ref{eq:Integrated_func_dPIE}) for the dPIE:
\begin{equation}
    \begin{split}
    f (r) &= \left\{ \left[ \frac{r}{r_S} \right] \left[ 1 + \frac{r}{r_S} \right]^{2} \right\}^{-1},\\
    g (r) &= r_S^3 \left[ \ln \left( 1 + \frac{r}{r_S}\right) - \frac{r}{r + r_S}  \right],\\
    h (r) &= - \frac{r_S^3}{r} \ln \left( 1 + \frac{r}{r_S}\right).
    \end{split}
    \label{eq:Integrated_func_NFW}
\end{equation}
Here $\rho_{0, m, i}$ of equation (\ref{eq:fully_integrated_hydrostatic_equation}) is $\rho_{S, m}$. 
In case of a NFW profile, we assume the total density to be represented by a single profile.
We shall write the resulting $n_e$ distribution \texttt{iNFW}. 

\subsection{Temperature}
\label{subsec:temperature_model}

In order to derive the ICM density and thermodynamic profiles using strong lensing constraints only, we need to adopt a general temperature model, independent of the specific observations of one cluster.
We shall consider polytropic models \citep[for example in][]{Capelo_2012} of the form :
\begin{equation}
    T_e(r) = T_0 \left( \frac{n_e (r)}{n_{0}} \right)^{\Gamma - 1},
\label{eq:polytropic_Te_cstindex}
\end{equation}
with $n_0$ the central electronic density, $T_0$ the temperature in the centre, and $\Gamma \approx 1.2 $, the polytropic index.

Following \citet{Ghirardini_2019_polyt}, we can extend this definition to a self-similar polytropic temperature model, with a varying polytropic index, $\Gamma (n_e)$:
\begin{equation}
    \begin{split}
    \frac{P_e}{P_{500,c}} &= \eta_P \left( \frac{n_e E(z)^{-2}}{\eta_n} \right)^{\Gamma (n_e)},\\
    \frac{T_e}{T_{500,c}} &= \eta_T \left( \frac{n_e E(z)^{-2}}{\eta_n} \right)^{\Gamma (n_e) - 1},
    \label{eq:pression_temperature_polytropic_Eckert}
    \end{split}
\end{equation}
where $E(z) = H(z) / H_0$ is the normalised Hubble factor 
assuming a $\Lambda$CDM cosmology with $\Omega_m = 0.3$ and $\Omega_{\Lambda} = 0.7$.
$\eta_{P}$ and $\eta_T$ are dimensionless proportionality constants, and $\eta_n$ the volume number density normalisation.
We write $T_0 (z) = \eta_T T_{500} (z)$ and $n_0 (z) = \eta_n E(z)^{2}$.
Section\,\ref{subsubsec:polytropic_index_model} presents a new model for the index $\Gamma (n_e)$, and also provides the quantitative values for the different constants presented here.

\subsection{X-ray surface brightness}
\label{subsec:X-ray_def}

In order to compare our results to observations, and therefore to evaluate the quality of our ICM reconstruction, we introduce the X-ray surface brightness $S_X$ \citep[see][for a review]{2010A&ARv..18..127B}.
In a band of wavelength (such as the $[0.7, 1.2]$\,keV band for \textit{XMM-Newton}) integrated over the line-of-sight, it reads:
\begin{equation}
    S_X (\Delta E) = \frac{1}{4 \pi (1 + z)^4} \frac{\mu_e}{\mu_H} \int_0^{\infty} n_e^2 (\boldsymbol{r}) \Lambda (\Delta E, T_e, Z) \mathrm{d}l.
\label{eq:X-ray_surface_brightness_fla}
\end{equation}
where $\mu_H \approx 1.35$ is the mean molecular weight of hydrogen, $\Lambda (\Delta E, T_e, Z)$ is the X-ray spectral emissivity (or \textit{cooling curve}), as a function of the X-ray energy band, $\Delta E$, the electron temperature of the gas, $T_e$, and the metallicity of the gas, $Z$. In this article, the metallicity is assumed to be constant for a given cluster.

\subsection{SZ effect}
\label{subsec:SZ_def}

Another observable, depending on the electron density and temperature is the SZ effect.
Given an observable frequency, $\nu$, we use the reduced frequency, $x$:
\begin{equation}
    x = \frac{h \nu}{k_B T_r},
\end{equation}
with $h$, the Planck constant, and $T_r \simeq 2.726$\,K, the temperature of the CMB.
We define the Compton parameter \citep[see][]{Rephaeli_95}:
\begin{equation}
    y (r) = \frac{k_B \sigma_T}{m_e c^2} \int_0^{\infty} T_e (r) n_e (r) \mathrm{d}l,
\label{eq:Compton_parameter}
\end{equation}
with $\sigma_T$, the Thomson cross-section, and $m_e$, the mass of the electron. 
The thermal SZ contrast then reads:
\begin{equation}
    \Theta_{\mathrm{SZ}} (r) = \frac{\Delta T}{T_r} = \left[ x \coth \left( \frac{x}{2} \right) - 4 \right] y (r).
    \label{eq:SZ_effect}
\end{equation}

\section{Quantitative models}
\label{sec:quantitative_results}

We have presented in the previous section a completely different manner to use the lensing study of galaxy clusters to predict their baryonic distributions. Assuming the hydrostatic equilibrium and a given temperature distribution, equation (\ref{eq:reverted_J_ne}) presents an \textit{analytical} ICM density prediction using lensing.
Moreover, in Appendix \ref{subsec:gas_fraction_presenting}, we present an alternative method, reducing the gas fraction to an analytical model, and using knowledge of $\rho_m$, constrained by lensing analysis. 
Although the general profile of the gas fraction is retrieved, this study is however not completely generalisable, and we do not use it here.

In this section, we provide the electron temperature models needed for the analytical method. We make quantitative estimates in order to yield quantitative predictions for ICM density profile from lensing data only.

\subsection{Polytropic index scaling}
\label{subsec:polytropic_index_scaling}

\subsubsection{Constant polytropic index}

As shown in Sect.\,\ref{subsec:fully_analytical_ne}, one can reconstruct the ICM profile using equation (\ref{eq:reverted_J_ne}) and an analytical expression for the function $\mathcal{J}$ which contains the information about the electron temperature profile.
Here, we assume the following form for $\mathcal{J}$, based on a self-similar polytropic temperature law with a constant index $\gamma > 1$ (see equation \ref{eq:pression_temperature_polytropic_Eckert}):  
\begin{equation}
    \mathcal{J}_z (n_e) = \frac{\gamma}{\gamma - 1} \left( \frac{n_e}{n_0 (z)} \right)^{\gamma - 1}.
    \label{eq:J_integral_polyt}
\end{equation}
We can write the electron density from equation (\ref{eq:fully_integrated_hydrostatic_equation}):
\begin{equation}
    n_e = n_0 (z)  \left[ \frac{\gamma - 1}{\gamma} \frac{\epsilon}{T_0(z)} \sum_i \rho_{0, m, i} h_i (r) \right]^{1 / (\gamma - 1)}.
    \label{eq:ne_polyt_bij}
\end{equation}
However, we find such a description to fail to describe the higher electron densities ($n_e > 10^{-2}$\,cm$^{-3}$), which is consistent with a constant $\gamma$ index fixed with the largest radii of clusters, i.e. the least dense regions.
We notice the specific case of a polytropic temperature density associated to a NFW profile has already been studied in \citet{2010ApJ...720.1038B}, where a self-normalisation in the centre is utilised.

\subsubsection{Polytropic index model}
\label{subsubsec:polytropic_index_model}

To describe the relation between ICM thermodynamic quantities ($n_e$, $P_e$, $T_e$), it is common practice to describe the stratification of the ICM using a polytropic equation of state $P(n_e)\propto n_e^\Gamma$ \citep[e.g.][]{2010ApJ...720.1038B, Capelo_2012, Tchernin_2018, Ghirardini_2019_polyt}. Analytic models assuming the ICM to be in hydrostatic equilibrium within a NFW potential predict that the polytropic index $\Gamma$ should be close to a constant value of $\sim1.2$ throughout the cluster volume \citep{Capelo_2012}. Observationally speaking, the measured values of the polytropic index closely match the NFW expectation in cluster outskirts ($R>0.2R_{500}$), but significantly deviate from it in the cluster core, where $\Gamma$ decreases down to $\sim0.8$ under the influence of radiative cooling \citep{Ghirardini_2019_polyt}. 
Using the data from the X-COP programme, \citet{Ghirardini_2019_polyt} showed that the $P(n_e)$ relation is nearly universal across the cluster population with a low scatter of $\sim 15\%$, independently of a system's dynamical state. Here we propose a new functional form to describe the self-similar polytropic model \citep[supported by e.g.][]{2019MNRAS.483.3390M}. We describe the relation with a smoothly varying polytropic index $\Gamma(n_e)$ as:
\begin{equation}
    \begin{gathered}
    \begin{aligned}
    \frac{\mathrm{d}\ln P_e}{\mathrm{d}\ln n_e} \equiv \Gamma (n_e) = \Gamma_0 \left[ 1 + \Gamma_S \arctan \left( \ln \frac{n_e E(z)^{-2}}{\eta_n} \right) \right],
    \end{aligned}
    \end{gathered}
    \label{eq:gamma_of_n}
\end{equation}
with $\eta_n$ the reference number density around which the transition between core (low $\Gamma$) and outskirts (NFW $\Gamma$) occurs, $\Gamma_0$ the polytropic index at $\eta_n$, and $\Gamma_S$ the slope of the transition. 

\begin{figure*}
    \centering
    \resizebox{\hsize}{!}{\includegraphics{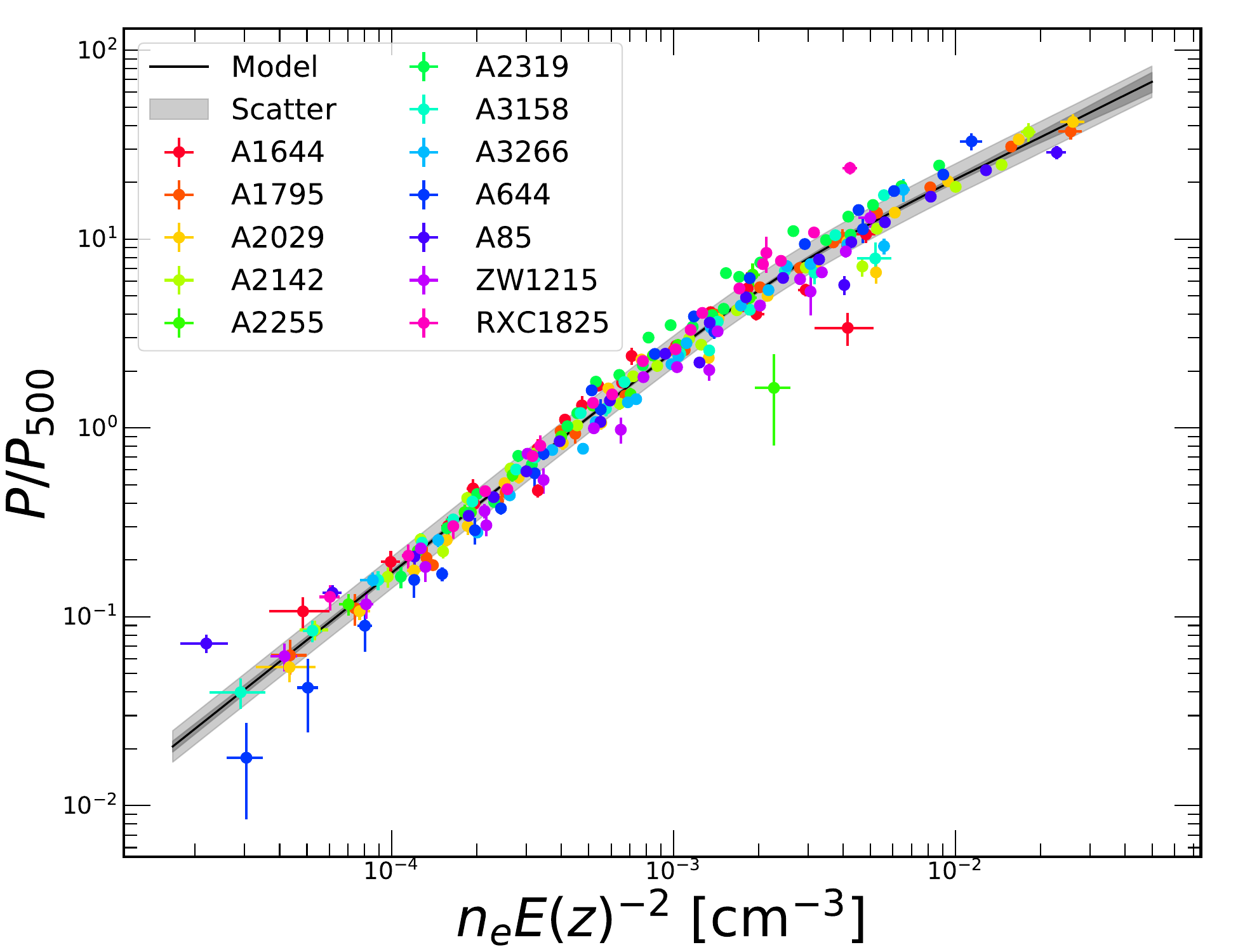}\includegraphics{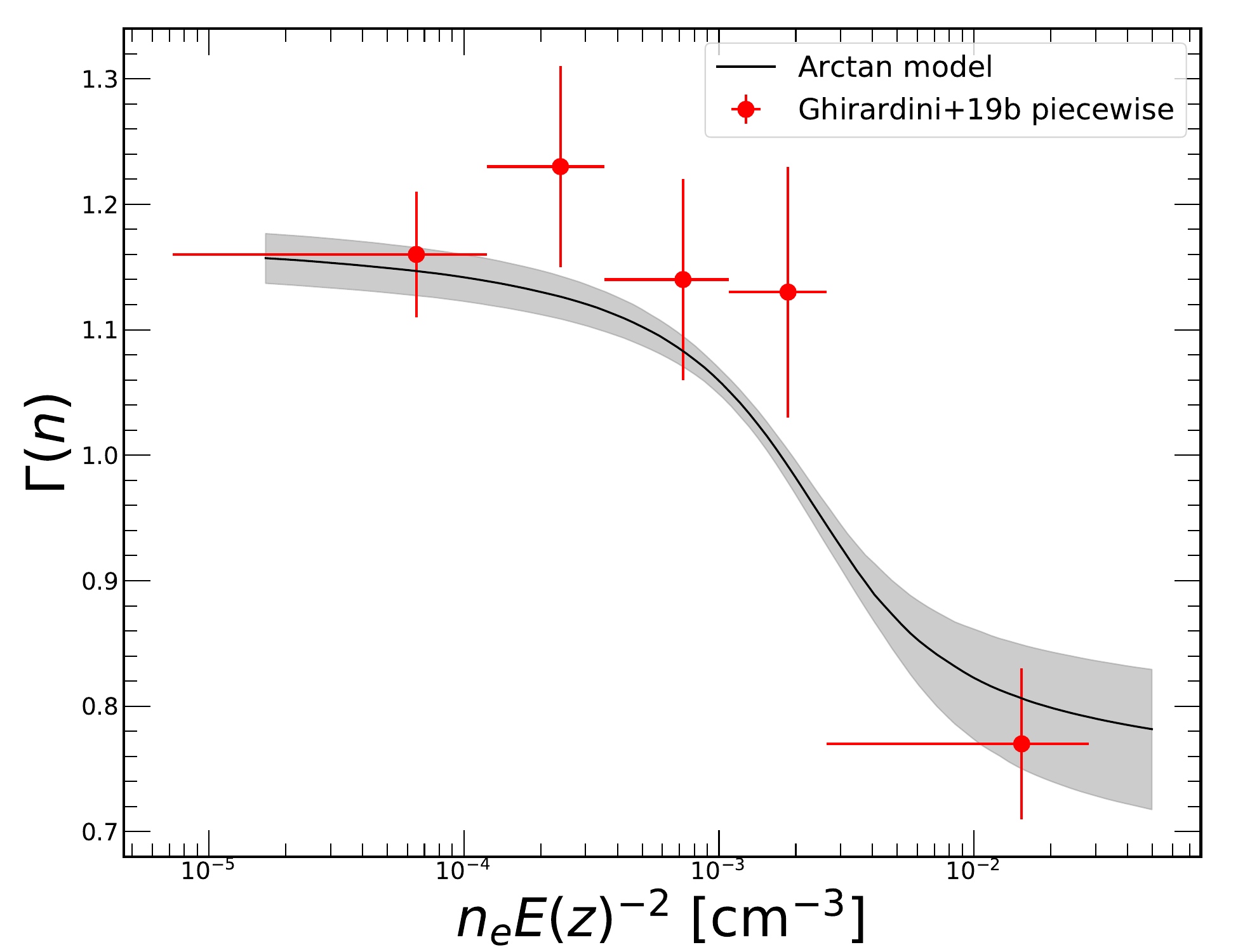}}
    \caption{Calibration of the variable $\Gamma(n_e)$ polytropic model (equation \ref{eq:gamma_of_n}) on X-COP data. \textit{Left:} Relation between self-similar scaled ICM pressure and electron density for the 12 X-COP clusters. The solid black line and the grey shaded area show the best-fit model and the intrinsic scatter around the model, respectively. \textit{Right:} Polytropic index $\Gamma = \mathrm{d}\ln P_e / \mathrm{d}\ln n_e$ as a function of electron density. The black line and shaded area shows the best-fit model with equation (\ref{eq:gamma_of_n}), whereas the red data points show the result of a piece-wise fit with constant polytropic index over several ranges in electron density.
    }
    \label{fig:polytropic_arctan}
\end{figure*}

We used the publicly available X-COP data, which provide high-quality observations of the ICM thermodynamic properties over a wide radial range ($\approx [0.01-2]R_{500}$), to calibrate the model and determine the parameters of equation (\ref{eq:gamma_of_n}). 
We fit the X-COP density and pressure data using the Bayesian analysis package \textsc{PyMC3} \citep{PyMC3}, including uncertainties on both axes, and a free log-normal intrinsic scatter. The observational data points on both axes are scaled by their respective self-similar scaling values \citep{Arnaud2010}. The result of this procedure is shown in Fig.\,\ref{fig:polytropic_arctan}. The model provides an excellent representation of the data over three decades in electron density, with a low intrinsic scatter of $\sigma_{\ln P} = 0.19 \pm 0.02$. The right-hand panel of Fig.\,\ref{fig:polytropic_arctan} also shows that the results obtained with the model defined in equation (\ref{eq:gamma_of_n}) are consistent with the values estimated by \citet{Ghirardini_2019_polyt} when fitting a piece-wise power law over several ranges in density. The fit parameters are included in Table\,\ref{tab:polytropic_Eckert_params}.

\begin{table*}
	\centering
	\caption{Parameters of the smoothly varying polytropic model defined in equations\,(\ref{eq:pression_temperature_polytropic_Eckert}) and (\ref{eq:gamma_of_n}).}
	\label{tab:polytropic_Eckert_params}
	\begin{tabular}{ccccc}
	    \hline
		\hline
		$\eta_P$ & $\eta_n$ [cm$^{-3}$] & $\Gamma_0$ & $\Gamma_S$ & $\sigma_{\ln P_e}$\\
		\hline
		$6.05 \pm 1.57$ & $(2.26 \pm 0.59) \times 10^{-3}$ & $0.97 \pm 0.04$ & $-0.15 \pm 0.03$ & $0.19\pm0.02$\\
		\hline
		\hline
	\end{tabular}
\end{table*}

Supposing the ICM to be an ideal gas, following \citet{2019A&A...621A..41G}, we write:
\begin{equation}
    \eta_T = \left(3.87 \times 10^{-4}\,\mathrm{cm}^{-3} \right) \frac{\eta_P}{\eta_n} \frac{f_b}{0.16} \frac{\mu_e}{1.14},
    \label{eq:Tref}
\end{equation}
with $f_b$ is the universal baryon fraction, taken to be $f_b = 0.158 \pm 0.002$ \citep{Planck_XIII_2016}. We find $\eta_T \approx 1.034$.

As for the full $T_{500} (z)$, we use the results of \cite{2019A&A...621A..41G}, which we reproduce here:
\begin{equation}
    T_{500} (z) = 8.85\,\mathrm{keV} \left( \frac{M_{500} E(z)}{h_{70}^{-1} 10^{15} \Msol} \right)^{2/3} \left( \frac{\mu_g}{0.6} \right),
    \label{eq:T_500_universal}
\end{equation}
with $h_{70} = h / 0.7 = 1$.
We therefore can normalise the temperature universally.

\subsubsection{Varying polytropic index}

\begin{figure}
    \centering
    \includegraphics[width=\columnwidth]{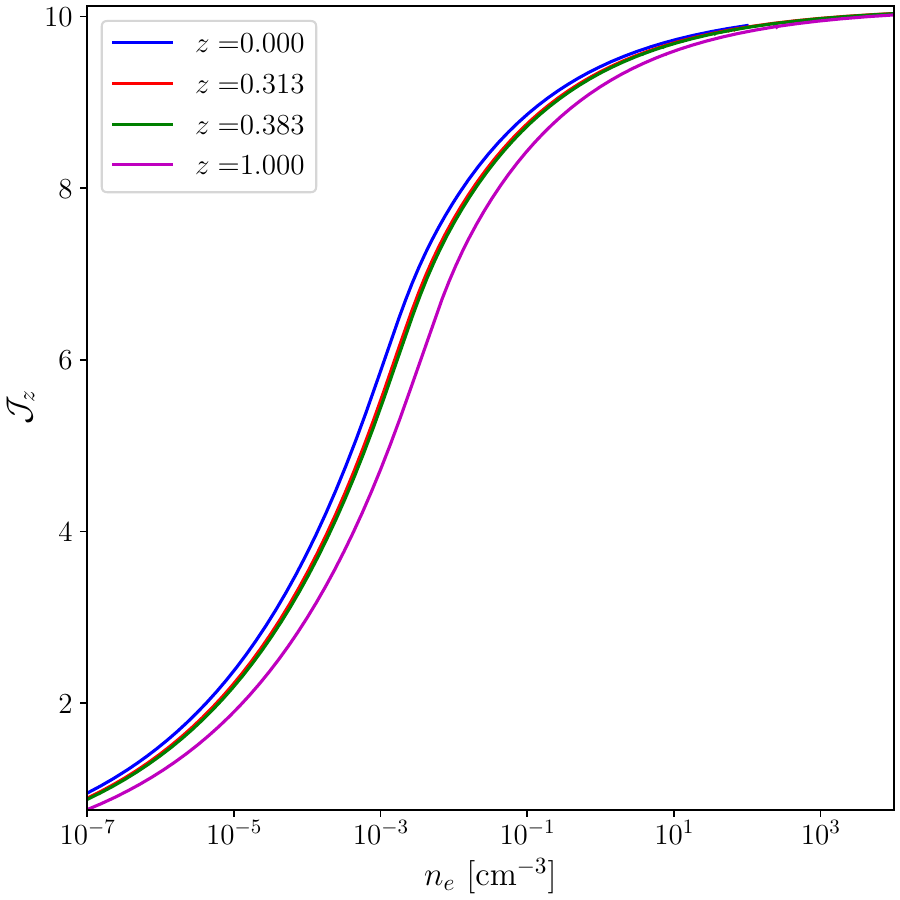}
    \caption{$\mathcal{J}_z (n_e)$ in a range of redshifts, including those of galaxy cluster MACS\,J0242 and MACS\,J0949 -- which are almost identical. The $\mathcal{J}_z$ function is therefore extremely sensitive, i.e. a small error on the potential is associated to a much larger error in the determination of $n_e$.}
    \label{fig:Jz_curve}
\end{figure}

At a given redshift, $z$, using the self-similar polytropic temperature described equations (\ref{eq:pression_temperature_polytropic_Eckert}) and (\ref{eq:gamma_of_n}), we can define the integral $\mathcal{J}_z$, i.e. a redshift dependent $\mathcal{J}$, defined equation (\ref{eq:J_gen_definition}):
\begin{equation}
\begin{split}
    \mathcal{J}_z (n_e) = \scaleint{9ex}_{\bs 0}^{n_e} \left[ \Gamma(n) + \frac{\Gamma_0 \Gamma_S \ln \left( \frac{n}{n_0 (z)} \right)}{1 + \left[ \ln \left(\frac{n}{n_{0} (z)} \right) \right]^2} \right] \left( \frac{n}{n_0 (z)} \right)^{\Gamma(n) - 1} n^{-1} \mathrm{d} n.
    \label{eq:J_integral_definition}
\end{split}
\end{equation}
In this case, $\mathcal{J}_z$ can be easily computed at a given redshift, and reverted. It is however not analytically solvable. 
An example of $\mathcal{J}_z$ is displayed in Fig.\,\ref{fig:Jz_curve}.

We find empirically $\mathcal{J}_z$ to be a monotonically increasing function, i.e. a bijection. Therefore we can take its inverse function, allowing to define $n_e$ as a function of the radius, as displayed in equation (\ref{eq:reverted_J_ne}).

\subsection{Relating all density models to lensing}
\label{subsec:relating_lensing_to_density_params}

We assume the metallicity of clusters MACS\,J0242 and MACS\,J0949 to be constant, and the plasma to be fully ionised $\mathcal{F}_I = 1$.
The X-ray data suggest metallicities in the range 0.2 -- 0.7\,$Z_{\sole}$ for the two strong lensing clusters. 
The metallicity profiles obtained are in agreement with the $Z = 0.3 Z_{\sole}$ in both clusters, and we therefore make this assumption. Moreover, the influence of metallicity on the cooling function is limited at these temperatures \citep[to justify this approximation, read e.g.][]{2016ApJ...826..124M}.

For the analytical density profiles (such as \texttt{idPIE} or \texttt{iNFW}), the parameter priors are directly given from the lensing analysis Sect.\,\ref{sec:lensing_models}.
Conversely, we can not assume the parameters of the $\beta$-profile \textit{a priori}.
However their optimisation requires priors, for which we take $\beta = 0.63$ \citep[in agreement with e.g.][]{Bohringer_2016}, $n_{e,0}$ is taken to be the normalisation of the DMH density, and $r_{c}$ to be $a_1$, i.e. the core radius of the DMH obtained with the lensing optimisation.

\section{Method: Measurable ICM predictions and optimisations}
\label{sec:Method_ThePredictions}

\begin{figure*}
    \centering
\begin{tikzpicture}[node distance=2.5cm]

\newcount\prevnode
\prevnode=0

\def\EDynkin#1{
\node (mod1) [Model, aspect=1.5, inner sep=-0.ex] {$T_e \propto n_e^{\Gamma(n_e)}$};
\node (data1) [Data, below of=mod1, xshift=2.5cm, yshift=1cm, text width=1cm] {X-COP sample};
\node (mod2) [Model, below of=data1, xshift=-2.5cm, yshift=1cm, text width=1.5cm, aspect=1.5, inner sep=0.ex] {Gas fraction models};

\node (pred1) [Prediction, right of=data1, text width=1.5cm] {General $T_e$ and $f_g$ models};

\node (pred3) [Prediction, above of=pred1, xshift=4.5cm, text width=2cm] {Predicted X-ray + SZ contrast};
\node (mod3) [Model, right of=pred3, text width=2cm, xshift=1cm, aspect=1.5, inner sep=-.8ex] {$n_e = \mathcal{J}_z^{-1} (\Phi)$ model};
\node (pred2) [Prediction, above of=pred1, text width=1.5cm] {Lensing models};
\node (data3) [Data, left of=pred2, text width=1.5cm] {\textit{HST} + MUSE data};

\node (pred4) [Prediction, right of=pred1, xshift=2cm, text width=2cm] {ICM-optimised potential $\Phi$};
\node (data2) [Data, right of=pred4, xshift=1cm, text width=2cm] {\textit{XMM-Newton} + ACT data};

\draw [arrow] (mod1) -- (data1);
\draw [arrow] (mod2) -- (data1);
\draw [arrow] (data1) -- (pred1);
\draw [arrow] (data3) -- (pred2);
\draw [arrow] (pred2) -- (pred3);
\draw [arrow] (mod3) -- (pred3);
\draw [arrow] (pred1) -- (pred3);
\draw [arrow] (pred1) -- (pred4);
\draw [arrow] (mod3) -- (pred4);
\draw [arrow] (data2) -- (pred4);
\foreach\kthweight[count=\k] in {#1}{
\global\advance\prevnode by1
}
}

\newcommand{\DrawHalo}[2][]{%
\foreach \Node[count=\i] in {#2}
{
\xdef\imax{\i}
\coordinate (AuxNode-\i) at ($(\theDynkinDiagram-\Node)$);
}
\ifnum\imax=3%
\draw[#1] ($(AuxNode-1)+(-0.50,0)$) -- ($(AuxNode-1)+(-0.50,0.50)$) -|  ($(AuxNode-2)+(-0.50,0.50)$)
    -- ($(AuxNode-2)+(0.50,0.50)$) |-
    ($(AuxNode-3)+(0.50,0.50)$)
    --($(AuxNode-3)+(0.50,-0.50)$) -- ($(AuxNode-1)+(-0.50,-0.50)$) -- cycle;
\else
\ifnum\imax=2%
\draw[#1] ($(AuxNode-1)+(-0.50,0.50)$) --  ($(AuxNode-2)+(0.50,0.50)$)
    -- ($(AuxNode-2)+(0.50,-0.50)$) -- ($(AuxNode-1)+(-0.50,-0.50)$)--
    cycle;
\else
\draw[#1] ($(AuxNode-1)+(-0.50,0.50)$) --  ($(AuxNode-1)+(0.50,0.50)$)
    -- ($(AuxNode-1)+(0.50,-0.50)$) -- ($(AuxNode-1)+(-0.50,-0.50)$)--
    cycle;
\fi 
\fi 
}

\EDynkin{}
\pgfdeclarelayer{bg}
\pgfsetlayers{bg,main}

\begin{pgfonlayer}{bg}
    \node (art1) [rectangle, draw, fill=purple!70, opacity=0.5, label=below:\textcolor{purple}{Lensing models: \citet{2023MNRAS.522.1118A}}, fit = (data3) (pred2)]{} ;
\end{pgfonlayer}

\end{tikzpicture}
    \caption{Full workflow diagram. Models are denoted by green diamonds, data by red rectangles, while results are in blue ellipses. The light magenta rectangle denotes work carried out within our previous companion article \citet{2023MNRAS.522.1118A}.}
    \label{fig:work_diagram}
\end{figure*}
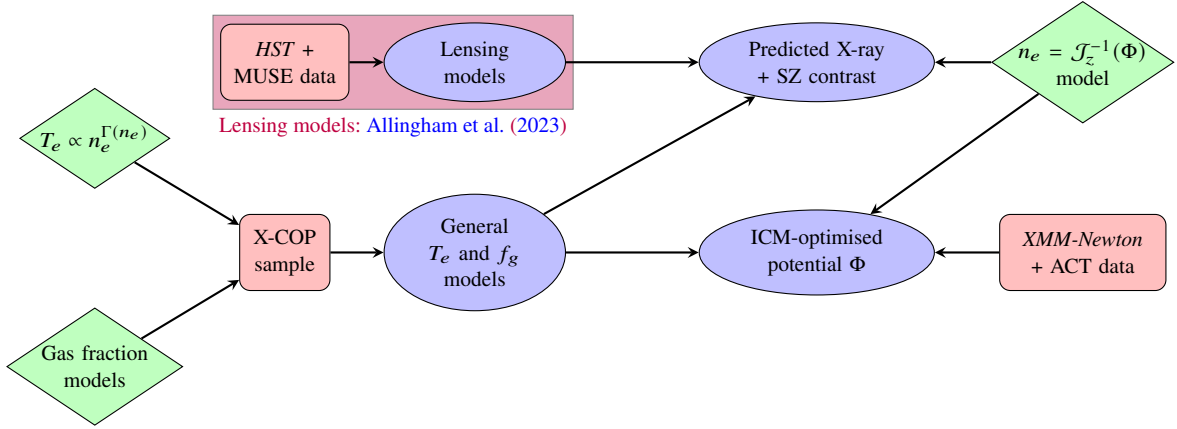

In this Section, we present the results of the method developed in Sections\,\ref{sec:Theory} and \ref{sec:quantitative_results} to convert the ICM predictions into ICM observables ($S_X$, $\Theta_r$), in order to compare them to observations.
Moreover, we contrast these predictions with a profile of the same type (e.g. \texttt{idPIE}) optimised with a MCMC using the ICM observations.
We summarise the whole process undertaken in the present article in Fig.\,\ref{fig:work_diagram}. 

\subsection{Point spread function of \textit{XMM-Newton}}

In order to analyse the \textit{XMM-Newton} observations, we reduce them to smaller maps centred around the cluster, of around 1\,Mpc width (respectively 88 and 78 pixels for clusters MACS\,J0242 and MACS\,J0949). 
We then need to take into account the point-spread function (PSF).
At first order, we define the PSF as:
\begin{equation}
PSF (r) = \left[ 1 + \left( \frac{r}{r_0} \right)^2 \right]^{-\alpha},
\label{eq:PSF}
\end{equation}
with $r$ the distance to the centre, $r_0 = 5.304$\,arcsec and $\alpha = 1.589$\footnote{\url{https://xmm-tools.cosmos.esa.int/external/xmm_calibration/calib/documentation/epic_cal_meetings/200111/PSF-MOS_Ghizzardi.pdf}}.
We note that the number of pixels of the PSF must be odd to account for the centre.
The measured surface brightness writes:
\begin{equation}
S_X^{\mathrm{conv}} (x, y) = (S_X^{\mathrm{model}} \circledast PSF) (x, y) 
\label{eq:apply_PSF_Xray}
\end{equation}
where $S_X^{\mathrm{model}}$ is provided in equation (\ref{eq:X-ray_surface_brightness_fla}). 

As the edge effect is quite important, we decide not to use the borders, and to cut 8 pixels on each side of the map.
The final comparison maps are respectively 72 and 62 pixels wide for MACS\,J0242 and MACS\,J0949.
We note that we still have to compute our model for the full width of the original maps, as they are needed in the convolution.

\subsection{Converting surface brightness into count signal}

We use the EPIC count maps of \textit{XMM-Newton}, masking point-like sources, including diffuse emission only. As we also have access to the time of exposure, $E$, and total background, $B$ (particle background, soft protons, PN chip out-of-time events), maps from surface brightness models, we can make predictions on the number count of detection:
\begin{equation}
    N_{X, c} =  C_{\mathrm{flux}}^{\mathrm{count}} S_X^{\mathrm{conv}} \times E + B + C_{\mathrm{sky}}
    \label{eq:number_counts_X_formula}
\end{equation}
where $C_{\mathrm{sky}}$ is the sky constant, measured in the empty regions of the raw count map.

We also take dust absorption into account, with the absorption ratio \citep[see e.g.][]{2000ApJ...542..914W}:
\begin{equation}
    \frac{I_{\mathrm{obs}}}{I_{\mathrm{em}}} = \exp \left[ - n_H^{\mathrm{gal}} \sigma (E_X) \right],
\label{eq:dust_absorption}
\end{equation}
where $n_H^{\mathrm{gal}}$ is the galactic hydrogen, and $\sigma$, the extinction cross-section of that same dust for a photon energy, $E_X$.
We report absorption factors of $0.9439$ and $0.9398$ for clusters MACS\,J0242 and MACS\,J0949 respectively.

\subsection{Cooling curve}

\begin{figure}
    \centering
    \includegraphics[width=\columnwidth]{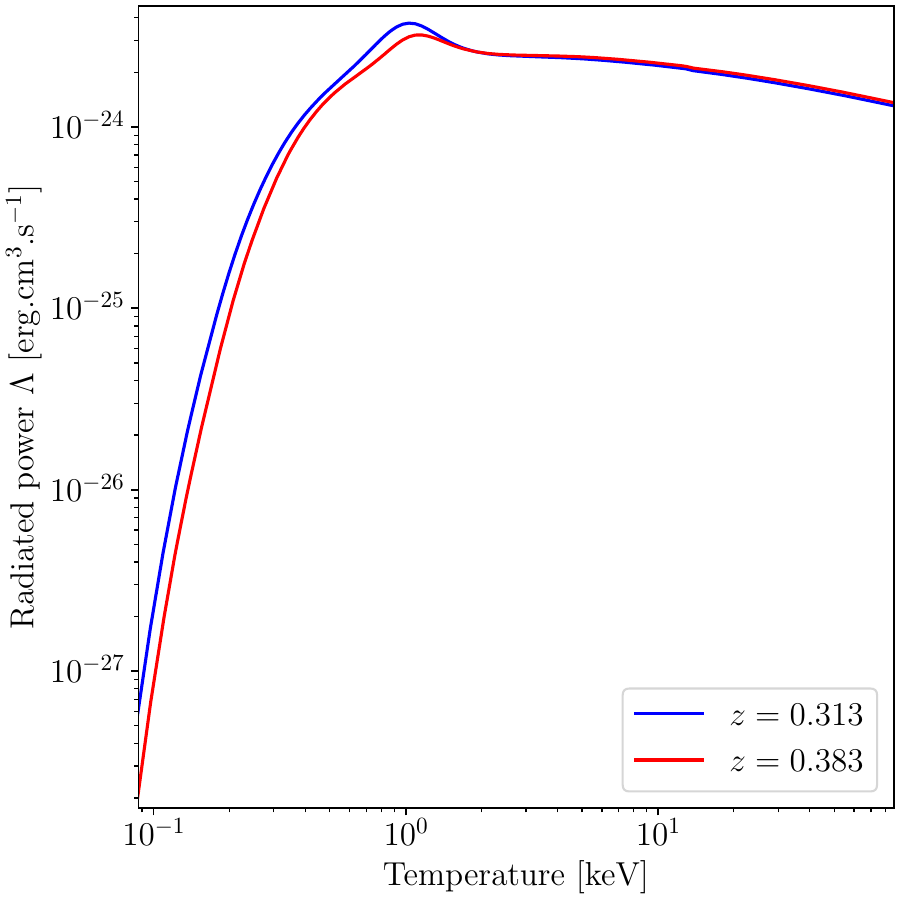}
    \caption{Radiated power (cooling curve) for a metallicity $Z = 0.3 Z_{\sole}$, using the metallicities described in \citet[][]{2009ARA&A..47..481A}, in the band $[0.7; 1.2]$\,keV, at the respective redshift of MACS\,J0242 (blue) and MACS\,J0949 (red).}
    \label{fig:cooling_curve_m0949}
\end{figure}

In order to access the X-ray spectral emissivity, $\Lambda (\Delta E, T_e, Z)$, mentioned in equation (\ref{eq:X-ray_surface_brightness_fla}), we use \href{https://atomdb.readthedocs.io/en/master/index.html}{\textsc{AtomDB}} \citep[see ][]{Foster_2012}.
With the metallicities of set \citet[][]{2009ARA&A..47..481A} adjusted to our $Z_{\mathrm{cl}} = 0.3 Z_{\sole}$, we can plot the cooling curve in the energy band of \textit{XMM-Newton} (adjusted for K-correction). The results at the redshifts of clusters MACS\,J0242 and MACS\,J0949 are displayed in Fig.\,\ref{fig:cooling_curve_m0949}.

\subsection{Sunyaev-Zel'dovich maps filtering}
\label{subsec:SZ_filtering}

We simply filter both ACT DR5 map $\Theta_{r,f}^{\mathrm{obs}}$ of MACS\,J0949 with a Gaussian filter of radius $0.05$\,degrees $\mathcal{G}(0.05\deg)$, with \textsc{nemo}\footnote{\url{https://nemo-sz.readthedocs.io}}. We then subtract it from the original map, thus high-pass Gaussian filtering the original map.
Although this does not allow to fully remove either the CMB signal or the atmosphere variability, at the scale of the cluster, it allows to smooth and attenuate the CMB variability.
In order to compare our SZ effect model to the filtered data maps, we convolve the modelled signal map $\Theta_{r,f}^{\mathrm{mod}}$ with the ACT beam $\mathcal{B}_f$ at the map frequency $f$. This allows to take the telescope PSF into account. We further apply the Gaussian filter, and compare the resulting maps:
\begin{equation}
\begin{split}
    \Theta_{r,f}^{\mathrm{obs, filt}} &= \Theta_{r,f}^{\mathrm{obs}} - \Theta_{r,f}^{\mathrm{obs}} \circledast \mathcal{G}(0.05\deg),\\
    \Theta_{r,f}^{\mathrm{mod, filt}} &= \left( \Theta_{r,f}^{\mathrm{mod}} \circledast \mathcal{B}_f \right) \circledast \mathcal{G}(0.05\deg).
\end{split}
\label{eq:SZ_maps}
\end{equation}
In a similar fashion to that of the X-ray PSF, we have to remove the borders of the SZ filtered image because of border effects. 
We therefore used model maps of $\sim 4$ Mpc initially (26 pixels), reduced to $\sim 2$\,Mpc (14 pixels), and compared these predictions to the ACT DR5 filtered data.

\begingroup
\renewcommand{\arraystretch}{1.5}
\begin{table*}
	\centering
	\caption{Best fit of all optimisation models for cluster MACS\,J0242.
    The columns are as follows: (i) The observation type used to constrain the profile. (ii) The profile type. (iii) $\rho_{0,1}$ denotes the DMH central density in the case of \texttt{idPIE} profile, $\rho_S$ in the \texttt{iNFW} case, and the central gas density $\rho_c$ in the case of a $\beta$-profile. All density values are displayed in g.cm$^{-3}$.
    (iv) $a_1$ denotes the DMH core radius in the \texttt{idPIE} model, $r_S$ the scale radius in the \texttt{iNFW} case, and $r_c$ in the $\beta$-model.
    All of these distances are displayed in kpc. 
    (v) $s_1$ denote respectively the cut radius of the DMH and of the BCG, in the case \texttt{idPIE}.
    (vi) $\rho_{0, 2}$ denotes the BCG central density in the \texttt{idPIE} case.
    (vii) $s_2$ is the cut radius of the BCG.
    (viii) $\beta$ is the power index of the $\beta$ profile.
    (ix) $-\ln \mathcal{L}_X$ is the negative log-likelihood evaluated on X-rays data.
	When a parameter set is out of the invertible range of function $\mathcal{J}_z$ (i.e. $\sim [0, 10]$), we denote the log-likelihood $\ln \mathcal{L}$ as infinite.
    For cluster MACS\,J0242, the optimisation is only performed with \textit{XMM-Newton} data. 
    The core radius of the BCG $a_2$ is fixed for both the lensing and ICM optimisations.
    Starred values were fixed.
    We find the optimisation of $s_2$ to be degenerate; we therefore fix this parameter.
	}
	\label{tab:m0242_ICM_opt_summary}
\makebox[\textwidth][c]{
    \begin{tabular}{llcccccccc}
\hline
\hline
Observation & Profile & $\rho_{0,1}$, $\rho_S$ or $\rho_c$ & $a_1$, $r_S$ or $r_c$ & $s_1$ & $\rho_{0,2}$ & $s_2$ & $\beta$ & $- \ln \mathcal{L}_X$\\
Unit & & [g.cm$^{-3}$] & [kpc] & [kpc] & [g.cm$^{-3}$] & [kpc] & & \\
\hline
Lensing & dPIE & $1.0_{-0.2}^{+3.4} \times 10^{-24}$ & $57.2_{-8.4}^{+6.0}$ & $1500^{\star}$ & $1.2_{-0.2}^{+0.3} \times 10^{-20}$ & $177.6_{-58.0}^{+32.2}$ & \_ & $\infty$\\
 & NFW & $3.4_{-0.4}^{+0.5} \times 10^{-25}$ & $209.9_{-15.8}^{+17.1}$ & \_ & \_  & \_ & \_ & $\infty$\\
\hline
ICM & $\beta$ & $2.6_{-1.8}^{+11.0} \times 10^{-23}$ & $17.9_{-13.3}^{+38.1}$ & \_ & \_ & \_ & $0.54_{-0.08}^{+0.20}$ & $0.68$\\
 (X-rays only) & \texttt{idPIE} & $5.3_{-2.4}^{+9.3} \times 10^{-25}$ & $73.0_{-23.2}^{+34.7}$ & $2720_{-890}^{+460}$ & $4.3_{-2.7}^{+3.9} \times 10 ^{-21}$ & $177.6^{\star}$ & \_ & $0.67$\\
& \texttt{iNFW} & $1.3_{-0.6}^{+15.9} \times 10^{-25}$ & $320.5_{-230.1}^{+99.0}$ & \_ & \_  & \_ & \_ & $0.70$\\
\hline
\hline
\end{tabular}
}
\end{table*}
\endgroup

\subsection{Working hypotheses on the density distribution}

We based our analysis on the mass models described in our previous work \citep{2023MNRAS.522.1118A}.
We model the total mass profile of MACS\,J0242 and MACS\,J0949 using strong lensing constraints of radii ranging in  $R \in [50, 200]$\,kpc.
While we acknowledge the limitations of the constraining power, we extrapolate the 3D density out to $R_{500,c}$, and recover a mass $M_{500,c} = 5.95_{-0.46}^{+0.40} \times 10^{14} \Msol$ and $11.48_{-3.42}^{+0.00} \times 10^{14} \Msol$ for clusters MACS\,J0242 and MACS\,J0949 respectively.
With equation (\ref{eq:T_500_universal}), we find temperature normalisations $T_0 = \eta_T T_{500} = 7.30$\,keV and $11.63$\,keV respectively.

Given the quality of the X-ray and SZ observations, and the dominant importance of the DMH and BCG in strong lensing models, we neglect the potentials associated to the individual galaxies, owing to the dynamical state of the clusters, analysed in \citet{2023MNRAS.522.1118A}. 
As they are not strongly perturbed, the ICM distribution should be governed by the large-scale potential. In Fig.\,\ref{fig:astro_images}, we can for instance see cluster MACS\,J0949 Southern halo mass contours to be undetectable on the X-rays -- see the large red circle in the South of the image, crossed by the dashed green line.

Moreover, we postulate that the ICM density distribution is ellipsoidally symmetric, and of the same ellipticity as the DMH potential. 
This simply follows the hypothesis `ICM traces potential'.
Therefore, as a natural consequence of the hydrostatic equilibrium, we expect the potential to be rounder than the total matter distribution (this is qualitatively verified on Fig.\,\ref{fig:astro_images}, and the potential ellipticity formula is presented in Appendix \ref{sec:ellipticity}).

Finally, as the line-of-sight ellipticity of the potential is assumed to be equal to the geometric average of the semi-major and semi-minor axes, $\sqrt{a b}$.
As the goal of this article is to present new methods to predict the density profile of the ICM using strong lensing analyses, we fix all geometric parameters to their best lensing values. The `semi-depth' of the cluster is unknown from lensing, but it is also degenerate with the density distribution. Therefore, we do not optimise this parameter.


\subsection{MCMC optimisations}

For the two different types of models of the electron density -- canonical and analytical -- represented by three different models -- $\beta$, \texttt{idPIE} and \texttt{iNFW} -- we let a number of parameters free. 
For the $\beta$ profile, we set all three parameters of the density distribution $\left\{\rho_0; r_c; \beta\right\}$ free.
For the \texttt{idPIE} profile, we initially let $\left\{\rho_{0,1}; a_1; s_1; \rho_{0,2}; s_2\right\}$ free, but as discussed in Sect.\,\ref{sec:ICM-Optimised_Results}, $s_2$ appears entirely degenerate in the optimisation, and we therefore fix it to its lensing value.
The two parameters characterising a NFW distribution, $\left\{\rho_{S}; r_S\right\}$, are set free for the \texttt{iNFW} optimisation.
We optimise them with the data for each galaxy cluster. For cluster MACS\,J0242, we only use the X-ray data, while for MACS\,J0949, we have the choice to use either X-rays, SZ, or both.

We define the log-likelihood for the X-ray data. As the photon counts are limited, the X-ray maps are following a Poissonian distribution.
We take them to follow the Cash statistic \citep{1979ApJ...228..939C}: 
\begin{equation}
    \ln \mathcal{L}_X (\Theta) = \frac{1}{N_X} \sum_i \left[ C_i - M_i (\Theta) - C_i \ln \left( \frac{C_i}{M_i (\Theta)} \right) \right],
\label{eq:llikelihood_Cash}
\end{equation}
where $C_i = N_{X,c,i}$ is the data count in the $i$-th pixel (see eq. \ref{eq:number_counts_X_formula}), $N_X$, the number of pixels, and $M_i (\Theta)$, the model prediction for the parameter vector, $\Theta$.

As for the SZ statistic, with $M_i$ now being the temperature contrast model, and $C_i$, its SZ measurement, we simply take the likelihood to be Gaussian:
\begin{equation}
\begin{split}
    \ln \mathcal{L}_{SZ} (\Theta) &= - \frac{1}{2 N_{SZ}} \sum_i \left[ \left( \frac{M_i (\Theta) - C_i}{\sigma_i} \right)^2 + \ln \sigma_i^2 \right],\\ 
    \sigma_i^2 &= M_i^2 (\Theta) + \sigma_{C,i}^{2},
\end{split}
\label{eq:llikelihood_SZ}
\end{equation}
where $\sigma_i$ is the standard deviation in the $i$-th pixel, and $N_{SZ}$, the number of SZ pixels. 
We take the model standard deviation to be the model itself, accordingly to a Gaussian model.
$\sigma_{C,i}^2$ is the instrument variance of ACT. 
This does not take into account the CMB variance nor the atmosphere, but these are smoothed out on the scale of a cluster by the top-hat filtering \citep[we follow][]{2018ApJS..235...20H,2021ApJS..253....3H}.
$C_i$ and $M_i$ represent here the data and model respectively, but for the SZ data.
In the case of cluster MACS\,J0949, we sum the log-likelihood of the 90 and 150\,GHz ACT DR5 bands. 

In the case of the joint optimisation of X-rays and SZ, the data are of the same type, i.e. detections in pixels.
We therefore define the joint likelihood as the weighted sum:
\begin{equation}
    \ln \mathcal{L}_{J} = \frac{N_X \ln \mathcal{L}_{X} + N_{SZ}\ln \mathcal{L}_{SZ}}{N_X + N_{SZ}},
\label{eq:llikelihood_joint}
\end{equation}
where $N_{SZ}$ must be understood as the sum of all SZ pixels, both in band f090 and f150. This takes into account the different pixelisations, and attributes equal weights to each pixel. The X-ray observations thus dominate, given the much better resolution (a \textit{XMM-Newton} pixel represents $2.5''$, and an ACT pixel $30''$).

We used the package \textsc{emcee}, took 100 walkers, iterated over 5000 steps, with a step type \texttt{\href{https://emcee.readthedocs.io/en/stable/user/moves/}{emcee.StretchMove}}.
We provide the cornerplots, realised with package \textsc{corner} \citep[][]{corner}, in Appendix \ref{sec:Cornerplots}.

\section{ICM-Optimised results}
\label{sec:ICM-Optimised_Results}

We present in this section the results of MCMC optimisations of each of the four ICM density models, for each galaxy cluster.

\subsection{Cluster MACS J0242}
\label{subsec:ICM_opt_m0242}

\begin{figure*}
    \centering
    \includegraphics[width=\textwidth]{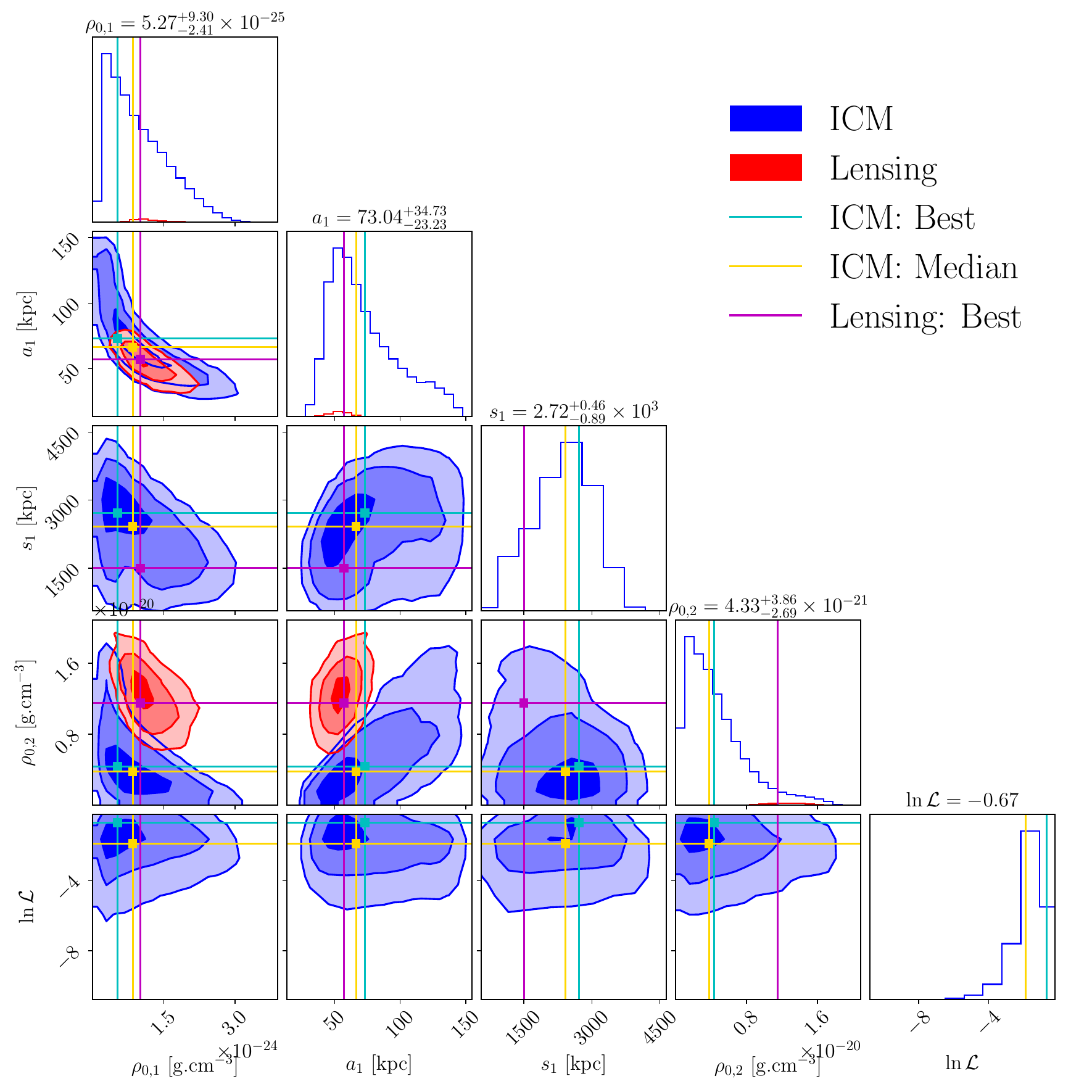}
    \caption{MCMC optimisation for \texttt{idPIE} model of the four relevant parameters for cluster MACS\,J0242: DMH central density $\rho_{0,1}$ in g.cm$^{-3}$, core radius $a_1$ and cut radius $s_1$ in kpc, and BCG central density $\rho_{0,2}$ in g.cm$^{-3}$.
    \textit{Blue:} Optimisation performed using the available ICM data (X-ray here).
    \textit{Red:} Strong lensing optimisation.
    \textit{Cyan:} Best-fit value of the ICM optimisation.
    \textit{Gold:} Median ICM optimisation.
    \textit{Magenta:} Best-fit strong lensing model (described in Table\,\ref{tab:best_model}).
    }
    \label{fig:m0242_MCMC_idPIEMD_v871}
\end{figure*}

\subsubsection{$\beta$ model}
\label{subsubsec:m0242_beta}

In order to compare our profiles inferred from lensing with the more popular profiles describing the ICM, we run a MCMC optimisation for the $\beta$ profile. 
We present the optimised parameters in Table\,\ref{tab:m0242_ICM_opt_summary}, alongside all other optimisations for MACS\,J0242. 
The associated cornerplot is presented Fig.\,\ref{fig:m0242_MCMC_beta_v872}.

We find the best likelihood to be $\ln \mathcal{L} = -0.68$, close enough from the best possible likelihood, $-0.5$, so that we can expect other canonical ICM profiles -- such as double-$\beta$ -- not to significantly improve the model.

\subsubsection{\texttt{idPIE}}
\label{subsubsec:m0242_idPIE}

\begin{figure*}
\begin{minipage}{0.48\textwidth}
\centerline{\includegraphics[width=1\textwidth]{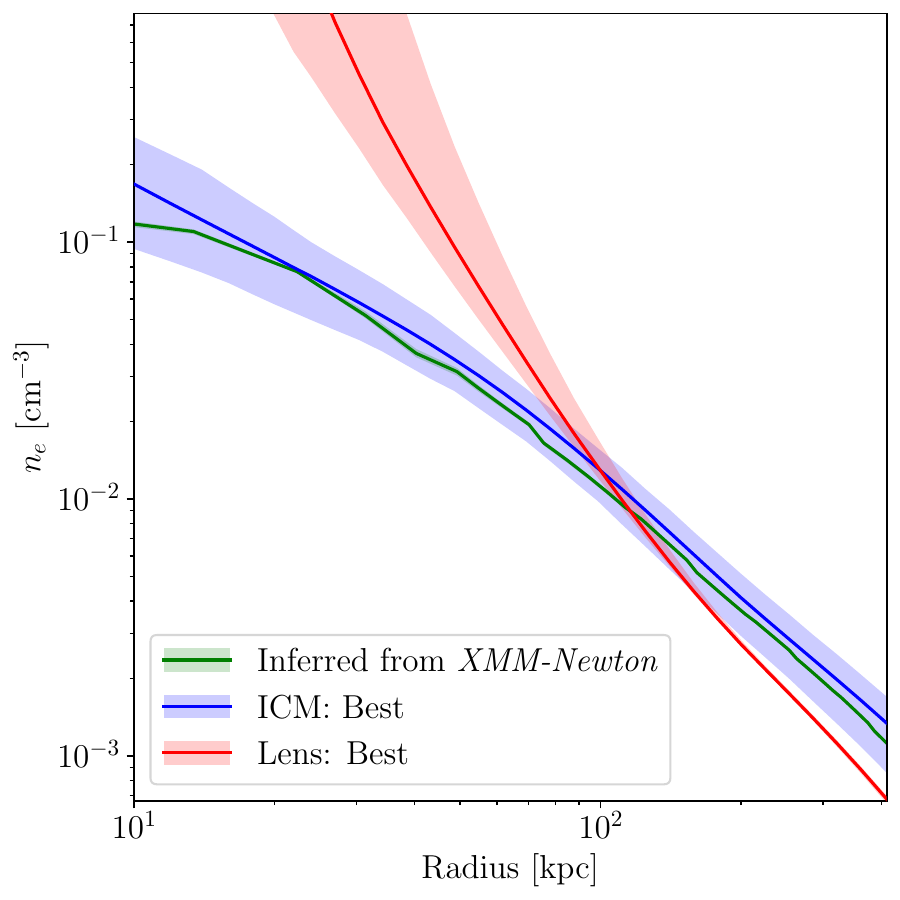}}
\end{minipage}
\hfill
\begin{minipage}{0.48\textwidth}
\centerline{\includegraphics[width=1\textwidth]{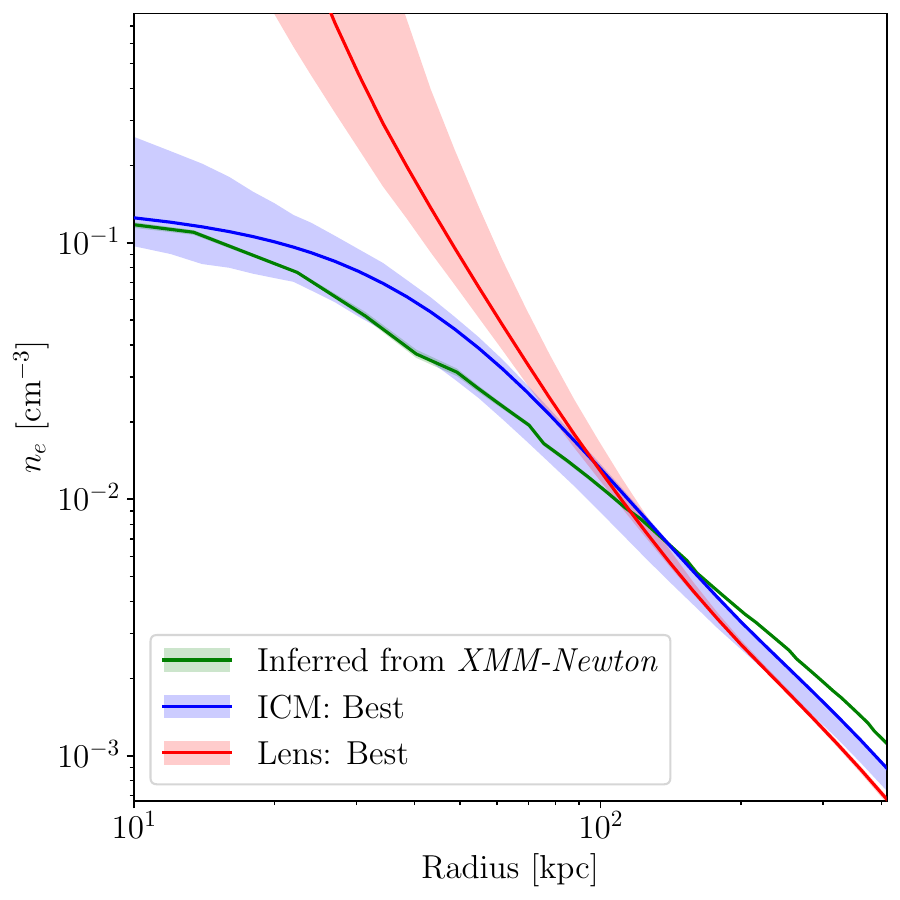}}
\end{minipage}
\caption{Electron density $n_e$ for \texttt{idPIE} model, for cluster MACS\,J0242.
\textit{Green:} X-ray surface brightness deprojected profile (assuming spherical symmetry).
\describecolours
\textit{Left:} In the case of the optimisation of parameters $\rho_{0,1}$, $a_1$, $s_1$ and $\rho_{0,2}$, as illustrated in Fig.\,\ref{fig:m0242_MCMC_idPIEMD_v871}. \textit{Right:} In the case of the optimisation of parameters $\rho_{0,1}$, $a_1$ and $\rho_{0,2}$, as illustrated in Fig.\,\ref{fig:m0242_MCMC_idPIEMD_v874}. 
This shows the lens-optimised ICM density model to be inconsistent with the X-ray data. As shown on Fig.\,\ref{fig:m0242_MCMC_idPIEMD_v871}, the X-ray optimised $\rho_{0,2}$ parameter yields lower values than the SL ones. This may be caused by ICM turbulence in the centre, or simply due to central total densities much larger than modelled through the polytropic temperature model (equation \ref{eq:gamma_of_n}).
More significantly, the offset in the ICM-optimised model on the right panel shows that
if $s_1$ is not optimised, the larger scales ($\gtrsim 100$\,kpc) $n_e$ densities can not be properly fitted.
}
\label{fig:ne_m0242_MCMC_idPIEMD_v871_v874}
\end{figure*}

\begin{figure*}
\begin{minipage}{0.48\textwidth}
\centerline{\includegraphics[width=1\textwidth]{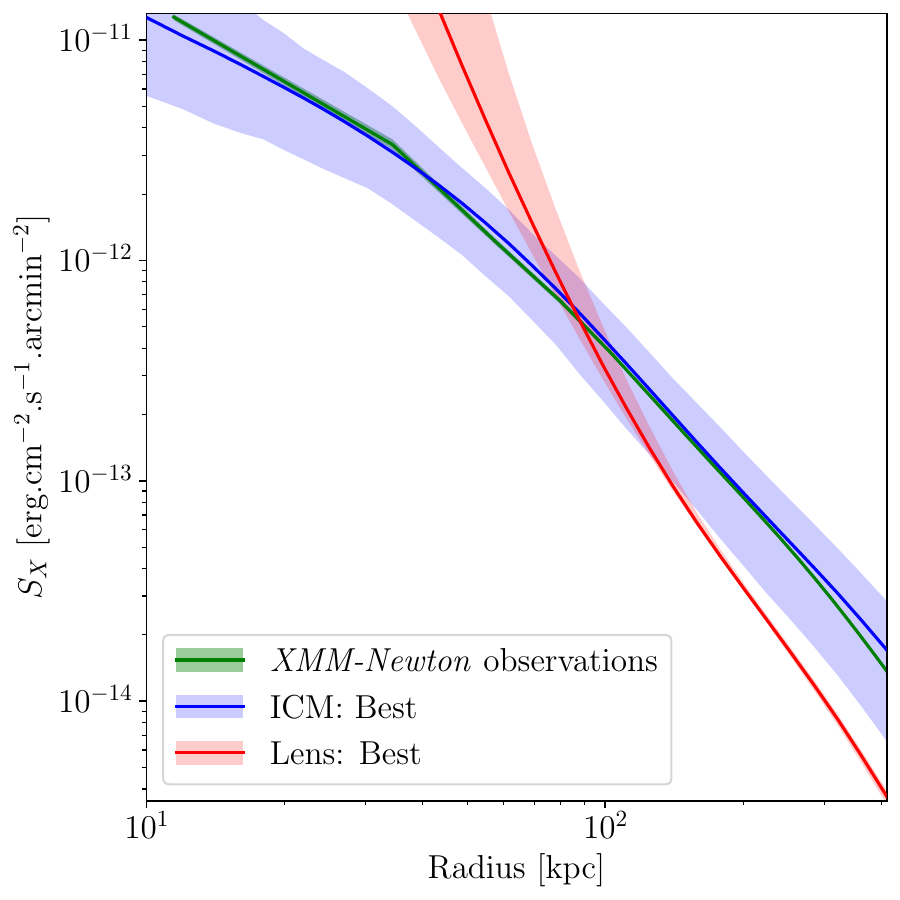}}
\end{minipage}
\hfill
\begin{minipage}{0.48\textwidth}
\centerline{\includegraphics[width=1\textwidth]{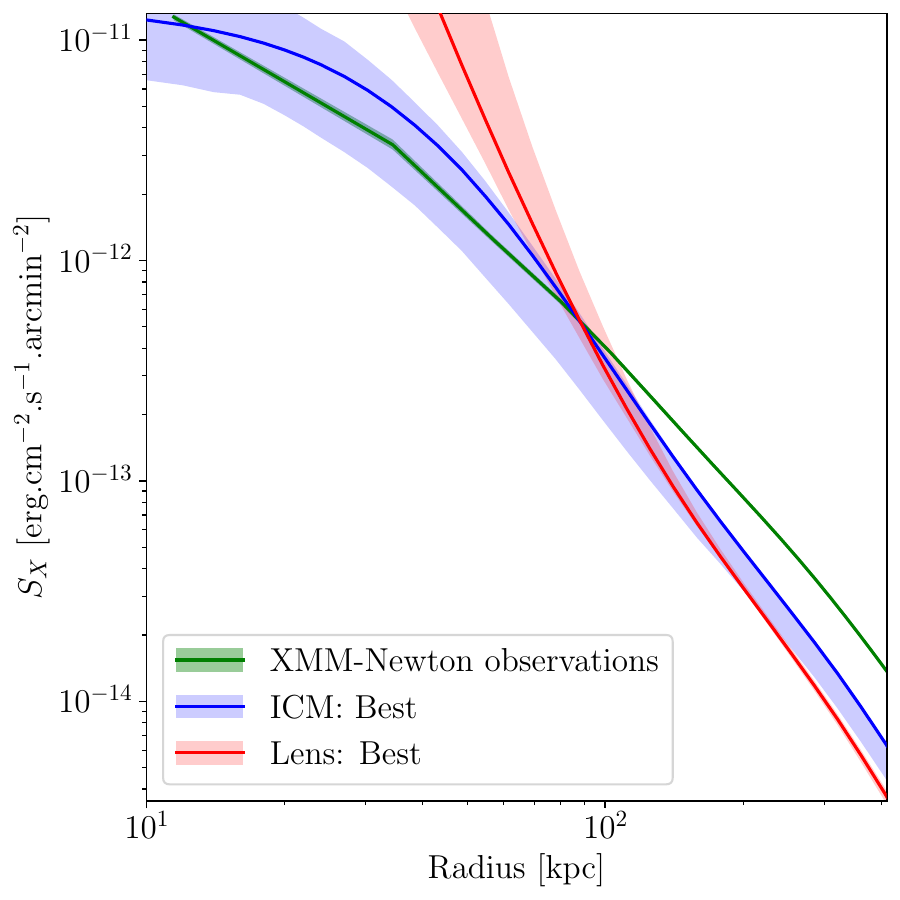}}
\end{minipage}
\caption{Expected X-ray surface brightness $S_X$ for \texttt{idPIE} model, for cluster MACS\,J0242. 
\textit{Green:} X-ray surface brightness deprojected profile (assuming spherical symmetry).
\describecolours
\textit{Left:} In the case of the optimisation of parameters $\rho_{0,1}$, $a_1$, $s_1$ and $\rho_{0,2}$, as illustrated in Fig.\,\ref{fig:m0242_MCMC_idPIEMD_v871}. \textit{Right:} In the case of the optimisation of parameters $\rho_{0,1}$, $a_1$ and $\rho_{0,2}$, as illustrated in Fig.\,\ref{fig:m0242_MCMC_idPIEMD_v874}.}
\label{fig:SX_m0242_MCMC_idPIEMD_v871_v874}
\end{figure*}

\begingroup
\renewcommand{\arraystretch}{1.5}
\begin{table*}
	\centering
	\caption{Best fit of all optimisation models for cluster MACS\,J0949. 
    The columns are as follows: (i) The observation type used to constrain the profile. (ii) The profile type. (iii) $\rho_{0,1}$ denotes the DMH central density in the case of \texttt{idPIE} profile, $\rho_S$ in the \texttt{iNFW} case, and the central gas density $\rho_c$ in the case of a $\beta$-profile. 
    (iv) $a_1$ denotes the DMH core radius in the \texttt{idPIE} model, $r_S$ the scale radius in the \texttt{iNFW} case, and $r_c$ in the $\beta$-model.
    All of these distances are displayed in kpc.
    (v) $s_1$ denote respectively the cut radius of the DMH and of the BCG, in the case \texttt{idPIE}.
    (vi) $\rho_{0, 2}$ denotes the BCG central density in the \texttt{idPIE} case.
    (vii) $s_2$ is the cut radius of the BCG.
    (viii) $\beta$ is the power index of the $\beta$ profile.
    (ix) $-\ln \mathcal{L}_J$ is the negative joint SZ-X-ray log-likelihood.
    The core radius of the BCG $a_2$ is model-dependent, and is thus it is not optimised here.
    Starred values were fixed.
	}
	\label{tab:m0949_ICM_opt_summary}
\makebox[\textwidth][c]{
\begin{tabular}{llcccccccc}
\hline
\hline
Observation & Profile & $\rho_{0,1}$, $\rho_S$ or $\rho_c$ & $a_1$, $r_S$ or $r_c$ & $s_1$ & $\rho_{0,2}$ & $s_2$ & $\beta$ & $- \ln \mathcal{L}_J$\\
Units & & [g.cm$^{-3}$] & [kpc] & [kpc] & [g.cm$^{-3}$] & [kpc] & & \\
\hline
Lensing & dPIE & $4.6_{-1.0}^{+3.6} \times 10^{-25}$ & $116.3_{-51.7}^{+24.1}$ & $1500^{\star}$ & $3.9_{-0.6}^{+8.3} \times 10^{-18}$ & $98.0_{-34.3}^{+153.7}$ & \_ & $3.54$\\  
 & NFW & $1.2_{-0.0}^{+1.6} \times 10^{-25}$ & $405.5_{-156.1}^{+0.0}$ & \_ & \_  & \_ & \_ & $1.42$\\
\hline
ICM & $\beta$ & $1.7_{-1.0}^{+1.6} \times 10^{-24}$ & $108.4_{-48.2}^{+175.0}$ & \_ & \_ & \_ & $0.50_{-0.13}^{+0.28}$ & $0.61$\\
(X-rays and SZ) & \texttt{idPIE} & $3.6_{-1.8}^{+8.2} \times 10^{-25}$ & $96.1_{-32.6}^{+60.0}$ & $3780_{-1600}^{+420}$ & $1.9_{-1.1}^{+9.5} \times 10^{-21}$ & $98.0^{\star}$ & \_ & $0.60$\\ 
& \texttt{iNFW} & $6.4_{-2.1}^{+8.9} \times 10^{-26}$ & $546.2_{-186.0}^{+110.3}$ & \_ & \_  & \_ & \_ & $0.61$\\
\hline
\hline
\end{tabular}
}
\end{table*}
\endgroup

In the strong lensing optimisation -- presented Table\,\ref{tab:best_model} -- parameters $\left\{\rho_{0,1}; a_1 ; \rho_{0,2}; s_2\right\}$ were optimised, and the DMH cut radius, $s_1$, to which  strong lensing is insensitive, was fixed to 1.5\,Mpc. 
Here, we optimise the \texttt{idPIE} density profiles with these density parameters using the \textit{XMM-Newton} observations. We display both the strong lensing and ICM optimisations for the $\{\rho_{0,1}; a_1; s_1; \rho_{0,2}\}$ and $\{\rho_{0,1}; a_1; \rho_{0,2}\}$ parameter spaces in Fig.\,\ref{fig:m0242_MCMC_idPIEMD_v871} and \ref{fig:m0242_MCMC_idPIEMD_v874} respectively.

The X-ray optimisation of the BCG cut radius, $s_2$, is degenerate (see Fig.\,\ref{fig:m0242_MCMC_idPIEMD_v879}).
We therefore fix $s_2$ to its lensing value, $s_2 = 177.6$\,kpc from now on.
Although the prediction from lensing is diverging in the cluster centre ($r \to 0$) -- due to the central matter density being out of the $\mathcal{J}_z$ function bijective range -- 
we find the ICM optimisation in the $\{\rho_{0,1}; a_1\}$ space to provide results in agreement with those from the lensing optimisation, despite the larger ICM error bars.
The density normalisation of the BCG, $\rho_{0,2}$, yields different results than strong lensing.
This may be explained by the high central matter density, which yields values out of the $\mathcal{J}_z$ function inversion range.
As the validity of equations (\ref{eq:pression_temperature_polytropic_Eckert}) and (\ref{eq:gamma_of_n}) paired together is not verified beyond $n_e > 10^{-1}$\,cm$^{-3}$, the ICM optimisation avoids regions of the parameter space yielding an ICM density above this value.
Moreover, since the X-ray signal in the centre carries a high variability, we can not conclude about its significance.

We note that fixing the parameter $s_1$ to a value of 1.5\,Mpc in the ICM optimisation results in a best log-likelihood $\ln \mathcal{L} < -1.06$. 
On the contrary, when this parameter is optimised, the best value gets closer to the `perfect fit' $-0.5$ value at $\ln \mathcal{L} < -0.67$ (i.e. as good as the $\beta$ model). 
In this case, we can only notice the X-ray optimisation is bound to diverge from the fiducial 1.5\,Mpc value, as the best optimisation yields $s_1 = 2.72_{-0.87}^{+0.44}$\,Mpc.
The electronic densities, $n_e$, are represented in Fig.\,\ref{fig:ne_m0242_MCMC_idPIEMD_v871_v874}, and the \textit{XMM-Newton} physical -- i.e. without PSF effects -- X-ray surface brightness, $S_X$, in Fig.\,\ref{fig:SX_m0242_MCMC_idPIEMD_v871_v874}. 
If $s_1$ is not optimised, we notice a discrepancy between the ICM best-fit and X-ray data profile at large radii ($R > 200$\,kpc). This emphasises the necessity to let this parameter free.
We conclude our method can properly fit the X-ray signal for this cluster, provided the potential is optimised, and notably $s_1$. 
We further discuss the the importance of $s_1$ and the large discrepancy between the lensing-inferred model and the observations in Sect.\,\ref{sec:optimise_rcut}.

We can moreover outline the much larger error bars in the ICM optimisation, compared to the lens optimisation, in Fig.\,\ref{fig:m0242_MCMC_idPIEMD_v871}. This is due to the inherent difference in the data quality.
The high sensitivity of the $\mathcal{J}_z$ function (represented in Fig.\,\ref{fig:Jz_curve}) could be described as a double-edged sword: on the one edge, this sensitivity means any imprecision in the determination of the parameters, or even in the hypotheses (temperature model, temperature normalisation, determination of the total mass density, etc.) would result into a magnified error, i.e. a prediction error on the ICM density much larger than the total matter density associated error.
On the other edge, this allows to finely tune certain parameters. 
In order to reach such a quality in the reconstruction, the strong lensing parameters should be very finely determined, and fixed. The parameters not fit with strong lensing ($s_1$ here) could then be optimised.

\subsubsection{\texttt{iNFW}}

The NFW distributions attributed to strong lensing are all reductions of dPIE \textsc{Lenstool} optimisations to NFW best fit. 
In Fig.\,\ref{fig:m0242_MCMC_iNFW_v873}, we compare these to the ICM-optimised \texttt{iNFW} parameters. 
The latter yields a very satisfactory best likelihood value again at $\ln \mathcal{L} = -0.70$, and although the $\{\rho_S; r_S\}$ values we find are different from those of the strong lensing reduction, they are compatible with the total density we found.

\subsection{Cluster MACS J0949} \label{subsec:ICM_opt_m0949}

Similarly to the study performed on the cluster MACS\,J0242, we present the results of the ICM optimisation of MACS\,J0949.
We primarily present the joint fit results (X-rays and SZ effect).

\subsubsection{$\beta$ model}
\label{subsubsec:m0949_beta}

Figure\,\ref{fig:m0949_MCMC_beta_v878} presents the optimisation with the ICM data from \textit{XMM-Newton}
and the SZ data taken with ACT.
Using both the X-rays and SZ data for this optimisation, we find the best likelihood to be $\ln \mathcal{L}_J = -0.61$, a value which supports the good quality of the fit.
In detail, the X-ray likelihood is $\ln \mathcal{L}_X = -0.58$, and the SZ likelihood is $\ln \mathcal{L}_{SZ} = -0.88$.

\subsubsection{\texttt{idPIE}}
\label{subsubsec:idPIE_m0949}

\begin{figure*}
    \centering
    \includegraphics[width=\textwidth]{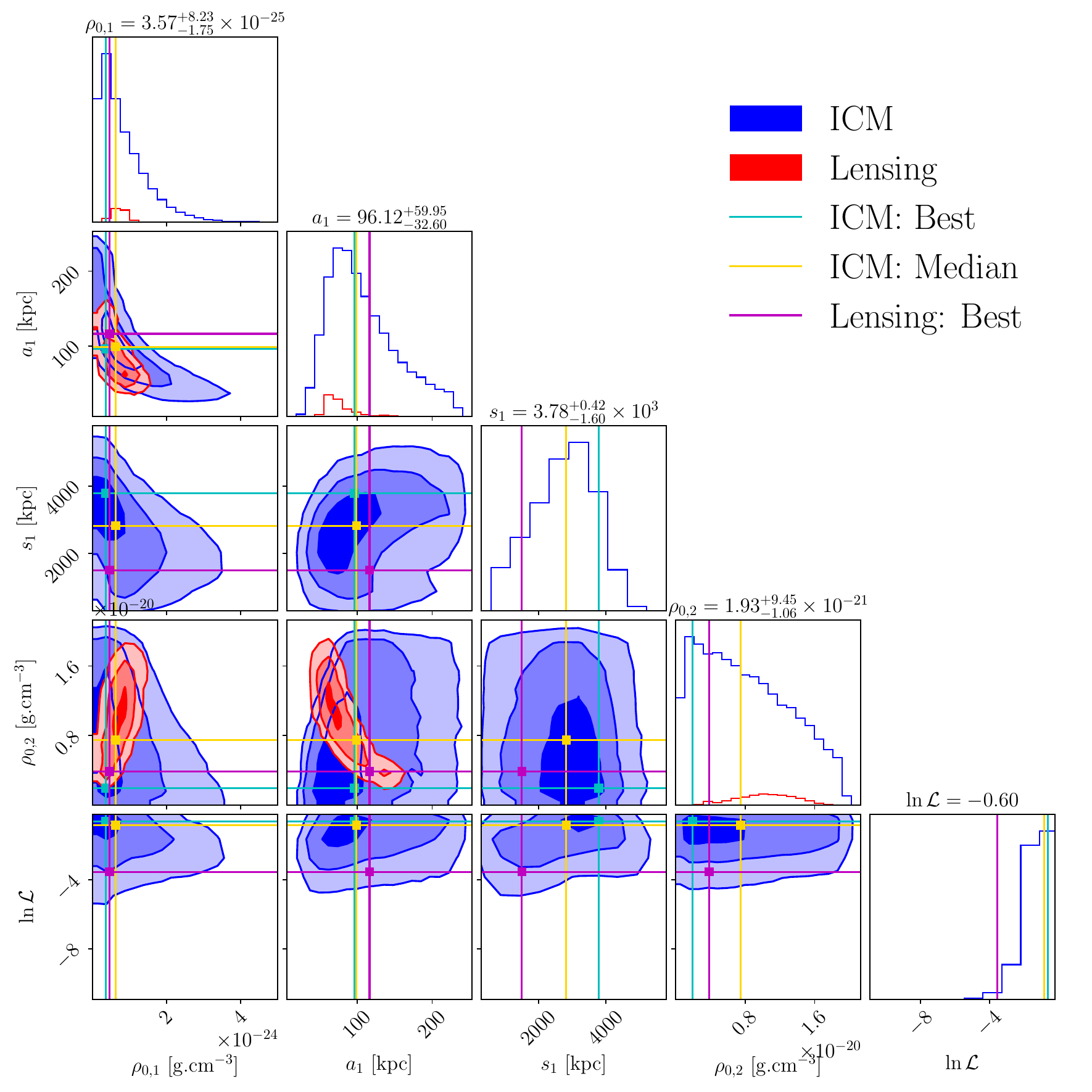}
    \caption{MCMC optimisation for \texttt{idPIE} model of the four relevant parameters for cluster MACS\,J0949: DMH central density $\rho_{0,1}$, core radius $a_1$ and cut radius $s_1$, and BCG central density $\rho_{0,2}$. The optimisation was performed with the X-ray and SZ data.
    \textit{Blue:} Optimisation performed using the available ICM data.
    \textit{Red:} Strong lensing optimisation.
    \textit{Cyan:} Best ICM optimisation.
    \textit{Gold:} Median of the ICM optimisation.
    \textit{Magenta:} Best strong lensing model (described in Table\,\ref{tab:best_model}).
    }
    \label{fig:m0949_MCMC_idPIEMD_v870}
\end{figure*}

\begin{figure*}
\begin{minipage}{0.48\textwidth}
\centerline{\includegraphics[width=1\textwidth]{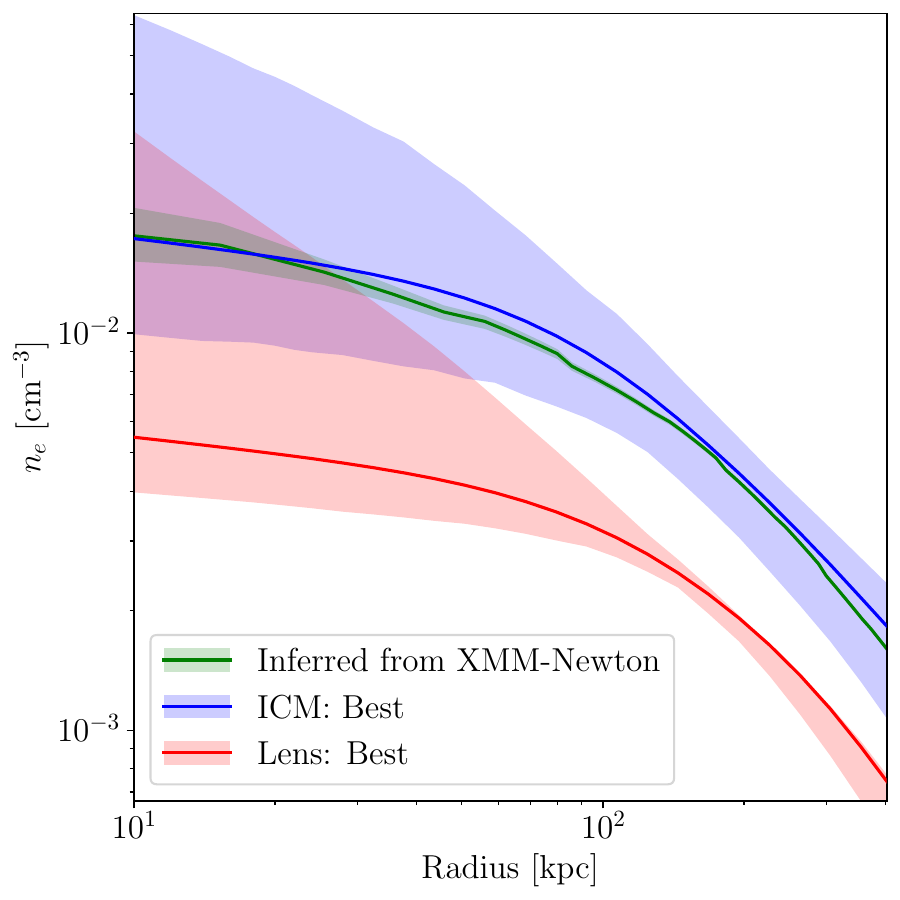}}
\end{minipage}
\hfill
\begin{minipage}{0.48\textwidth}
\centerline{\includegraphics[width=1\textwidth]{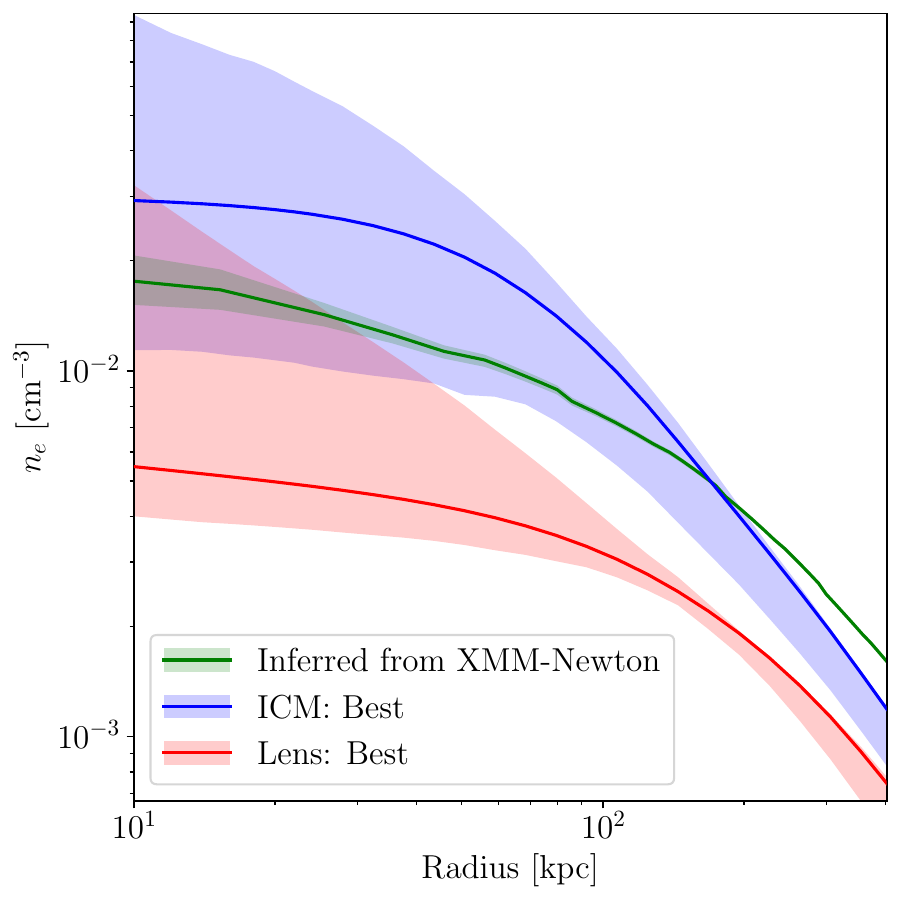}}
\end{minipage}
\caption{Electron density $n_e$ for \texttt{idPIE} model, for cluster MACS\,J0949. 
\textit{Green:} X-ray surface brightness deprojected profile (assuming spherical symmetry).
\describecolours
\textit{Left:} In the case of the optimisation of parameters $\rho_{0,1}$, $a_1$, $s_1$ and $\rho_{0,2}$, as illustrated in Fig.\,\ref{fig:m0949_MCMC_idPIEMD_v870}. 
\textit{Right:} In the case of the optimisation of parameters $\rho_{0,1}$, $a_1$ and $\rho_{0,2}$, as illustrated in Fig.\,\ref{fig:m0949_MCMC_idPIEMD_v875}.
Similarly to Fig.\,\ref{fig:ne_m0242_MCMC_idPIEMD_v871_v874}, we observe that not optimising $s_1$ prevents our \texttt{idPIE} model to fit the ICM density properly for $r \gtrsim 200$\,kpc.
}
\label{fig:ne_m0949_MCMC_idPIEMD_v870_v875}
\end{figure*}

As for MACS\,J0242, the optimisation of the BCG cut radius, $s_2$, is degenerate. We therefore choose to fix this parameter to its strong lensing value, $98.0$\,kpc.
We display in Fig.\,\ref{fig:m0949_MCMC_idPIEMD_v870} the strong lensing and ICM optimisations in the $\{\rho_{0,1}; a_1; s_1; \rho_{0,2}\}$ parameter space. 
The best-fit value to the ICM data is $\ln \mathcal{L}_J = -0.60$. 
The overlap between the strong lensing and ICM optimised spaces is obvious in this cornerplot. 
However, the quality of the reconstruction with the ICM does not converge as efficiently as that of strong lensing -- a result to be expected given the difference in methods and quality of data.

We present the comparison between the observations, the theoretical prediction using the lens model and the ICM-optimised model for observables, $n_e$, $S_X$, and $\Theta_r$ 
on Fig.\,\ref{fig:ne_m0949_MCMC_idPIEMD_v870_v875}, \ref{fig:SX_m0949_MCMC_idPIEMD_v870_v875} and \ref{fig:SZ_m0949_MCMC_idPIEMD_v870_v875} respectively.
Again, not optimising $s_1$ leads to a worst ICM optimisation (best log-likelihood $-1.06$, see Fig.\,\ref{fig:m0949_MCMC_idPIEMD_v875}).
In Fig.\,\ref{fig:ne_m0949_MCMC_idPIEMD_v870_v875}, the right panel shows the discrepancy between the X-rays inferred electron density and the best optimisation with a fixed DMH cut radius, $s_1 = 1.5$\,Mpc.
This demonstrates the importance of this parameter optimisation to recover the X-ray measured density profile. We further discuss this in Sect.\,\ref{sec:optimise_rcut}.

\subsubsection{\texttt{iNFW}}

In Fig.\,\ref{fig:m0949_MCMC_iNFW_v877}, we compare the best strong lensing optimisation fit to the ICM-optimised \texttt{iNFW} parameter space.
We find a best log-likelihood of $\ln \mathcal{L}_J = -0.61$.
The strong lensing fit NFW parameters and the ICM-optimised values are compatible within $1 \sigma$.

\subsection{Optimising the DMH cut radius with ICM core data and SL priors} 
\label{sec:optimise_rcut}


As strong lensing is only efficiently probing the most central regions of galaxy clusters ($R < 200$\,kpc), it only allows to effectively constrain the `central' parameters amongst those presented in \texttt{idPIE} optimisations, i.e. $\rho_{0,1}, a_1, \rho_{0,2}$ and $s_2$ -- leaving $s_1$ aside.
Even these strong lensing constrained parameters are not perfectly determined in the case of our study as we are limited in the number of multiple images and spectroscopic redshifts detected \citep[][]{2023MNRAS.522.1118A}.

In this context, for the \texttt{idPIE} ICM optimisations, we notice the efficiency of the optimisation of the dark matter halo cut radius, $s_1$.
For MACS\,J0949\footnote{For MACS\,J0242, we did not perform this optimisation, as the central pixels present a total density out of the invertible range of function $\mathcal{J}_z$, and optimising $s_1$ only would not suffice to bring the central density into the invertible range.}, we can use the \texttt{idPIE} model introduced in Sect.\,\ref{subsubsec:idPIE_m0949} and Fig.\,\ref{fig:m0949_MCMC_idPIEMD_v870}, and fix the parameters constrained through strong lensing, i.e. $\rho_{0, 1}, a_1$ and $\rho_{0,2}$. 
This new model, only performing the $s_1$-optimisation, reaches the best fit value, $s_1 = 3500_{-740}^{+520}$\,kpc, at $\ln \mathcal{L}_J = -0.63$.
This is much better than optimising all the other parameters excluding $s_1$ ($\ln \mathcal{L}_J = -1.06$).
Thus optimising $s_1$ could suffice to fit the ICM density.

To verify this $s_1$-optimisation does not affect the lens optimisation, we used the $s_1$ value for the best ICM optimisation of the \texttt{idPIE} model presented in Table\,\ref{tab:m0242_ICM_opt_summary} and \ref{tab:m0949_ICM_opt_summary} for respectively clusters MACS\,J0242 and J0949.
Fixing the DMH cut radius to (respectively) 2720 and 3780\,kpc, we perform a new lens optimisation for each cluster, following the procedure detailed in \citet{2023MNRAS.522.1118A}.
We find that, for both clusters, this does not change significantly any of the optimised parameters. 
We find a close match between the respective $rms$ of $0.37\arcsec$ and $0.16\arcsec$, compared to 0.39\arcsec and 0.15\arcsec $rms$ with $s_1$ fixed to 1.5\,Mpc.
We conclude that using the ICM-optimised $s_1$ values does not strongly influence the strong lensing reconstruction.

\section{Discussion}
\label{sec:Discussion}

\subsection{Importance of the DMH cut radius}
\label{subsec:importance_s1}

We remind the reader that the full \texttt{idPIE} ICM optimisation for clusters MACS\,J0242 and MACS\,J0949 respectively yielded $s_1 = 2720_{-870}^{+440}$\,kpc and $3780_{-1540}^{+390}$\,kpc -- this latter value being compatible with the $s_1$-optimisation only.
Amongst all \texttt{idPIE} profile parameters, $s_1$ is the only one not to be optimised with strong gravitational lensing.
Given the uncertainty in the determination of the lensing parameters due to observational limitations,
the stiffness of function $\mathcal{J}_z^{-1}$ magnifies small errors in the potential profile into significant ones in the ICM density.
Thus, these errors may contaminate the ICM-optimised $s_1$ value. If we could fix all other parameters (density parameters, geometry, and use a measured ICM temperature profile), the ICM-optimised value for $s_1$ should yield a physical result, with respect to the dPIE profile choice and the hydrostatic equilibrium hypothesis.
In spite of this constraint on the data quality, we have presented the importance of the DMH cut radius parameter, $s_1$, in the ICM density prediction, as its optimisation modifies the electron density, $n_e$, and thus the surface brightness, $S_X$, \textit{at all radii}. 
Indeed, this DMH cut radius parameter, $s_1$, is related to the total matter density density in the cluster outskirts, but has a direct influence in the central X-ray surface brightness.
This noticeable change in the ICM central density due to the $s_1$ DM halo cut radius is represented in Fig.\,\ref{fig:ne_m0949_MCMC_idPIEMD_v876}.
In the hydrostatic hypothesis, we can understand this as the effect of faster clustering of (baryonic) matter due to a larger dark matter halo, thus increasing the central baryonic density.
In other words, if the gravitational potential at large radii is more important, a cluster should have accreted gas faster, and thus the ICM should be denser in the centre.

In order to securely compare the lens model to the ICM observables using the method presented in this article, future works should focus on relaxed, strong lensing galaxy clusters with a well-constrained lens model, including a large number of multiply-lensed systems.
Given the utmost importance to know the limits of strong lens modelling \citep[as demonstrated in e.g.][]{2023MNRAS.526.2776L}, one should not conclude on the dark matter halo properties without solid evidence.
Moreover, weak lensing or a number of galaxy-galaxy strong lensing events far from the cluster centre should be coupled to the strong lens model, in order to lift the degeneracy on the potential constraints at large radii. In the settings of a similar parametric lens model as presented here, this would allow to constrain $s_1$ and to verify our optimisation results.

We may compare the cut radius to the \textit{splashback radius}.
This radius is defined as the largest distance from the cluster centre connected to the cluster dynamics, i.e. the largest orbital apocentre at which matter is accreted to the DMH.
The cut radius values we find are larger than splashback radii measured by \citet{2018ApJ...864...83C}.
Their typical values provided in a redshift range corresponding to our lensing clusters are in the $1.5$ -- $2$\,Mpc range.
These values are averaged for clusters of typical mass $M_{200, m} \simeq 2.5 \times 10^{14} \Msol$, to compare to the (extrapolated) $5.6$ and $10.4 \times 10^{14} \Msol$ for MACS\,J0242 and MACS\,J0949 respectively.
As we expect more massive clusters to be larger, we do not find any contradiction to our $s_1$ values in this broad comparison.
For a more quantitative assessment, we compare the logarithmic derivatives of the dPIE radial densities to the \citet{2014ApJ...789....1D} density profiles (hereafter DK14).
We find the transition between a logarithmic derivative of $-2$ to $-4$ to occur for the dPIE profile (with an optimised $s_1$ cut radius) from $0.5 R_{\rm vir}$ to $\sim 5 R_{\rm vir}$. This is in clear opposition to the DK14 predictions, for which the heaviest clusters present a steepening of the derivative (see Fig. 2), and reaches a -4 logarithmic derivative for $\sim R_{\rm vir}$.

We can also compare the cut radii obtained to the predicted \textit{edge radius}, defined as the smallest radius at which no more orbiting galaxies can be found. According to \citet[][]{2020MNRAS.499.1291T, 2023MNRAS.521.3981A}, the edge radius can be approximated to $r_e \sim 2 R_{200, m}$. 
Using the ICM-optimised values for $R_{200, m}$, we can compare the optimised cut radius values to the edge radii for both clusters. 
We obtain for MACS\,J0242, $s_1 = 2720_{-870}^{+440}$ and $2 R_{200, m} = 4632$\,kpc.
In the case of MACS\,J0949, we measure $s_1 = 3780_{-1540}^{+390}$ and $2 R_{200, m} = 4751$\,kpc.
Computing the ratios $s_1 / r_e = 0.59$ and $0.80$ for the two clusters respectively, we obtain values of the same order.
It should be noted that the cut, splashback and edge radii are not necessarily assumed to describe the same physical limit. 
As MACS\,J0242 presents a more cored profile than MACS\,J0949, we expect a higher DM concentration in the cluster centre. As a consequence, for a comparable bounded mass (e.g. $M_{200,m}$), we expect a lower $s_1$ value. This is compatible with the smaller $s_1 / r_e$ ratio found here.
This serves to illustrate that, although of comparable orders of magnitude, $s_1$ and $r_e$ do not exactly represent the same physical observables.
The shock radius, describing the size of a shock of gas falling into the ICM of a cluster, is predicted by \citet{2021MNRAS.508.1777B} to be equally of order $\sim 2 R_{200, m}$, using SZ profiles issued from the Three Hundred Project hydrodynamic simulations. 
Although not quite observable with the ACT resolution, future SZ surveys will be able to assist detecting this critical ICM large scale cluster radius.
We conclude that the optimised cut radii is commensurable to these various radii attempting to measure the `size' of galaxy clusters, and we find $s_1$ to fall between the splashback and the edge radii.
Using our analytical model, the thorough comparison of these different cluster-size radii should be greatly assisted by the combination of strong lensing and ICM observations.

\subsection{ICM and DMH geometries} 

In order to understand the effect of the density profile parameters optimisation, one must decorrelate these from the geometric differences between the DMH and the ICM.
Our measurements suggest that the ICM presents rounder shapes than the DMH (see e.g. Fig.\,\ref{fig:astro_images}). Indeed, in the case of clusters MACS\,J0242 and MACS\,J0949, the X-ray measured ellipticities are negligible, but the strong lensing optimisation of the DMH presents ellipticities in the range $0.2 - 0.3$.
As studied in e.g. \citet{2008ApJ...681.1076D, 2012ApJ...755..116L}, the geometry of the ICM of (relaxed) galaxy clusters may differ from that of their DMH.
Beyond baryonic effects associated to the ICM, the inability of CDM to dissipate kinetic energy tends to favour more elliptical DMH.
According to \citet{2011ApJ...734...93L}, simulations show both fluids' ellipticity also varies importantly depending on the radius in which it is measured. As the ICM ellipticity is measured in a significantly larger radius, it is expected to be smaller.
In this article, we modelled the ICM ellipticity to be that of the DMH measured through strong lensing. Any attempt to optimise the ICM density with spherical profiles yielded equivalent density profile parameters ($\rho_0$, $a$, $s$) and best-fit likelihoods as using an elliptical profile (for both clusters, the Mahalanobis distance between the best-optimisations is of $0.3\,\sigma$, and the $\ln \mathcal{L}$ difference $< 0.01$).

At last, line-of-sight projection effects were entirely neglected in this paper, and in the strong lensing reconstruction. 
The combination of SZ effect and X-ray observations may allow to inform asymmetries on the line-of-sight. However, given the SZ effect data resolution and the adoption of a self-similar temperature profile in this work, this is beyond the scope of this article.
Such effects may nonetheless affect the quality of the reconstruction. For instance, \citet{2015ApJ...806..207U} display a case of apparent mismatch between ICM and lensing observations due to the presence of a line-of-sight asymmetry.

\subsection{Relationship between DM and ICM densities}

A number of additional observational effects limit the analytical ICM density reconstruction method.
In general, relaxed galaxy clusters are expected to present a cool core, due to the radiated power in X-rays at high ICM densities. 
However, the observed central temperatures tend to be higher than the expected values considering only X-ray cooling \citep[][]{2001A&A...365L.104P, 2002A&A...382..804B}. 
This is generally explained by feedback effects, and most notably active galaxy nuclei \citep[AGN, see e.g.][]{2010MNRAS.406..936P} and the cooling of the plasma. 
The change produced in the temperature profile was here taken into account by using the varying index, self-similar polytropic temperature model, which represents the physically measured temperatures, without implementing AGN feedback itself.
However, this self-similar model is imperfect to take into account the feedback specific to each cluster.
This may decorrelate the central ICM density from the DM in the outskirts. 
For instance, \citet[Fig.\,7]{2019A&A...621A..41G} present up to a $70 \%$ intrinsic scatter in the central density.
The inclusion of a theoretical model for such baryonic effects would represent an important improvement to our models. 
Modelling the ICM fluid velocity evolution would also allow to understand the evolution of clusters, thus challenging the hydrostatic hypothesis.

Nevertheless, with precise lensing constraints and temperature model measurements in a cluster, we could directly compare the ICM data to the parametric lensing reconstruction, given our analytical models.
The difference between the observations and the model, with respect to systematic errors, could yield constraints on the ICM velocity field and on dark matter models.

\section{Conclusion}
\label{sec:Conclusion}

Using a comparison between a parametric strong lensing mass reconstruction model and the ICM observations (X-rays and SZ) on two non-perturbed galaxy clusters, we have shown it is possible to use a unique model to describe both the total and ICM density profiles. 
In fact, under the hydrostatic hypothesis and assuming self-similar electron temperature profiles, the bijection $\mathcal{J}_z$ (equation\,\ref{eq:fully_integrated_hydrostatic_equation}) we have established between the total matter density and the ICM density allows to describe both the electron and dark matter density with the same parameters.
We applied this technique to the dPIE and NFW potentials, and convincingly found the parameters optimised through lensing to be either compatible with their ICM optimisation, or to be degenerate and thus difficult to optimise with the ICM.

Given the sensitivity of this $\mathcal{J}_z$ function, and as strong lensing does not allow to probe regions out of the cluster inner core, we paired our models to ICM observations.
This allows to probe the \textit{cut} or \textit{scale radii} of relaxed clusters.
Indeed, the ICM central density appears to be bounded to these matter density large-scale parameters. 
The method exposed in this article differs from traditional ways of accounting for ICM data in conjunction with strong lensing models, where the inferred gas potential is added to the lens model as a fixed \citep{2016A&A...594A.121P, 2017ApJ...842..132B} or evolving \citep{2024MNRAS.527.3246B} component.
Our new technique should be verified using clusters' outskirts surveys, such as weak lensing, and could only be efficiently applied with stringent constraints on the strong lensing parameters.

We can also reverse this perspective. If we had a perfect description of the full density mass model, e.g. including the cut radius of the DM halo through weak lensing, we could then compare the predicted ICM signal to that detected. 
If our model were satisfactory enough, we could then probe possible discrepancies, associated to other physical phenomena.

We here summarise the main results of this analysis:
\begin{itemize}
    \item[1.] We have proposed a self-similar polytropic temperature model with a varying index, using the X-COP sample of clusters. This allows to predict the ICM temperature for any cluster of measured mass $M_{500,c}$.
    \item[2.] As a major result, we exhibited an \textit{analytic} relationship between the ICM density and that of DM, assuming hydrostatic equilibrium. We have further shown this relationship to allow to predict the ICM density using strong lensing, as a proof of concept.
    \item[3.] We have demonstrated that the strong lensing ICM predictions are compatible with data through the ICM optimisation. We expect the strong lensing prediction to yield convincing results as long as: (i) the lensing galaxy cluster is not strongly perturbed, and (ii) we are able to properly predict the large-scale total density profile.
    \item[4.] This requirement to probe the large scales demonstrates the limitations of our current analysis. We however foresee weak lensing constraints as a mitigating solution to adjust our models to large scale variations, thus allowing us to make precise predictions.
    \item[5.] Reverting the perspective, this means the combination of X-rays or SZ data with strong lensing could allow to probe the dark matter profile of relaxed galaxy clusters far from their centres.
\end{itemize}

We have presented a proof of concept for the possibility to tie strong lensing constraints to the ICM. With higher-quality data and more observations on the large-scale profile, this should lead to powerful constraints on galaxy clusters physics.

\subsection*{Acknowledgements}
The authors want to thank Alastair Edge, Jose Maria Diego and Benjamin Beauchesne for fruitful discussions.
JA is supported by the Postgraduate Research Scholarship in Astroparticle Physics/Cosmology in the University of Sydney.
JA acknowledges support by the Israel Science Foundation (grant no. 2562/20) and by the Center of Excellence of the Israel Science Foundation (grant No. 1937/19).  
MJ and DL are supported by the United Kingdom Research and Innovation (UKRI) Future Leaders Fellowship (FLF), 'Using Cosmic Beasts to uncover the Nature of Dark Matter' (grant number MR/S017216/1).
The authors acknowledge the Sydney Informatics Hub and the use of the University of Sydney high performance computing cluster, Artemis.

\section*{Data Availability}

The lens models and the lensing-ICM optimisation programmes are available upon reasonable request to the corresponding author.


\bibliographystyle{mnras}
\bibliography{bibli}



\appendix

\section{Ellipticities}
\label{sec:ellipticity}

We define the mass ellipticity $e$ as:
\begin{equation}
    e = \frac{a^2 - b^2}{a^2 + b^2},
\end{equation}
where $a$ and $b$ are the semi-major and semi-minor axes respectively.

Due to the Poisson equation, the ellipticity of the gravitational potential, i.e. here that of the ICM, differs from that of the mass ellipticity\footnote{This convention is provided by \citet{1993ApJ...417..450K}. We tested it in practice, and we find this relationship to be appropriate, but it must be multiplied by a factor $\sqrt{2}$ when the ellipticity of the potential is measured in a 2D plane.}: 
\begin{equation}
    \mathcal{E} = \frac{1 - \sqrt{1 - e^2}}{e} = \frac{a - b}{a + b}.
\end{equation}

Upon defining an ellipsoidal radius, in order to take into account the potential ellipticity, assuming the ellipticity to be $e$, we take it to be:
\begin{equation}
    r = \sqrt{ \left(\frac{x \cos \theta + y \sin \theta}{a} \right)^2 + \left(\frac{y \cos \theta - x \sin \theta}{b} \right)^2 + \left(\frac{z}{c} \right)^2 },
\end{equation}
in the Cartesian coordinates, with $\theta$ the rotation angle on the sky plane. $c$ is the semi-axis along the line-of-sight, which we take to be the geometric average of $a$ and $b$ here.

\section{Alternative density distributions}
\label{sec:Additional_densities}

Beyond the NFW and dPIE density distributions, presented in Section\,\ref{subsec:full_matter_density_profiles}, alternative models may be used to compute the ICM thermodynamic parameters using relationship (\ref{eq:reverted_J_ne}).

\subsection{Generalised Navarro-Frenk-White profile}
In the case of a generalised NFW potential \citep[proposed as early as][]{1990ApJ...356..359H}, we compute the different integrals given equation (\ref{eq:def_g_h_func}):
\begin{equation}
    \begin{split}
    f (r) &= x^{-\gamma} \left( 1 + x^{\alpha} \right)^{-\frac{\beta - \gamma}{\alpha}},\\
    g (r) &= r_S^3 \frac{x^{3 - \gamma}}{3-\gamma} {}_2F_1 \left( \mu, \xi, 1+\mu, -x^{\alpha}\right),\\
    h (r) &= r_S^2 \Bigl\{ x^{2 - \gamma} \Bigl[ \frac{{}_2F_1 \left( \nu, \xi, 1+\nu, -x^{\alpha}\right)}{2 - \gamma}\\&~~~ - \frac{{}_2F_1 \left( \mu, \xi, 1+\mu, -x^{\alpha}\right)}{3-\gamma} \Bigr] - \frac{\Gamma(\nu) \Gamma(\xi - \nu)}{\alpha \Gamma (\xi)} \Bigr\},
    \end{split}
    \label{eq:Integrated_func_gNFW}
\end{equation}
where $x = r / r_S$, ${}_2F_1$ is the Gauss hypergeometric function, $\Gamma$ the extended factorial function (i.e. the complete gamma function), and $\mu$, $\nu$ and $\xi$ are simple reformulations of the three indices $\alpha$, $\beta$, $\gamma$:
\begin{equation}
    \mu = \frac{3-\gamma}{\alpha} ~~;~~ \nu = \frac{2-\gamma}{\alpha} ~~;~~ \xi = \frac{\beta - \gamma}{\alpha}.
\end{equation}
This integration takes constants into account, but requires $\alpha > 0$, $\beta > 2$ and $\gamma < 2$.
The generalised NFW density is normalised by $\rho_S$, similarly to NFW: $\rho_{\rm gNFW} (r) = \rho_S f(r)$.

\subsection{Einasto profile}
The Einasto potential was proposed in \citet[][]{1965TrAlm...5...87E}.  Following \citet{2005MNRAS.358.1325C, 2012A&A...540A..70R}, the different functions equation (\ref{eq:def_g_h_func}) write:
\begin{equation}
    \begin{split}
    f (r) &= \exp \left( - s^{n^{-1}} \right),\\
    g (r) &= l^3 n\left[ \Gamma(3n) - \Gamma \left( 3n, s^{n^{-1}} \right) \right],\\
    h (r) &= - \frac{n l^2}{s} \left[ \Gamma (3n) - \Gamma \left(3n, s^{n^{-1}}\right) + s \Gamma\left(2n, s^{n^{-1}}\right) \right],
    \end{split}
    \label{eq:Integrated_func_Einasto}
\end{equation}
where $n$ is the inverse index of the density slope at large radii, $s = (2n)^n r/r_{-2}$ the reduced scale radius, with $r_{-2}$ a transition radius, $l = r_{-2} / (2n)^n$ the scale length and $\rho_0$ the central density.
$\Gamma (\alpha)$ is the complete gamma function, and $\Gamma (\alpha, x)$ the incomplete upper gamma function:
\begin{equation}
    \Gamma (\alpha, x) = \int_x^{\infty}\, \mathrm{d}t t^{\alpha - 1} e^{-t}.
\end{equation}
The Einasto density writes $\rho_{\rm E} (r) = \rho_{0} f(r)$.

\section{Self-normalised ICM density distribution}
\label{sec:self-normalised_ne}
Following equation\,(\ref{eq:J_gen_definition}), assuming $\mathcal{J}$ to be a bijection, we can also invert it and normalise the distribution at a given radius $\Delta$ (e.g. 100\,kpc, where the strong lensing signal is strongly constraining the total density profile, or $R_{500, c}$), if we happen to know $n_{e, \Delta}$. We can therefore write:
\begin{equation}
    n_e (r) = \mathcal{J}^{-1} \left[ \mathcal{J} (n_{e, \Delta}) \frac{\Phi (r)}{\Phi (R_{\Delta, c})} \right].
    \label{eq:ne_self-norm_dPIE}
\end{equation}
We call this latter expression \textit{self-normalised}.

We did not include any self-normalised models in the optimisations, as these imply to use X-ray data both as input (as a normalisation) and to perform the optimisation, which would make one or several parameters degenerate.

\section{Gas fraction study}
\label{subsec:gasfra_study}

\subsection{Gas fraction definition}
\label{subsec:gas_fraction_presenting}

We define the gas fraction as the ratio of the gas mass to the total mass.
The gas mass includes all baryons except stars.
We distinguish the local gas fraction, $f_g (r) = \rho_g (r) / \rho_{m} (r)$, considered in this article to be a radial function, and the cumulative gas fraction, $F_g$, given within a radius $r$:
\begin{equation}
    \begin{split}
    F_g (r) &= \frac{\int_0^r \mathrm{d}s s^2 \rho_g (s)}{\int_0^r \mathrm{d}s s^2 \rho_m (s)} = \frac{M_g (<r)}{M_m (<r)},\\
    f_g (r) &= \frac{\mathrm{d} F_g}{\mathrm{d} r} (r) \frac{\int_0^r \mathrm{d}s s^2 \rho_m (s)}{r^2 \rho_m (r)} + F_g (r).
    \end{split}
\label{eq:definition_gas_fraction}
\end{equation}
The full knowledge of either of these gas fractions would provide a bijective relationship between the gas and matter content of galaxy clusters.
We will therefore name \textit{gas density} reconstruction our electron density prediction using an empirical gas density model.

We here present an alternative attempt to model the hot gas distribution using the gravitational potential. Using the local gas fraction $f_g = \rho_g / \rho_m$, a general model for $f_g$ coupled with the lensing constraints on $\rho_m$ would yield a $\rho_g$ prediction in each lensing cluster.

In order to derive a quantitative model for the gas fraction, we use the `X-COP+2' sample analysis, i.e. the X-COP (\textit{XMM} Cluster Outskirts Project) sample, complemented with similar analyses for our two strong-lensing clusters, MACS\,J0242 and MACS\,J0949.
We compare the cumulative gas fraction reconstruction (as defined in eq. \ref{eq:definition_gas_fraction}) in each of the 14 clusters in the sample, and propose two new \textit{ad hoc} models.

\subsection{Proposed models}

With respect to the data analysed in Sect.\,\ref{sec:gas_fraction_data_analysis}, we propose the following models.
First, we attempt to describe the increasing cumulative gas fraction, $F_g$, as a power law:
\begin{equation}
    F_g (r) = f_g^0 \left( 1 + \frac{r}{r_c} \right)^{\zeta},
\label{eq:powlaw_fgas_description}
\end{equation}
where $r_c$ is a pivot, or core radius, $f_g^0 = F_g (r=0)$, the central gas fraction, i.e. the baryonic fraction excepted the stellar fraction, and $\zeta$, the power exponent to find.

However, for all clusters, the integrated gas fraction presents a transition between the inner and the outer regions of the cluster, as represented on \citet[Fig.\,1]{2019A&A...621A..40E}.
We propose to analytically describe this transition with a \textit{transitive} model:
\begin{equation}
    F_g (r) = a \left[ \frac{2}{\pi} \arctan \left( \exp{ \frac{r - r_c}{r_f} } \right) - \frac{1}{2} \right] + b,
    \label{eq:arctan_fgas_description}
\end{equation}
where $a$ and $b$ are defined with the expected gas fraction at two given radii, respectively at $R_E$, a radius where the potential is typically constrained through strong gravitational lensing (e.g. the Einstein radius), and at $R_{500,c}$, where the gas fraction should tend towards the Universal gas fraction. We respectively write these gas fractions $F_g^{E}$ and $F_g^{500}$:
\begin{equation}
    \begin{split}
    \upsilon (r) &= \frac{2}{\pi} \arctan \left[ \exp \left( \frac{r - r_c}{r_f} \right) \right],\\
    a &= \frac{ F_g^{500} - F_g^{E} }{\upsilon (R_{500}) - \upsilon (R_E)},\\
    b &= F_g^{500} - a \upsilon (R_{500}),
    \end{split}
    \label{eq:params_arctan_fgas_model}
\end{equation}
and $r_p$ and $r_f$ are the pivot radius and flattening distance respectively, i.e. the radii of the transition inflexion point, and the characterisation of the slope of the transition.

In the simplest model, i.e. assuming a cluster to be dynamically relaxed, the local gas fraction pivot scale should be related to the pivot parameter in the total matter density $\rho_m$. For instance, in a NFW description, this parameter would be the scale radius $r_S$, as $f_g (r) = \rho_g (r; r_S) / \rho_m (r; r_S)$, using the hydrostatic relationship (\ref{eq:reverted_J_ne}). 
Therefore, we can take $a_1$, a typical BCG-to-DMH transition radius, to be a reasonable prior on both the pivot radius $r_p$ and the flattening distance $r_f$.
Important limitations nonetheless prevent from identifying these values to be identical. Indeed, multiple parametrisations of the typical pivot radius exist in the potential, and are not identical to a unique pivot scale.
More importantly, the distribution of hot gas is dominated by non-gravitational phenomena (AGN feedback, plasma turbulence, etc.), whose characteristic radii have no \textit{a priori} reason to match the potential's.
For $F_g^{500}$, we assume $F_g^{500} \simeq f_g^{500}$, and we use the universal gas fraction $\Omega_b / \Omega_m = 0.1580 \pm 0.0021$ \citep{Planck_XIII_2016}, corrected for the stellar fraction $f_{\star}$, not accounted for in the intracluster gas, and the baryon depletion factor $Y_b$ \citep[see][]{2019A&A...621A..40E}:
\begin{equation}
    f_{g}^{\mathrm{th}} (r) = Y_b (r) \frac{\Omega_b}{\Omega_m} - f_{\star} (r).
    \label{eq:Universal_gas_fraction}
\end{equation}
\citet{2019A&A...621A..40E} provide $Y_{b, 500} = 0.938^{+0.028}_{-0.041}$, which we use here. Following the same study, we take the error on $f_{\star, 500}$ to be $5\times 10^{-3}$. 
The study of the lensing galaxy clusters gives stellar fraction of $f_{\star, 500} = (1.92 \pm 0.21) \times 10^{-2}$ and $(1.87 \pm 0.36) \times 10^{-2}$ for MACS\,J0242 and J0949 respectively.

\subsection{X-COP+2 study \label{sec:gas_fraction_data_analysis}}

With the power-law and the transitive models, equations (\ref{eq:powlaw_fgas_description}) and (\ref{eq:arctan_fgas_description}) respectively, we conduct a study for the 14 galaxy clusters of the X-COP+2 sample.

All these clusters were tested with both the power-law and transitive models, optimised for all respective 3 and 4 parameters ($\{$ $f_g^0$, $r_c$, $\zeta$ $\}$ and $\{$ $F_g^E$, $F_g^{500}$, $r_p$, $r_f$ $\}$) with a MCMC, with package \textsc{emcee} \citep[see][]{Foreman_Mackey_2013}. We arbitrarily took $R_E = 50$\,kpc, as both lensing clusters present strong constraints in this region, and the gas fraction at this radius is significantly different from that at $R_{500,c}$ for all clusters.
We define the log-likelihood function as Gaussian, with an underestimated variance of fractional amount, $f$:
\begin{equation}
    \ln \mathcal{L}_{f_g} (\Theta) = -\frac{1}{2} \sum_i \left[ \left( \frac{F_{g,i}^{\mathrm{val}} - F_{g, i}^{\mathrm{pred}} (\Theta)}{\sigma_i} \right)^2 + \ln \sigma_i^2 \right],
\end{equation}
where $F_{g,i}^{\mathrm{val}}$ are the values of the cumulative gas fraction in radius bins (the gas mass being measured through X-ray deprojection, and the total mass obtained using the hydrostatic equilibrium), $F_{g, i}^{\mathrm{pred}}$ are the predictions in the same bins, and:
\begin{equation}
    \sigma_i^2 = \left( \sigma_i^{\mathrm{err}} \right)^2 + \left[ F_{g, i}^{\mathrm{pred}} (\Theta) \right]^2 f^2,
\end{equation}
where $\sigma_i^{\mathrm{err}}$ are the $F_{g,i}^{\mathrm{val}}$ measured standard deviation error.
In practice, the model scatter $f$ is optimised.

Out of the 14 clusters, 5 were found to be better modelled with the power-law, and 9 with the transitive relationship -- including MACS\,J0242 and MACS\,J0949. 
We performed the optimisation over all radii accessible in the X-ray data range, except for MACS\,J0242, where non-statistically significant perturbations exist in the gas fraction reconstruction.
To avoid these, the optimisation was performed in $r \in [20; 350]$\,kpc for this specific cluster.

Overall, we find the transitive model to be consistently better.
Indeed, even the clusters which were better modelled by a power-law are well fit by a transitive model. 
The largest model scatter on the power-law model reaches $f = 17\%$, to be compared with the maximum of $f = 9\%$ for the transitive model.
Moreover, by construction, the transitive model can use physical parameters as priors for $F_g^{500}$, $r_p$ and $r_f$, and converges at large radii, which is expected from the Universal hot gas fraction.
We give in Tables\,\ref{tab:best_params_fgas_powlaw} and \ref{tab:best_params_fgas_arctan} the optimised parameters for respectively the power-law and the transitive models. 
These are the averages of the best parameters found by the MCMC for each individual cluster.

\begin{table}
	\centering
	\caption{Average over all clusters of the optimised parameters for the gas fraction power law model. $r_c$ is in kpc.}
	\label{tab:best_params_fgas_powlaw}
	\begin{tabular}{ccccc}
	    \hline
		\hline
		$f_g^0 (\%)$ & $r_c$ [kpc] & $\zeta$\\
		\hline
		$1.26 \pm 1.12$ & $1.01 \pm 0.62$ & $0.41 \pm 0.14$\\
		\hline
		\hline
	\end{tabular}
\end{table}

\begin{table}
	\centering
	\caption{Average over all clusters of the optimised parameters for the gas fraction transitive model. $r_p$ and $r_f$ are in kpc.
	}
	\label{tab:best_params_fgas_arctan}
	\begin{tabular}{cccc}
	    \hline
		\hline
		$F_g^E (\%)$ & $F_g^{500} (\%)$ & $r_p$ [kpc] & $r_f$ [kpc]\\
		\hline
		$5.59 \pm 2.21$ & $13.6 \pm 2.1$ & $20.4 \pm 65.6$ & $259.8 \pm 135.8$\\
		\hline
		\hline
	\end{tabular}
\end{table}

Trying to relate these parameters to physical values, we notice that $f_g^0$ is the gas fraction at the centre of the clusters. 
The X-COP+2 sample does not precisely provide the hot gas fraction in the centre of clusters ($r < 20$\,kpc), due to the stellar effects, turbulence, feedback and resolution of X-ray surveys.
For these reasons, we do not directly use a physically measurable value for $f_g^0$, and simply use a fit of this parameter across all radii.
Conversely, $F_g^{E}$ is well measured for all the clusters of the sample, but we can not generalise this value overall.

The exponent $\zeta$ of the power-law model is purely empirical. 
As for the power-law pivot radius, $r_c$, its relative error bars are quite important. 
However, discarding the two clusters coming from lensing yields $r_c = 0.96 \pm 0.65$\,kpc, i.e. a result very close the X-COP+2 one.
The same process on $\zeta$ and $f_g^0$ gives results similar to those presented in Table\,\ref{tab:best_params_fgas_powlaw}, for respectively $\zeta = 0.41 \pm 0.14$ and $f_g^0 = (1.2 \pm 1.1) \times 10^{-2}$.
We can therefore propose a model:
\begin{equation}
    F_g (r) = 1.3 \times 10^{-2} \left[ 1 + \frac{r}{1.0\,\mathrm{kpc}} \right]^{0.41}.
\end{equation}
However, given the important error bars found on all parameters of this model, its inability to predict accurately the gas fraction for most clusters, and the lack of theoretical motivation for its parameters, we conclude to the ineffectiveness of this model.

\begin{figure*}
\centering
\begin{minipage}{0.48\textwidth}
\centerline{\includegraphics[width=\columnwidth]{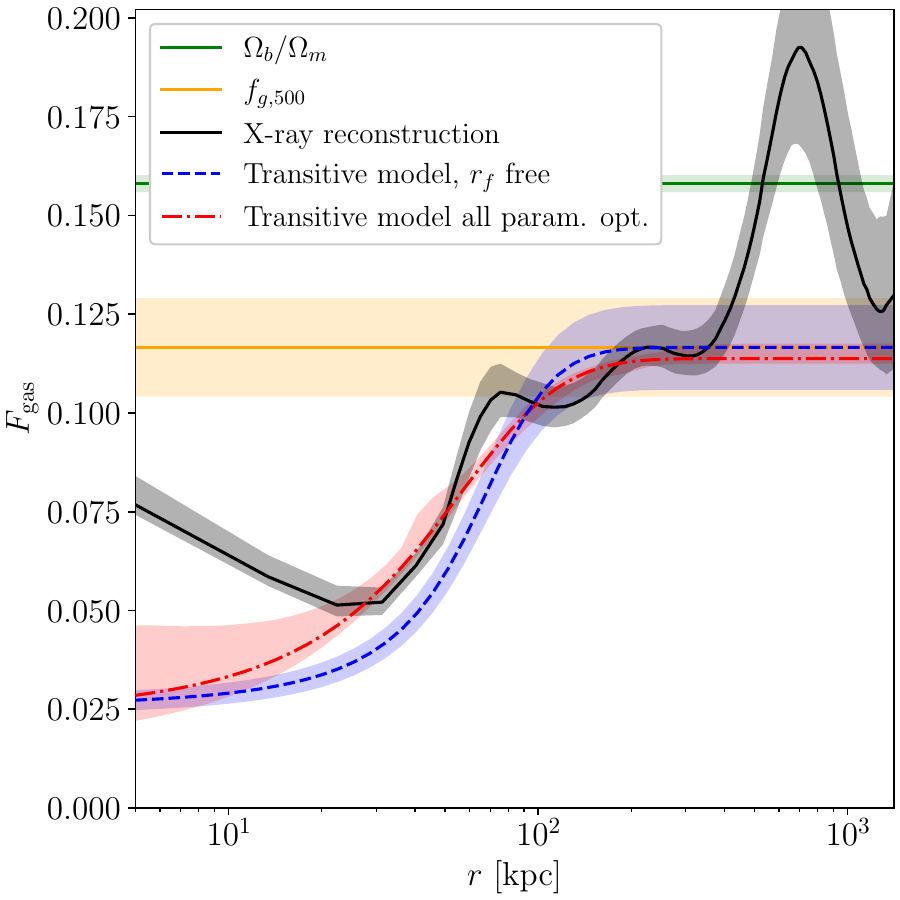}}
\end{minipage}
\hfill
\begin{minipage}{0.48\linewidth}
\centerline{\includegraphics[width=\columnwidth]{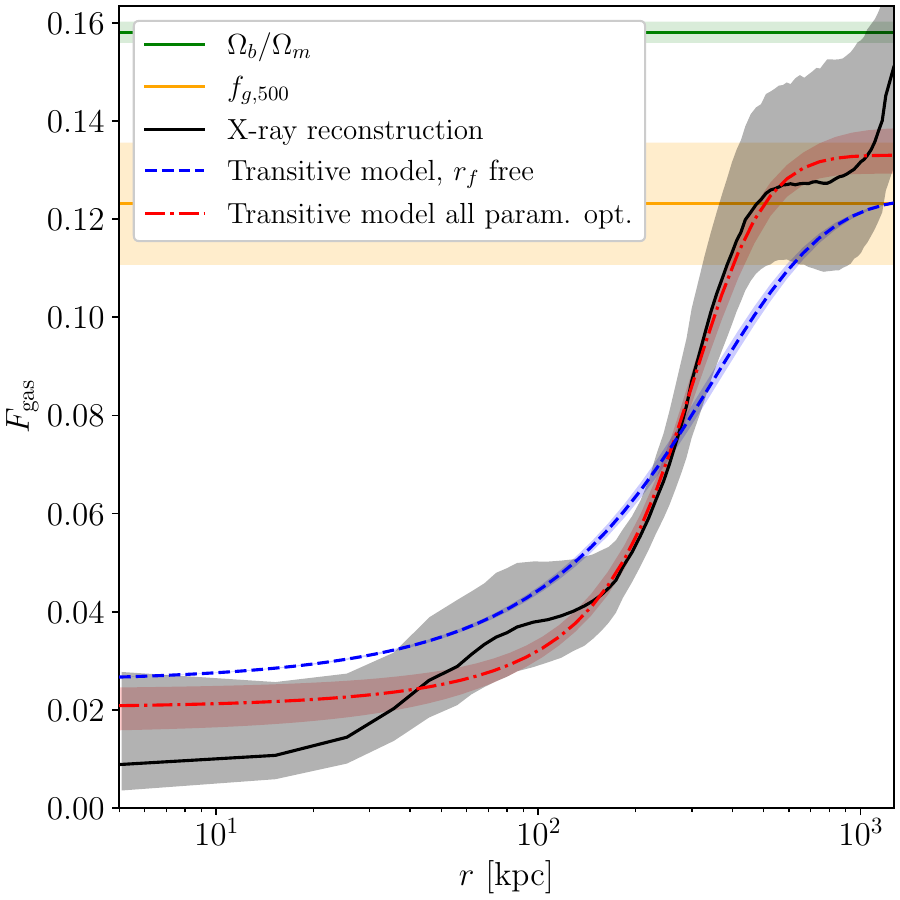}}
\end{minipage}
\caption{Cumulative gas fraction $F_g (r)$ in lensing galaxy clusters MACS\,J0242 (left) and MACS\,J0949 (right).
\textit{Black:} Reconstructed gas fraction from \textit{XMM-Newton} data analysis. Errors are $1 \sigma$. \textit{Green:} Universal gas fraction $\Omega_b / \Omega_m$. 
\textit{Orange:} Gas fraction at $R_{500,c}$ (see eq.\,\ref{eq:Universal_gas_fraction}).
\textit{Red:} Transitive gas fraction model (see equation \ref{eq:arctan_fgas_description}) optimised for the specific cluster. 
\textit{Blue:} Transitive model with the cluster lensing parameter $a_1$ as $r_p$, and $F_g^0$ and $F_g^{500}$ determined using the general reduction over X-COP+2 and the theoretical gas fraction $f_g^{500}$ respectively.
As presented in e.g. \citet{2022A&A...662A.123E}, the baryon fraction is approximately constant out of the central regions of clusters. Baryons are distributed between stars and the ICM, with a higher concentration of the former in the centre, and of the latter in the outskirts ($r > 0.2 R_{500}$). This matches the trend presented here.
}
\label{fig:F_gas}
\end{figure*}

As for the transitive model, all parameters may be physically interpreted.
The quantity $F_g^{500}$ not only can be found quite precisely, but is also in agreement with equation (\ref{eq:Universal_gas_fraction}).
Indeed, for all X-COP clusters, the stellar fraction is considered to be unknown. We could use the X-COP average value $f_{\star, 500} = 0.015 \pm 0.005$, but prefer to use the results of simulations presented in \citet{2023A&A...675A.188A}. Given a $M_{500}$, which we have for X-COP clusters, we can draw a value for $f_{\star, 500}$.
Completing the set with the stellar fractions found in \citet{2023MNRAS.522.1118A} for the lensing clusters, we can fix the $F_g^{500}$ value for each cluster. 
These values are in remarkable agreement with Table\,\ref{tab:best_params_fgas_arctan}.
As for $r_p$ and $r_f$, we can not study the X-COP clusters and our lensing clusters jointly.
Indeed, we are using $a_1$, the core radius of the DMH, as a prior on both these parameters. However this radius is unknown for the X-COP clusters due to the lack of lensing data.
Moreover, the best fit values of $r_p$ for the X-COP clusters are all $< 2$\,kpc, while for MACS\,J0242, $r_p = 22.9_{-16.2}^{+48.0}$\,kpc and for MACS\,J0949, $r_p = 256.1_{-29.2}^{+16.7}$\,kpc.

Therefore, for the dynamically relaxed X-COP clusters, we can set $r_p = 0$. After performing an optimisation on the X-COP clusters only of this restricted model, we find the average best parameters summarised in Table\,\ref{tab:best_params_fgas_arctan_updated}. The largest model scatter across all 12 clusters is still of $f = 9\%$. If we chose to constrain $F_g^{500}$ using eq. (\ref{eq:Universal_gas_fraction}), the largest $f$ would be of $36\%$.

\begin{table}
	\centering
	\caption{Updated parameters and model for the gas fraction transitive model, for the X-COP clusters only. $r_f$ is in kpc.}
	\label{tab:best_params_fgas_arctan_updated}
	\begin{tabular}{ccc}
	    \hline
		\hline
		$F_g^E (\%)$ & $F_g^{500} (\%)$ & $r_f$ [kpc]\\
		\hline
		$5.40 \pm 1.83$ & $13.7 \pm 1.9$ & $290.7 \pm 122.1$ \\
		\hline
		\hline
	\end{tabular}
\end{table}

Given the results in the lensing clusters transitive model optimisation, we choose to fix $r_p$ to the $a_1$ prior for both MACS\,J0242 and MACS\,J0949. This is an effective model, which should be further informed using observations on other lensing clusters. We then find the model error to be respectively $f = 1.7\%$ and $1.8\%$.
On all 14 clusters, we may prolong the transitive function into a central gas fraction $F_g^0$, which is predicted to be $F_g^0 = 2.6 \pm 1.9 \%$, averaging on all clusters, with the fixed $r_p$ models.
We may thus fix all parameters (using formula \ref{eq:Universal_gas_fraction} for $F_g^{500}$), except $r_f$. This final model is represented and compared to the transitive model with all parameters let free on Figure\,\ref{fig:F_gas}.
We find a maximum scatter of $f = 40\%$, but the average scatter at $8.7\%$.
As expected, the local models appear to be better fits, but we notice the reduced model ($r_f$ free only displayed in blue is a good approximation (the X-ray reconstruction is never distant of more than $2 \sigma$ from the `reduced' models, in the fitting radii range). 
For the lensing clusters, we can not conclude on the scale to set for $r_p$, but found the approximation $r_p \approx a_1$ to be empirically reasonable. 
We can not conclude absolutely on the flattening distance $r_f$, with the final model yielding $r_f = 26.5_{-2.8}^{+4.0}$\,kpc for MACS\,J0242 and $r_f = 267.3_{-16.4}^{+19.9}$\,kpc for MACS\,J0949. A larger study would be necessary to conclude to the general validity of a reduced transitive model without any free parameter.
For the X-COP clusters, $r_f$ ranges from $80$ to $750$\,kpc, and we can thus not conclude either. Studying the morphology and dynamical state parameters may allow to determine $r_f$ using observable data.

As we could not extract a general, Universal prediction for the gas fraction simply using lensing-determined parameters, we can not conclude to the success of this technique for the moment.
The transitive model is however encouraging, and a more general study coupling X-ray and lensing data may manage to generalise the resisting parameter $r_f$.

\section{Cornerplots}
\label{sec:Cornerplots}

We display here the cornerplots of the MCMC optimisations of the potential of the ICM data, as described Sect.\,\ref{sec:Discussion}.
In the following graphs, densities $\rho$ are displayed in g.cm$^{-3}$, while core, scale and cut radii in kpc.

\begin{figure}
    \centering
    \includegraphics[width=\columnwidth]{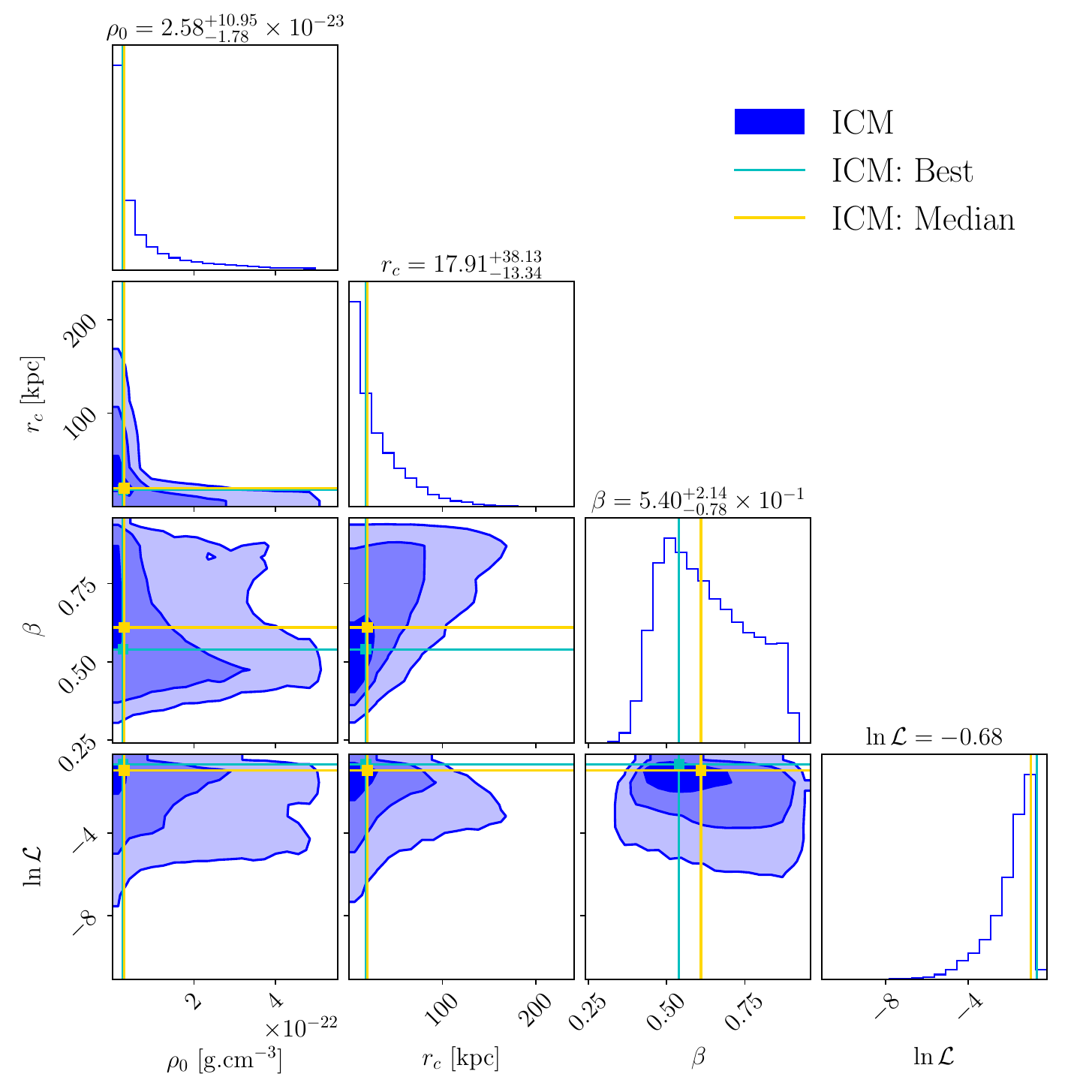}
    \caption{MCMC optimisation for $\beta$ model of the three relevant parameters for cluster MACS\,J0242: central density $\rho_{0,1}$, core radius $r_c$ and density index $\beta$.
    As per all other cornerplots, the density $\rho_{0,1}$ is here displayed in g.cm$^{-3}$ and the core radius $r_c$ in kpc.
    \textit{Blue:} Optimisation performed using the available ICM data (X-ray here).
    \textit{Gold:} Median of the ICM optimisation.
    \textit{Cyan:} Best ICM optimisation (described in Table\,\ref{tab:best_model}).
    These best ICM-optimisation values are displayed over the histogramme distributions.
    }
    \label{fig:m0242_MCMC_beta_v872}
\end{figure}


\begin{figure*}
    \centering
    \includegraphics[width=\textwidth]{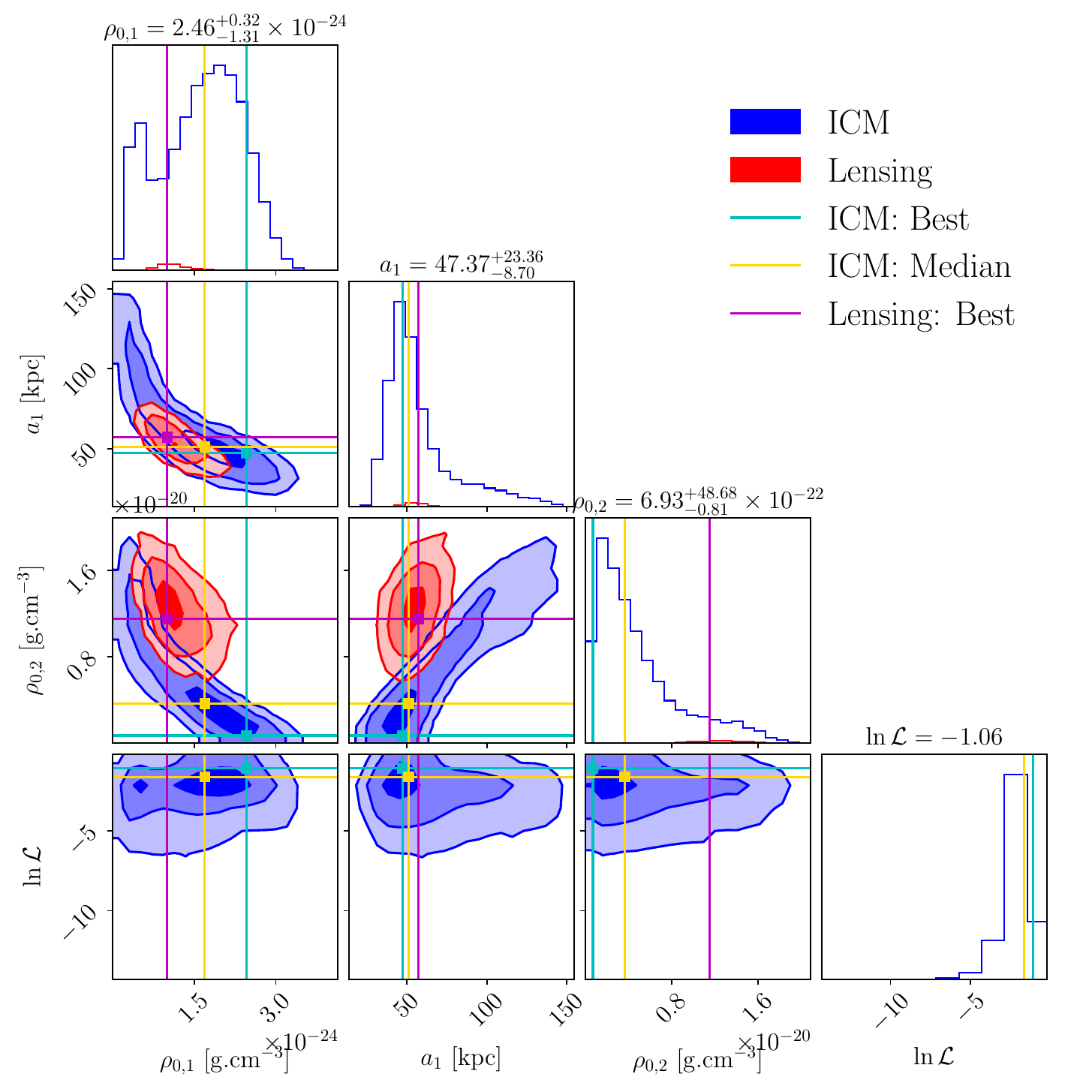}
    \caption{MCMC optimisation for idPIE model, for cluster MACS\,J0242: DMH central density $\rho_{0,1}$ and core radius $a_1$, and BCG central density $\rho_{0,2}$. The DMH cut radius is fixed to the fiducial value of 1.5 Mpc.
    Densities are displayed in g.cm$^{-3}$, distances in kpc.
    \textit{Blue:} Optimisation performed using the available ICM data (X-ray here).
    \textit{Red:} Strong lensing optimisation.
    \textit{Cyan:} Best ICM optimisation.
    \textit{Gold:} Median of the ICM optimisation.
    \textit{Magenta:} Best strong lensing model (described in Table\,\ref{tab:best_model}).
    Comparing this to Fig.\,\ref{fig:m0242_MCMC_idPIEMD_v871}, we notice the importance of the optimisation of parameter $s_1$, as the ICM best-fit likelihood here is $-1.06$, i.e. the optimisation is of much worse quality than with this optimisation.
    }
    \label{fig:m0242_MCMC_idPIEMD_v874}
\end{figure*}

\begin{figure*}
    \centering
    \includegraphics[width=\textwidth]{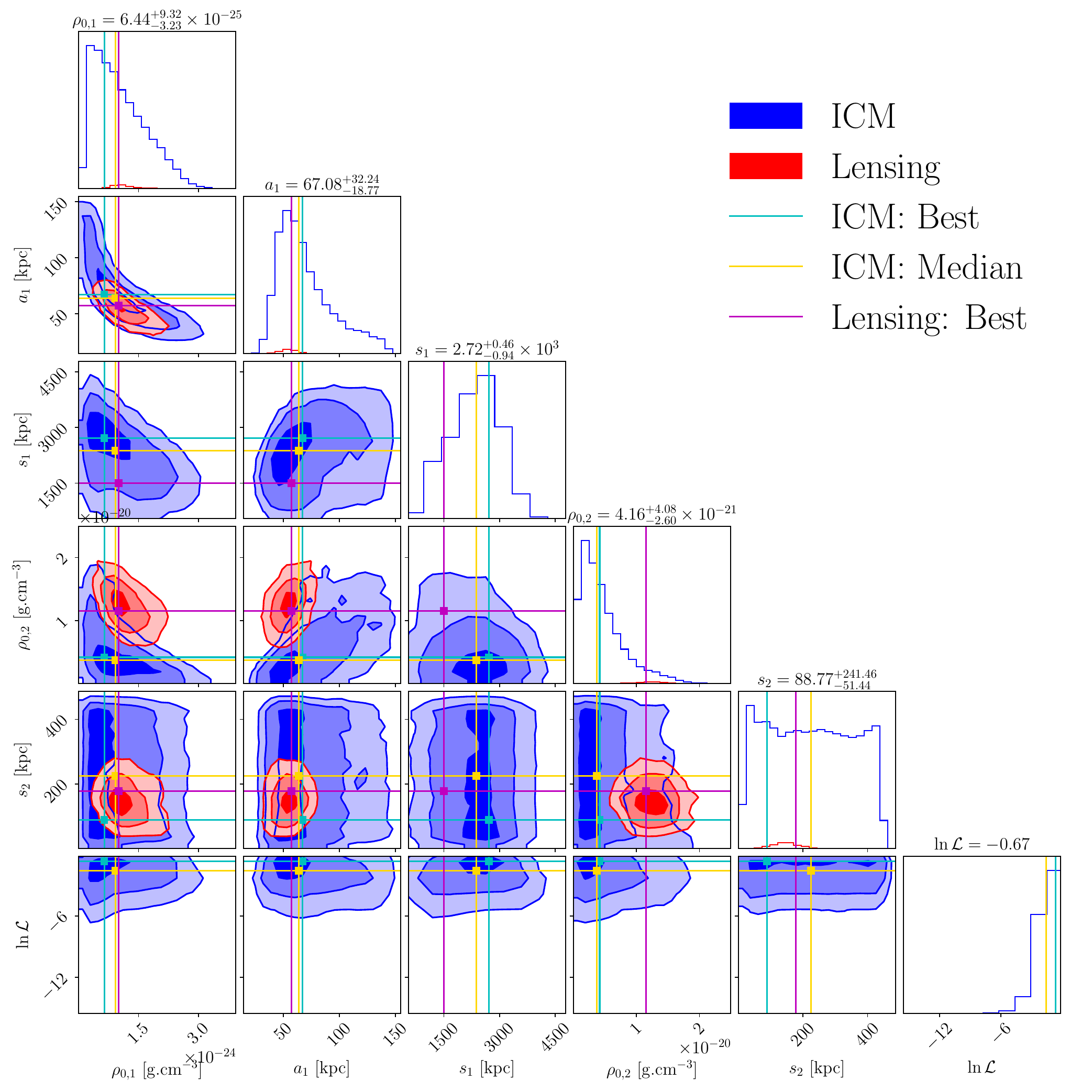}
    \caption{MCMC optimisation for idPIE model, for cluster MACS\,J0242: DMH central density $\rho_{0,1}$, core radius $a_1$, cut radius $s_1$, and BCG central density $\rho_{0,2}$ and cut radius $s_2$.
    Densities are displayed in g.cm$^{-3}$, distances in kpc.
    \textit{Blue:} Optimisation performed using the available ICM data (X-ray here).
    \textit{Red:} Strong lensing optimisation.
    \textit{Cyan:} Best ICM optimisation.
    \textit{Gold:} Median of the ICM optimisation.
    \textit{Magenta:} Best strong lensing model (described in Table\,\ref{tab:best_model}).
    Comparing this to Fig.\,\ref{fig:m0242_MCMC_idPIEMD_v871}, we can see that the $s_2$ optimisation is degenerated, and therefore not necessary. This explains why we fixed the $s_2$ value to that of lensing in Section\,\ref{sec:Discussion}.
    }
    \label{fig:m0242_MCMC_idPIEMD_v879}
\end{figure*}

\begin{figure}
    \centering
    \includegraphics[width=\columnwidth]{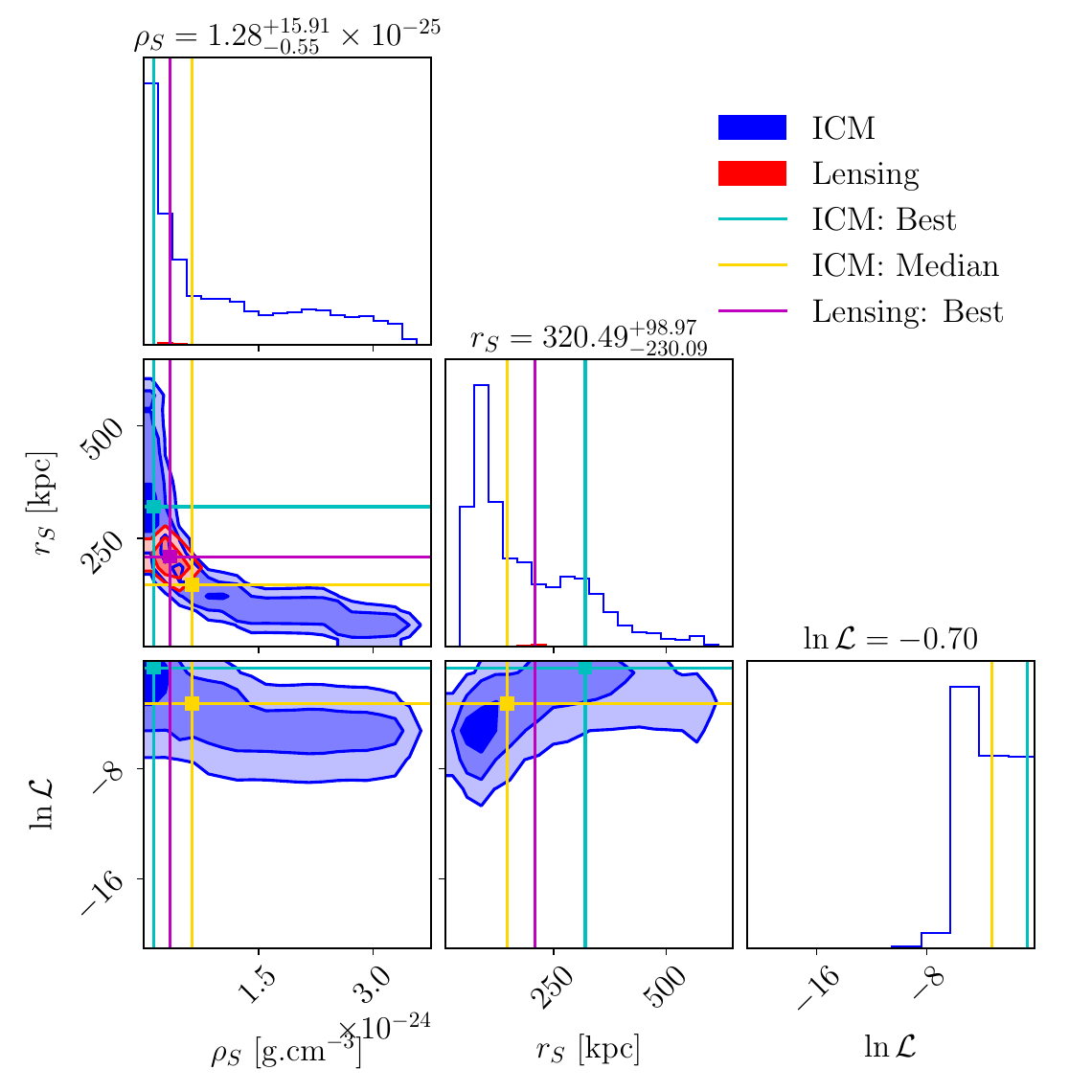}
    \caption{MCMC optimisation for iNFW model of the two relevant parameters for cluster MACS\,J0242: the density normalisation $\rho_S$ and the scale radius $r_S$.
    Densities are displayed in g.cm$^{-3}$, distances in kpc.
    \textit{Blue:} Optimisation performed using the available ICM data (X-ray here).
    \textit{Red:} Strong lensing optimisation.
    \textit{Cyan:} Best ICM optimisation.
    \textit{Gold:} Median of the ICM optimisation.
    \textit{Magenta:} Best strong lensing model (described in Table\,\ref{tab:best_model}).
    }
    \label{fig:m0242_MCMC_iNFW_v873}
\end{figure}

\begin{figure}
    \centering
    \includegraphics[width=\columnwidth]{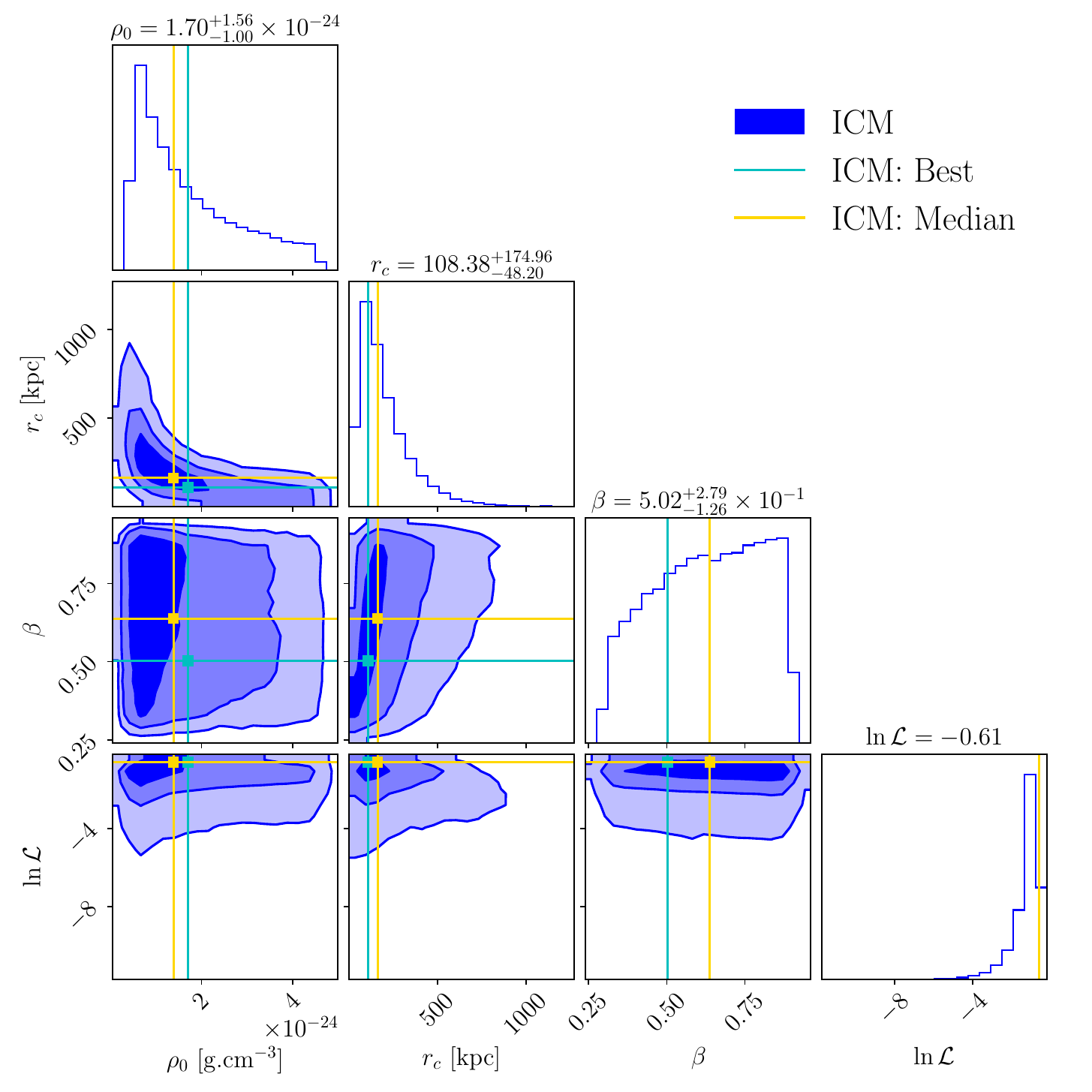}
    \caption{MCMC joint optimisation for the $\beta$ model, for cluster MACS\,J0949.
    Densities are displayed in g.cm$^{-3}$, distances in kpc.
    \textit{Blue:} Optimisation performed using the available ICM data (X-ray and SZ here).
    \textit{Cyan:} Best ICM optimisation.
    \textit{Gold:} Median of the ICM optimisation.
    }
    \label{fig:m0949_MCMC_beta_v878}
\end{figure}

\begin{figure}
    \centering
    \includegraphics[width=\columnwidth]{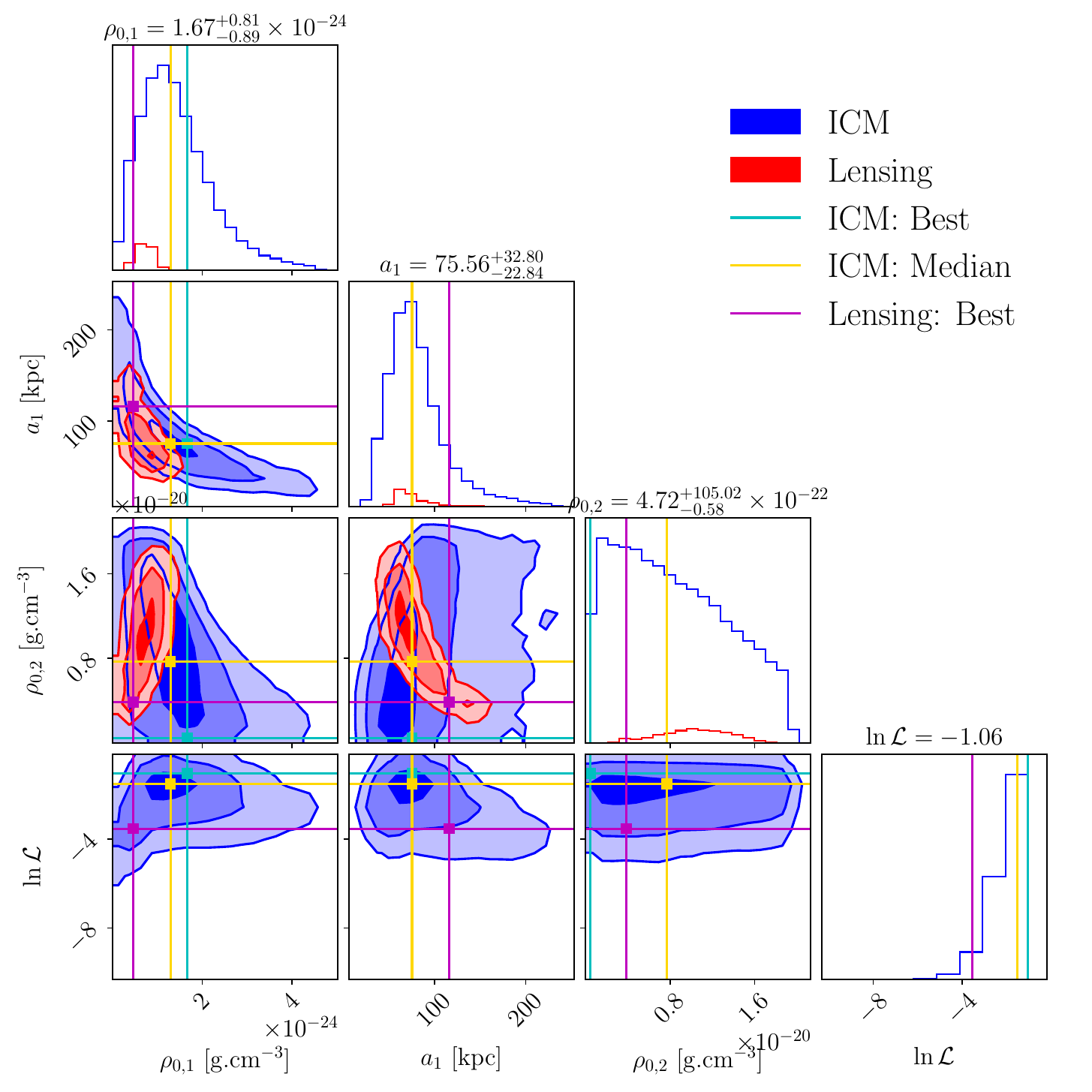}
    \caption{MCMC joint optimisation for \texttt{idPIE} model, for cluster MACS\,J0949: DMH central density $\rho_{0,1}$ and core radius $a_1$, and BCG central density $\rho_{0,2}$. The DMH cut radius is fixed to the fiducial value of 1.5\,Mpc.
    Densities are displayed in g.cm$^{-3}$, distances in kpc.
    \textit{Blue:} Optimisation performed using the available ICM data (X-ray and SZ here).
    \textit{Red:} Strong lensing optimisation.
    \textit{Cyan:} Best ICM optimisation.
    \textit{Gold:} Median of the ICM optimisation.
    \textit{Magenta:} Best strong lensing model (described in Table\,\ref{tab:best_model}).
    Similarly to Fig.\ref{fig:m0242_MCMC_idPIEMD_v874} for MACS\,J0242, the comparison between this optimisation excluding $s_1$ and Fig.\,\ref{fig:m0949_MCMC_idPIEMD_v870} displays the importance of the $s_1$ optimisation.
    }
    \label{fig:m0949_MCMC_idPIEMD_v875}
\end{figure}

\begin{figure*}
\begin{minipage}{0.48\textwidth}
\centerline{\includegraphics[width=1\textwidth]{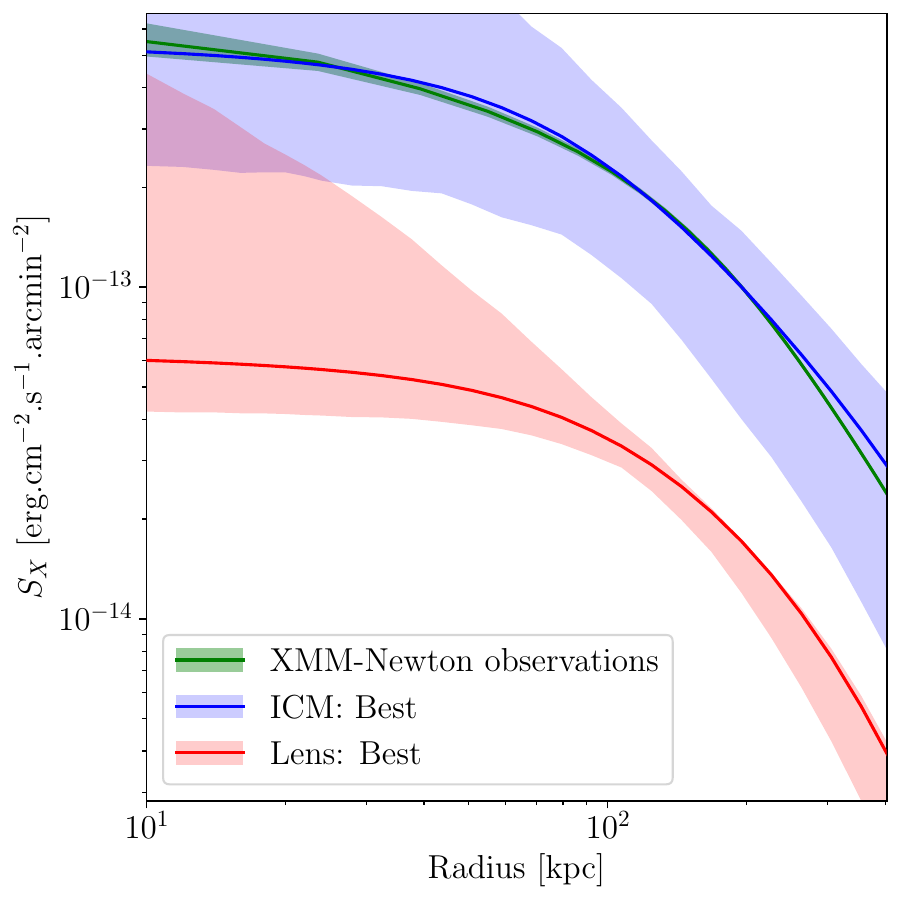}}
\end{minipage}
\hfill
\begin{minipage}{0.48\textwidth}
\centerline{\includegraphics[width=1\textwidth]{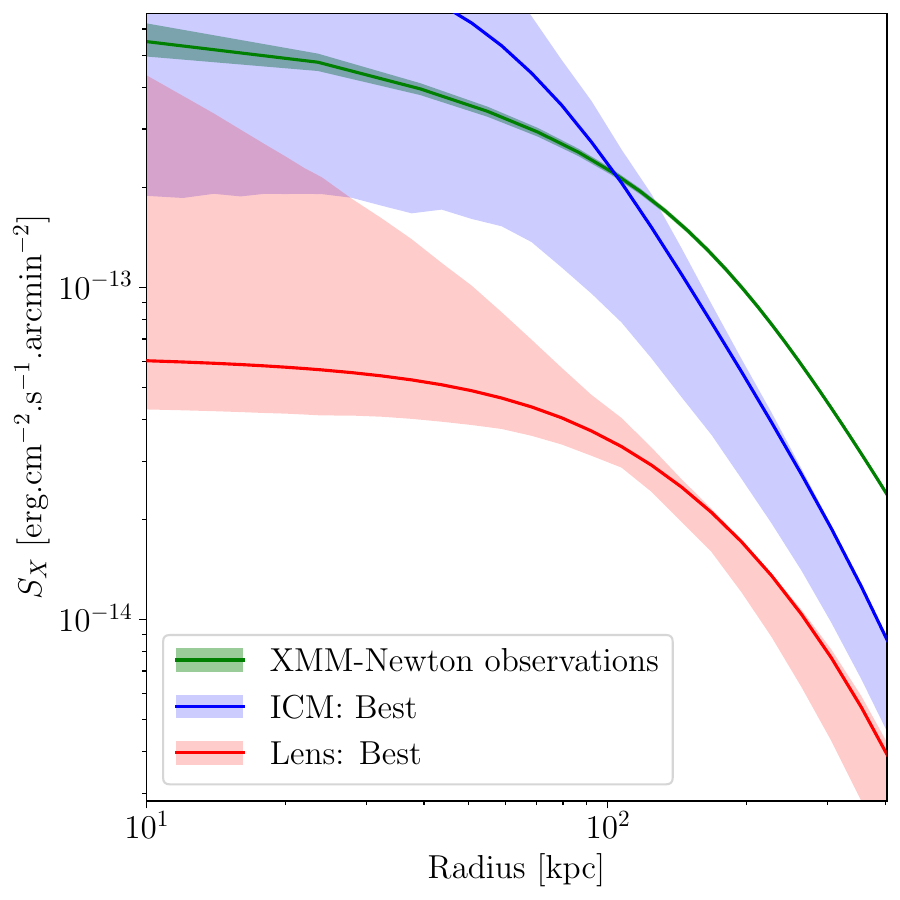}}
\end{minipage}
\caption{X-ray surface brightness $S_X$ for \texttt{idPIE} model, for cluster MACS\,J0949.
\textit{Green:} X-ray surface brightness deprojected profile (assuming spherical symmetry). 
\describecolours
 \textit{Left:} In the case of the optimisation of parameters $\rho_{0,1}$, $a_1$, $s_1$ and $\rho_{0,2}$, as illustrated in Fig.\,\ref{fig:m0949_MCMC_idPIEMD_v870}. \textit{Right:} In the case of the optimisation of parameters $\rho_{0,1}$, $a_1$, and $\rho_{0,2}$, as illustrated in Fig.\,\ref{fig:m0949_MCMC_idPIEMD_v875}.
}
\label{fig:SX_m0949_MCMC_idPIEMD_v870_v875}
\end{figure*}

\begin{figure*}
\begin{minipage}{0.48\textwidth}
\centerline{\includegraphics[width=1\textwidth]{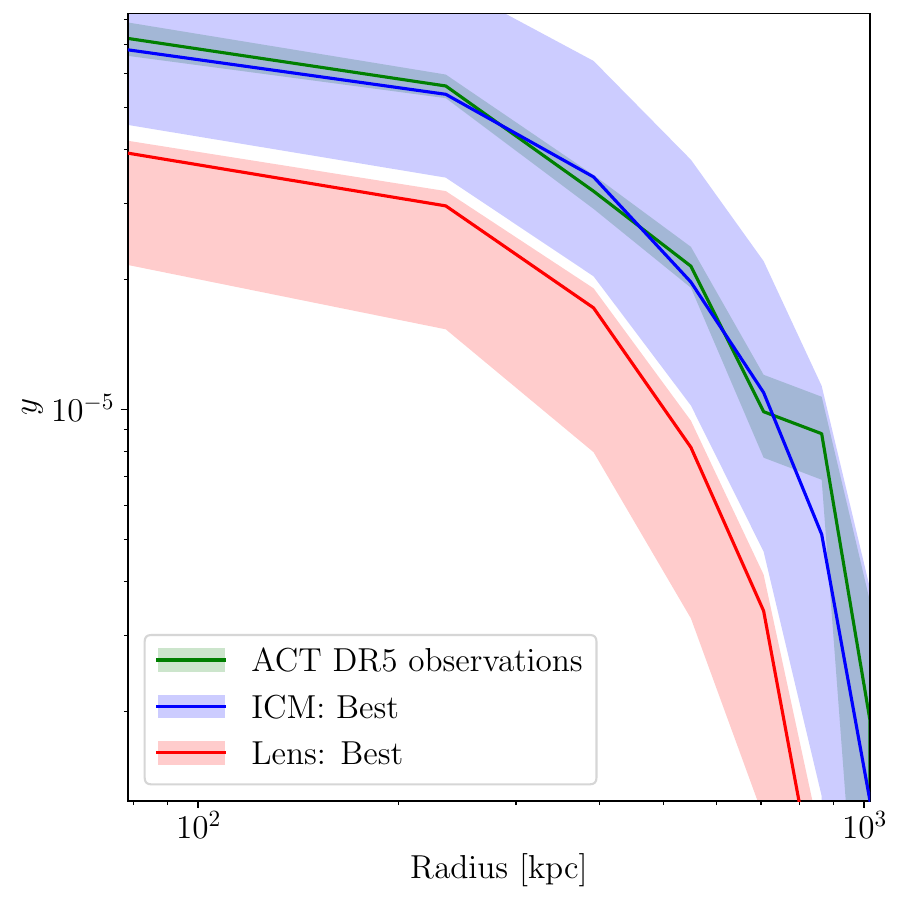}}
\end{minipage}
\hfill
\begin{minipage}{0.48\textwidth}
\centerline{\includegraphics[width=1\textwidth]{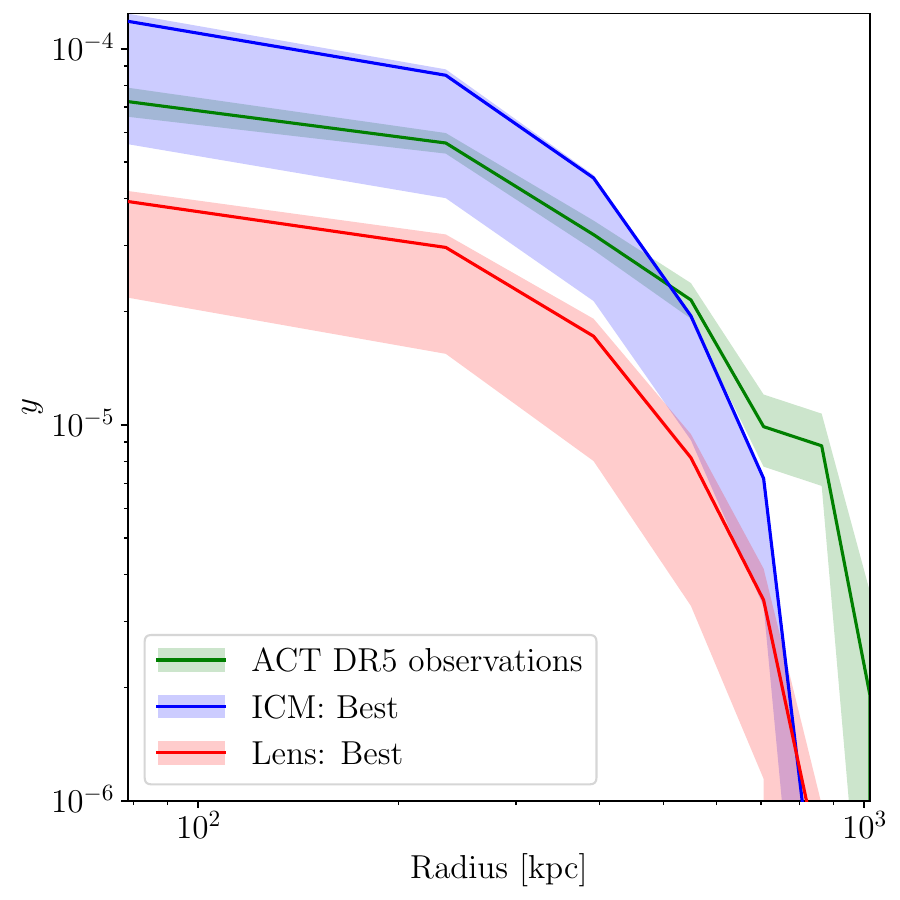}}
\end{minipage}
\caption{Compton parameter $y$ for \texttt{idPIE} model, for cluster MACS\,J0949.
\textit{Green:} SZ effect Compton $y$ parameter observed profile with ACT.
\describecolours
\textit{Left:} In the case of the optimisation of parameters $\rho_{0,1}$, $a_1$, $s_1$ and $\rho_{0,2}$, as illustrated in Fig.\,\ref{fig:m0949_MCMC_idPIEMD_v870}. \textit{Right:} In the case of the optimisation of parameters $\rho_{0,1}$, $a_1$, and $\rho_{0,2}$, as  illustrated in Fig.\,\ref{fig:m0949_MCMC_idPIEMD_v875}.
}
\label{fig:SZ_m0949_MCMC_idPIEMD_v870_v875}
\end{figure*}

\begin{figure}
    \centering
    \includegraphics[width=\columnwidth]{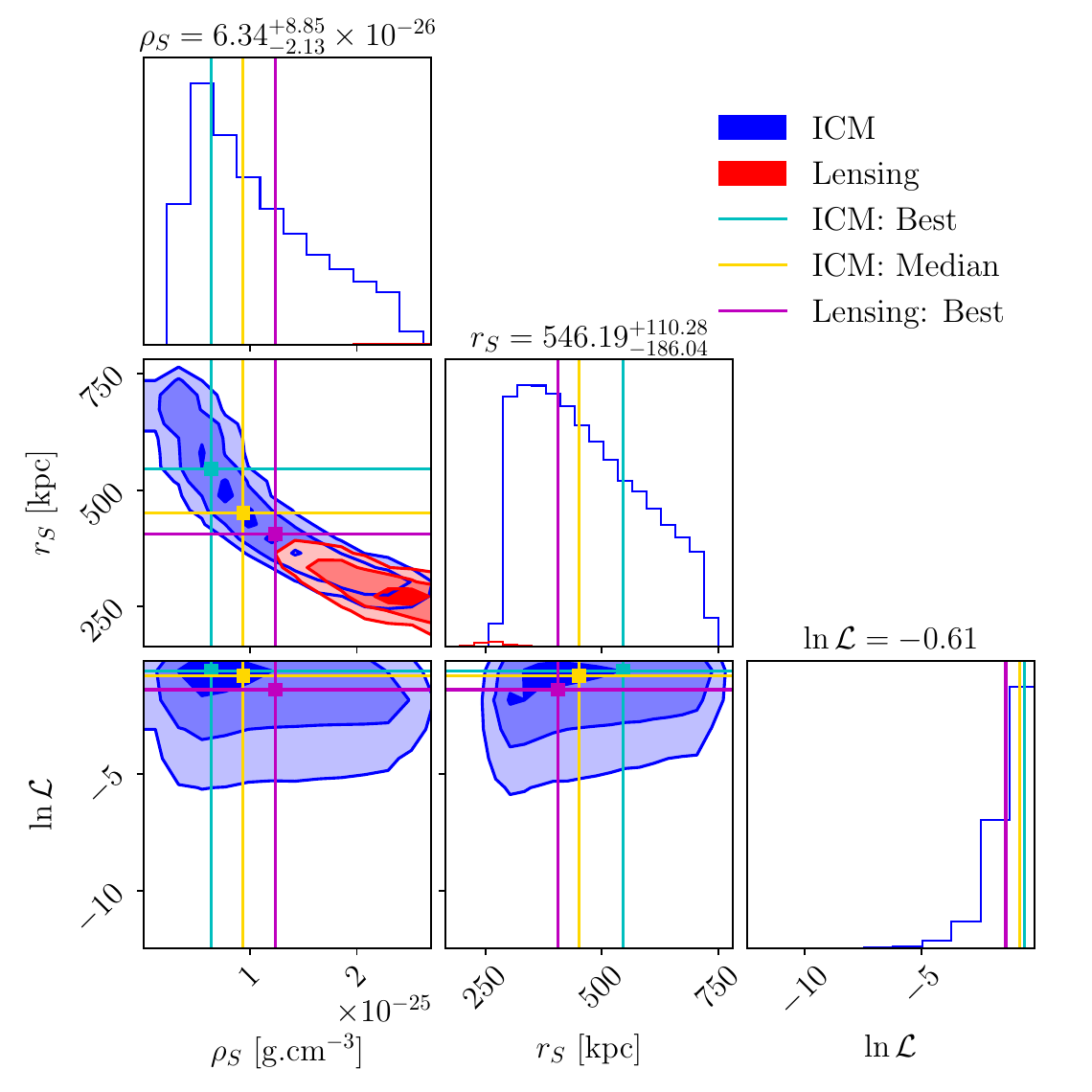}
    \caption{MCMC joint optimisation for \texttt{iNFW} model.
    The individual values for the best optimisation here presented are: $\ln \mathcal{L}_X = -0.58$ and $\ln \mathcal{L}_{SZ} = -0.90$.
    Densities are displayed in g.cm$^{-3}$, distances in kpc.
    \textit{Blue:} Optimisation performed using the available ICM data (X-ray and SZ here).
    \textit{Red:} Strong lensing optimisation.
    \textit{Cyan:} Best ICM optimisation.
    \textit{Gold:} Median of the ICM optimisation.
    \textit{Magenta:} Best strong lensing model (described in Table\,\ref{tab:best_model}).
    }
    \label{fig:m0949_MCMC_iNFW_v877}
\end{figure}

\begin{figure}
\centerline{\includegraphics[width=1\columnwidth]{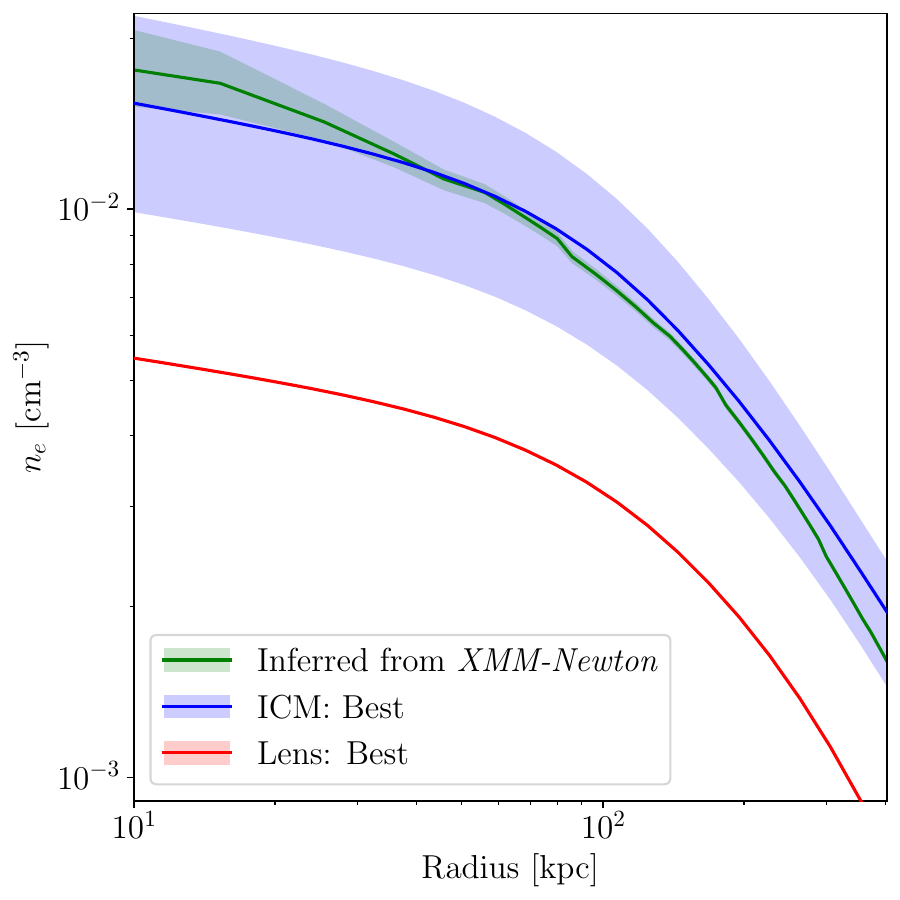}}
\caption{Electron density $n_e$ of the ICM for the \texttt{idPIE} model, for cluster MACS\,J0949.
\textit{Green:} X-ray surface brightness deprojected profile (assuming spherical symmetry). 
\describecolours
The ICM optimisation was here only performed on parameter $s_1$, setting the other profile parameters to their strong lensing value. The comparison between the red and blue outlines the importance of the DMH cut radius optimisation. Indeed, the $n_e$ density does not only change dramatically in the outskirts, but also in the centre of the cluster. This is due to the sensitivity of the function $\mathcal{J}^{-1}_z$ and of the shape of the potential $\Phi$, and is discussed in Sec.\,\ref{subsec:importance_s1}.
}
\label{fig:ne_m0949_MCMC_idPIEMD_v876}
\end{figure}


\bsp	
\label{lastpage}
\end{document}